\documentclass[a4paper,10pt, twocolumn]{article}

\usepackage{graphicx}
\usepackage{amsmath}
\usepackage{amssymb}
\usepackage{authblk}
\usepackage[colorlinks=true, allcolors=blue]{hyperref}
\usepackage{fancyhdr}
\usepackage[titletoc]{appendix}
\usepackage{subcaption}

\usepackage[switch,columnwise]{lineno} 
\usepackage{color}

\newcommand{\micron}{{\upmu\mathrm{m}}}

\usepackage{graphicx}
\usepackage{amsmath}
\usepackage{amssymb}
\usepackage{authblk}
\usepackage[colorlinks=true, allcolors=blue]{hyperref}
\usepackage{fancyhdr}
\usepackage[titletoc]{appendix}

\usepackage{bm}
\usepackage{amsfonts}
\usepackage{epstopdf}
\usepackage{xcolor}
\usepackage{color}
\usepackage{wasysym}
\usepackage[english]{isodate}
\usepackage[switch,columnwise]{lineno} 
\usepackage{ulem}
\usepackage[capitalise]{cleveref}
\usepackage{authblk}
\usepackage{upgreek}

\usepackage{makecell}
\usepackage{hhline}
\usepackage{setspace}
\usepackage{etoolbox}

\usepackage{afterpage}
\usepackage{float}
\usepackage{abstract}

\title{ Advances in laser-plasma interactions \\ using intense vortex laser beams}

\author[1]{Yin Shi}
\author[2]{Xiaomei Zhang}
\author[3]{Alexey Arefiev}
\author[2]{Baifei Shen}

\affil[1]{Department of Plasma Physics and Fusion Engineering, University of Science and Technology of China, Hefei 230026, China}

\affil[2]{Department of Physics, Shanghai Normal University, Shanghai 200234, China}

\affil[3]{Department of Mechanical and Aerospace Engineering, University of California at San Diego, La Jolla, CA 92093, USA}

\begin{document}

\twocolumn[
\maketitle
\vspace{-1cm}
\begin{onecolabstract}
Low-intensity light beams carrying Orbital Angular Momentum (OAM), commonly known as vortex beams, have garnered significant attention due to promising applications in areas ranging from optical trapping to communication. In recent years, there has been a surge in global research exploring the potential of high-intensity vortex laser beams and specifically their interactions with plasmas. This paper provides a comprehensive review of recent advances in this area. Compared to conventional laser beams, intense vortex beams exhibit unique properties such as twisted phase fronts, OAM delivery, hollow intensity distribution, and spatially isolated longitudinal fields. These distinct characteristics give rise to a multitude of rich phenomena, profoundly influencing laser-plasma interactions and offering diverse applications. The paper also discusses future prospects and identifies promising general research areas involving vortex beams. These areas include low-divergence particle acceleration, instability suppression, high-energy photon delivery with OAM, and the generation of strong magnetic fields. With growing scientific interest and application potential, the study of intense vortex lasers is poised for rapid development in the coming years. 

\end{onecolabstract}]

\tableofcontents



\section{Introduction}\label{sec:1}

In the past two decades, there has been a significant surge in the development of high-power, high-intensity laser systems employing chirped-pulse amplification~\cite{cpa1985}, a growth trend that continues unabated. This expansion is largely fueled by the diverse applications facilitated by these laser systems. Ongoing efforts are dedicated to enhancing various laser parameters, aiming to explore novel realms of light-matter interactions. Notably, there is a growing interest in methods for generating high-intensity laser beams with angular momentum. Theoretical and simulation studies in the literature unmistakably demonstrate that these beams have the potential to qualitatively transform laser-plasma interactions. This article provides a comprehensive review of methods suitable for creating high-intensity laser beams with angular momentum, along with the associated regimes of laser-plasma interactions that these lasers enable exploration into. We next introduce different types of angular momentum of light that have been effectively used at lower intensities (\cref{sec 1.1}), define the concept of a vortex beam, and set the stage for the discussion of ultra-intense vortex beams and their interactions with plasmas (\cref{sec 1.2}). 

\subsection{Orbital angular momentum \& optical vortex beam}
\label{sec 1.1}

\begin{figure}[H]
\centering
\includegraphics[scale=0.7]{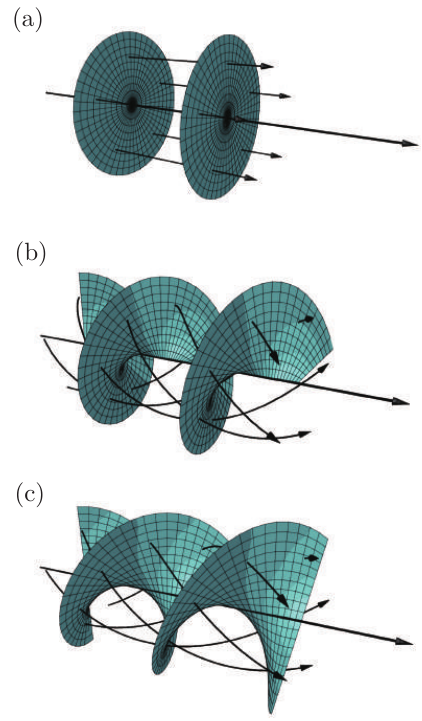}
\caption{Wavefronts or surfaces of constant phase for beams with (a) $l$=0, (b) $l$=1, and (c) $l$=2. The arrows show the direction of the Poynting vector. The beams with azimuthal dependence are referred to as vortex or twisted beams. }\label{fig:twisted_light}
\end{figure}

The angular momentum (AM) of polarized light has two contributions~\cite{Jackson1999, Allen1992, Yao2011, Shen2019_Light}: the spin angular momentum (SAM) and the orbital angular momentum (OAM). The distinction is qualitatively similar to the distinction between the rotation of the Earth around its own axis and its rotation around the Sun. The SAM of light is associated with its polarization. The photons of linearly polarized light carry no SAM. On the other hand, each photon of circularly polarized (CP) light carries SAM equal to $\pm \hbar$. The OAM is associated with the azimuthal phase dependence of wavefronts, so it is completely independent of the polarization. The OAM per photon is $l\hbar$ for a beam with an azimuthal dependence of $\exp \left( -il \phi \right)$. Figure~\ref{fig:twisted_light}provides several examples. The beam in Fig.~\ref{fig:twisted_light}(a) has no azimuthal dependence. The beams in Figs.~\ref{fig:twisted_light}(b) and \ref{fig:twisted_light}(c) have wavefronts with azimuthal dependence. Such beams are usually referrred to as vortex beams or twisted beams. We will adopt the same terminology for this review.

Vortex beams are solutions of the paraxial wave equation that can be conveniently described by the Laguerre-Gaussian (LG) modes.
A unique feature of vortex beams is that their AM can significantly exceed that of a regular CP beam. As shown in \cref{fig:twisted_light}, the presence of the OAM manifests itself in two important ways: the Poynting vector has an azimuthal component and the beam intensity has a hollow profile. Since Allen {\it et al.} published their pioneering work in 1992~\cite{Allen1992}, such vortex light beams have found various applications like entanglement of OAM states in quantum optics~\cite{Mair2001}, information encoding and communication through an added dimension associated with OAM~\cite{Mann2018}, and optical manipulation of particles~\cite{Padgett2011}. 

Techniques for generating low-intensity laser light with OAM do exist and they have even been used to generate extreme ultraviolet (XUV)  beams with OAM~\cite{Zurch2012}. However, they are not applicable to higher intensities because they rely on laser propagation through an optical crystal without inducing damage. The optical crystal is usually called spiral phase plates or forked diffraction gratings~\cite{Yao2011}.  A fundamentally different approach not involving crystals has been proposed for generating XUV beams with OAM~\cite{Primoz2014}. Nonetheless, it would not be suitable for generating high-intensity optical laser beams of $\upmu$m wavelength.

In addition to the vortex beams mentioned above, vortex beams with a special structure can be synthesized by space-time vortex beams in which the topological charge scales (usually linearly) with frequency. The space-time vortex beam of this type was originally termed the light spring by G. Pariente~{\it et al.}~\cite{Pariente15}. Its peak intensity describes a coil structure in space and resembles a spring. 

The OAM carried by a vortex beam can be parallel or transverse to the beam propagation direction. Conventionally, the OAM carried by a vortex beam is aligned with the beam propagation direction and is referred to as the longitudinal OAM.
One can also create a beam with only transverse OAM. Such a beam is called a spatiotemporal optical vortex (STOV). It also has a helical phase, but in the spatiotemporal domain. As a result, its structure is different from the twisted structure of a vortex beam with only longitudinal OAM. The electric field of an STOV rotates around an axis perpendicular to its plane of propagation, just like a photonic wheel~\cite{Aiello2015}.
 
\subsection{Intense vortex in laser-plasma interactions}
\label{sec 1.2}

As mentioned earlier, there has been a remarkable advancement in the field of high-intensity lasers -- the lasers that are sufficiently intense to turn laser-irradaited material into a plasma.
High-intensity laser beams have enabled studies of new physics~\cite{Mourou2006} and development of promising applications like inertial confinement fusion, compact laser wakefield accelerators~\cite{Esarey2009}, ion acceleration~\cite{Macchi2013}, high harmonic generation~\cite{Teubner2009}, and compact laser-driven X-ray sources~\cite{Corde2013}. Motivated by these applications, numerous petawatt (PW) class lasers have been constructed around the world~\cite{danson2015, elinp10pw2022, Li2022}. 
Currently, there are plans for several 10-100~PW high-power laser systems, including SEL(Station of Extreme Light)~\cite{Wang2022sel}, ELI(\href{www.eli-laser.eu.}{Extreme light infrastructure european project}), XCELS(\href{www.xcels.iapras.ru.}{Exawatt center for extreme light studies}), Apollon(\href{www.polytech nique.edu.}{Apollon multi-pw laser users facility}), Vulcan(\href{www.clf.stfc.ac.uk.}{The vulcan 10-pw project}), SULF(the Shanghai Superintense Ultrafast Laser Facility)~\cite{Liang2020sulf}. 
With the construction of these laser systems, the peak light intensity is likely to approach $10^{23-24} \rm{W/cm^2}$ in the foreseeable future. 

Most of the effort has been focused on increasing the power, on-target intensity, total beam energy, and the contrast of the compressed pulse. However, laser light can carry OAM in addition to the linear momentum and energy. 
A high-power high-intensity vortex beam, with its ability to carry OAM, has the potential to provide an additional `control knob', enhancing the versatility of these sophisticated laser systems.
It is important to emphasize that none of the operational laser facilities produce high-intensity vortex beams with OAM. Therefore, it is imperative to develop techniques that make it possible to convert conventional laser beams without OAM into vortex beams with OAM. The use of such OAM-carrying vortex beams offers a way into a relatively unexplored realm of \emph{laser-plasma interactions} where a significant exchange of AM may take place between the laser beam and the plasma and where collective effects play a promienet role. In addition, as a kind of structured light~\cite{Forbes2021}, vortex beams have other properties that can also modify the interaction processes. For example, the hollow shaped intensity of vortex beams can drive hollow wakefields in the plasma.
In the remainder of this review, we show how intense vortex beams can be generated and discuss novel physics regimes that can be explored with these beams. 

\section{Generation of intense vortex beams: theory and experiment}\label{section:vortex-generation}

This section discusses several methods for generating intense vortex beams. 
The term `intense' refers to laser intensity and implies that the laser intensity is sufficiently high, so that the laser can easily ionize atoms and transform a laser-irradiated matter into a plasma. We will focus more on the relativistic vortex, where the electron speed is close to the speed of light and many nonlinear effects occur in laser-plasma interactions. 
Although various optical elements can be used to convert conventional beams into different types of vortex beams, producing intense vortex beams remains a significant challenge. After providing a comprehensive review, this section concludes with a summary and discussion of various methods.

\subsection{Mode conversion before or after focusing}

\begin{figure}[htp]
\centering
\includegraphics[width=8cm]{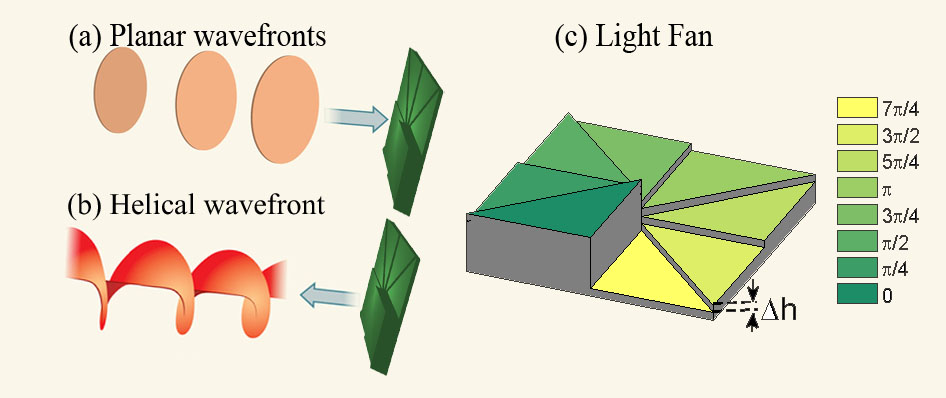}
\caption{A setup for generating a laser beam with helical wavefronts (b) by reflecting a beam with planar wavefronts (a) off a so-called `light fan' (c). The target (`light fan') used in the simulations consists of eight segments that are longitudinally offset by $\Delta h = \lambda/16$ to mimic a spiral phase plate~\cite{Shi2014}. Copyright 2014 by the American Physical Society.}
\label{fig:lightfan}
\end{figure}

A simple and effective method for generating a relativistic vortex beam and demonstrating the OAM effect was proposed in 2014 by Y. Shi~{\it et al.}~\cite{Shi2014}. It is called a `light fan' and involves a structured reflective surface. This method is illustrated in \cref{fig:lightfan} and can be summarized the following way: a relativistic laser pulse (with very high photon density) impinges on a spiral foil (the fan); the pulse is reflected back; during the reflection both the fan and the reflected pulse achieve a large net OAM (\cref{fig:lightfan}). The effectiveness of the 'light fan' method has been confirmed using three-dimensional (3D) particle-in-cell (PIC) simulations. The study detailed in Ref.~\cite{Shi2014} demonstrated, for the first time, that a reflective fan structure is applicable in the relativistic (high intensity) regime. More importantly, the dynamic process of such a structure exhibits new and unique features in the relativistic intensity regime.

The functionality of the `light fan' shown in \cref{fig:lightfan} can be illustrated by approximating the structured plasma surface as a set of eight perfectly reflecting mirrors.     
It is convenient to represent the fields of a laser beam with helical wavefronts as a superposition of orthogonal Laguerre-Gaussian (LG) modes, denoted as ${\rm LG}_{nm}$. In cylindrical coordinates $(r, \theta, x)$, each mode is defined as
\begin{align} 
 {\rm LG}_{nm}(\rho, \phi, x) &= \frac{C_{nm}}{w(x)} \left[ \frac{\rho \sqrt{2} }{w(x)} \right]^{|n-m|} \exp \left( - \frac{\rho^2}{w^2 (x)} \right) \notag\\
 & \times L_{min(n,m)}^{|n-m|} \left( \frac{2\rho^2}{w^2(x)} \right) \nonumber \times  (-1)^{min(n,m)} \notag\\
 & \times \exp \{-i[k\rho^2/R(x)/2 \notag\\
 & +(n+m+1)\psi(x)+(n-m) \phi] \}, 
 \label{eq:LGnm}
\end{align}
with $R(x) \equiv (x_{R}^{2} + x^2)/x$, $kw^2(x)/2 \equiv (x_{R}^{2} + x^2)/x_R$, $\psi(x) \equiv \arctan(x/x_R)$. 
Here $C_{nm}$ is a normalization constant, $k \equiv 2\pi/\lambda$ is the wave number, $x_R$ is the Rayleigh range that is treated as an input parameter, and $L_{p}^{l}(x)$ is a generalized Laguerre polynomial. 
When it comes to OAM, the last term in~\Cref{eq:LGnm} is the most important one. The multiplier $\exp \left[ -i(n-m) \theta \right]$ sets the twist of the wavefronts and, as a result, it determines the OAM of an $\rm{LG}_{nm}$ mode and the hollow shape of the transverse intensity profile. In fact, the OAM per photon for this mode is $l \hbar$, where $l \equiv n-m$ can be appropriately referred to as the twist index.
The Gaussian mode of the incident laser beam is the ${\rm LG}_{00}$ mode. After the reflection off the fan, the structure of the beam changes, as it acquires new LG modes. 
Each LG$_{nm}$ mode in the reflected beam is characterized by an expansion coefficient
$a_{nm} = \int_{0}^{ 2\pi} \rm{LG}_{nm} \rm{exp}(-i\Delta \phi ) \rm{LG}_{00} d \phi$,
where $\Delta \phi = \Sigma_{n=0}^{7}\rm{H}(\phi - n\pi/4)\rm{H}(n \pi/4 + \pi/4 - \phi)n\pi/4$. 
Here $\rm{H}(x)$ is the Heaviside function. Due to the fact that $\Delta \phi \approx \phi$, only the modes with $l = n - m =1$ will contribute most in the $\phi$ integral. The number of intertwined helices can be found to be $l$ = 1. 
The numerical calculation assumes that the Rayleigh range and the waist of the reflected light are equal to those of the incident beam, respectively (i.e., $R_{nm} = R, w_{nm} = w$). The relative weight of the modes is given by $I_{nm} = |a_{nm}|^2$. The calculations show that $I_{10} \approx  64.8\%$ and $I_{21} \approx  13.6\%$. 
The electric field of the reflected light can be approximated by $\sqrt{0.65}\rm{LG}_{10} + e^{i\theta \sqrt{0.14}} \rm{LG}_{21}$, where $\theta$ is the relative phase chosen to make the approximation closer to the simulation results (in our case, $\theta = \pi/2$).

Utilizing the LG representation, it can be shown that the peak OAM density of an LG beam is given by $u = l I_0 /(c\omega)$, where $I_0$ represents the peak laser intensity. 
Consequently, the OAM density of a relativistic twisted beam can become remarkably high due to the high intensity of the relativistic beam.
Addressing a common issue in experiments, which involves the precision of the focal region on the target, simulations in Ref.~\cite{Shi2014} were specifically designed to account for deviations. It was then observed that the distinctive characteristics of twisted light and the associated OAM effects remain prominent as long as the deviations stay within specific range. The theoretical calculation of the relative weight of $\rm{LG}_{10}$ indicates that a larger laser spot size leads to higher tolerance with respect to deviations. In actual experiments, a small sacrificial `light fan' mirror can be placed close to the laser target.

\begin{figure*}
\centering
\vspace{-5pt}
\includegraphics[scale=0.5]{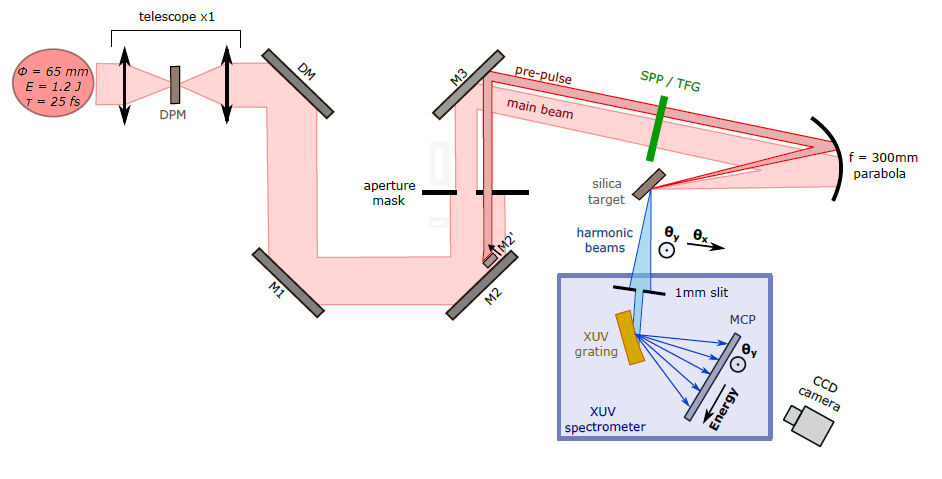}
\vspace{-5pt}
\caption{A scheme for creating an intense vortex beam using a spiral phase plate (SPP). The figure is adopted from Ref.~\cite{Denoeud2017}, where A. Denoeud {\it et al.} used 
an 80-mm-diameter silica SPP that consisted of 8 stair-steps to induce a helical phase
with a 2$\pi$ variation per azimuthal turn ($l$ = 1) at the central laser wavelength (800 nm). Reprinted figure with permission from  A. Denoeud {\it et al.}~Phys. Rev. Lett. 118, 033902(2017) . Copyright 2017 by the American Physical Society.}
\label{fig:Denoeud2017vortex}
\end{figure*}

There has been a concerted effort to generate intense vortex beams experimentally~\cite{porat2022spiral, Denoeud2017, Longman2017, BAE2020103499, wang2020hollow, Longman2022, Burger:23}. In 2022, E. Porat {\it et al}.~\cite{porat2022spiral} demonstrated the feasibility of generating a twisted laser beam of relativistic intensity using a plasma mirror with the `light fan' structure. The structure was realized using 3D direct laser
writing (3D-DLW) fabrication.
In the experiment, an intense plane wave was obliquely incident on the `light fan' to obtain the reflected vortex beam. 
Another approach is to insert a spiral phase plate in the collimated compressed beam just before focusing (in front of an off-axis focusing parabola)~\cite{Denoeud2017,Longman2017,BAE2020103499, wang2020hollow,Longman2022}.
The setup is schematically shown in \cref{fig:Denoeud2017vortex}. This setup was successfully used by A.~Denoeud {\it et al}~\cite{Denoeud2017} who employed an 80~mm diameter silica transmission spiral phase plate with 8 steps to generate an intense vortex beam.
The resulting doughnut-shaped focal spot however did not exhibit perfect azimuthal symmetry. Instead, it had  a double-lobe structure. 
It is shown in Ref.~\cite{Denoeud2017} that beam propagation simulations that account for the spectral width of the laser beam (close to 80
nm), finite number of stair-steps of the phase plate, and the intensity inhomogeneities ($\pm 20\%$) and residual wavefront aberrations ($\pm \lambda=15$) reproduce the shape of the measured focal spot. Furthermore, the simulations show that neither the limited number of stair-steps nor the width of the laser spectrum significantly affect the shape of the focal spot. Rather, the deviation from a perfect doughnut shape is due to residual intensity and phase inhomogeneities of the laser beam before focusing.
In another experiment by W. P. Wang~{\it et al} ~\cite{wang2020hollow}, a reflective phase plate with a size of 230 mm × 170 mm and 32 stair-steps was used to produce an intense vortex beam.
The relativistic vortex beams produced by A.~Denoeud {\it et al}~\cite{Denoeud2017} and W. P. Wang~{\it et al} ~\cite{wang2020hollow} were used for experiments with 
overdense plasmas that are discussed later in~\Cref{subsection:hhg-oam} and~\Cref{subsection:hollow}. Although the spiral phase plates with large size have been used in some proof-of-principle experiments, this may be a fail-safe solution for ultra-high power laser beams, such as the upcoming 10~PW laser systems. The large size of the optical mirror is expensive and fragile. 

Recently, a method that exploits synergy of
adaptive optics and genetic algorithm-guided feedback has been proposed and experimentally demonstrated in the wavefront optimization of intense vortex by M.~Burger~{\it et al}~\cite{Burger:23}. The intensity fluctuations along the perimeter of the target ring-shaped profile can be reduced to $\sim 15 \%$.

By initiating plasma expansion on a flat solid target with a holographic prepulse beam focus, a new type of plasma optics called plasma holograms is introduced by A. Leblanc~{\it et al}~\cite{Leblanc2017}. The basic idea is that the expansion velocity of the plasma on the target surface can be spatially modulated after ionization. During expansion into vacuum, this distribution of expansion velocity leads to a structured preplasma. 
Such structures can act as plasma-based fork gratings that diffract and spatially shape ultra-intense laser beams for several picoseconds. 
A reference beam (Gaussian beam) and an object beam (here a vortex beam) are used to obtain the initial interference pattern on the flat surface of the target. In the picoseconds following ionization, modulation of the plasma expansion velocity results in the growth of a plasma fork grating. The grating can then be used to generate an intense vortex beam by diffracting an incident femtosecond laser beam. 
Experiments have confirmed that the forked plasma gratings induce optical vortices in a femtosecond laser beam, as well as its high-order harmonics, with intensities exceeding $10^{19}\rm{W /cm^{2}}$. It should be pointed out that this technique may suffer from low efficiency and chromatic angular dispersion.

\subsection{Vortex amplification directly in chirped pulse amplification systems}

Back in 2004, K. Sueda {\it et al}~\cite{Sueda2004} at the Institute of Laser Engineering were already considering vortex generation in a high intensity laser system. 
In 2012, C. Brabetz {\it et al}~\cite{Brabetz2012} at PHELIX inserted a spiral phase element into the laser amplifier to produce a hollow, doughnut-like laser focus profile on the target surface. They obtained an intensity of $10^{18}~\rm{W/cm^2}$. 
In 2015, an experiment by the same group at PHELIX has used a vortex laser with FWHM pulse duration of 650 fs and beam energy of 70~J~\cite{Brabetz2015}. The focused laser intensity was in the range between $10^{18}~\rm{W / cm^{2}}$ and $10^{20}~\rm{W /cm^{2}}$. 
Their motivation for generating the hollow beam (which is a vortex beam) was to demonstrate laser focus shaping at a high intensity laser facility. They claim that their simulations show that the hollow beam can reduce the initial angular divergence of a proton beam produce via the target normal sheath acceleration process. 
Since the primary focus of the work was on the hollow shaping of the focused spot, some details regarding the intense vortex beam and laser-plasma interaction have been inadvertently omitted from the publication.

Several papers have explored the generation of terawatt helical lasers through the utilization of chirped pulse amplification (CPA) systems~\cite{chen2022forty,pan2020generation,Feng2023}. As discussed earlier, converting conventional high-power laser beams into vortex beams can be achieved using mode-converting devices. However, the broad bandwidth associated with high-power lasers poses a challenge. The vortex purity diminishes across a wide spectral range as maintaining the same topological charge for different wavelengths is difficult. A potential solution is to directly amplify the vortex seed within CPA systems. Nevertheless, this approach faces a hurdle due to the non-uniform amplification caused by the hollow intensity profile of the vortex. This non-uniformity has the potential to distort both the amplitude and phase of the amplified beam, leading to a laser spot with a lobe structure instead of the desired perfect azimuthal symmetry.
Recently, Z. Chen {\it et al}~\cite{chen2022forty} have proposed another method that enabled them to develop a vortex laser CPA system that delivers pulses with a peak power of 45 TW and a pulse duration of 29~fs. The approach  is implemented on an existing PW system at the Shanghai Institute of Optics and Fine Mechanics (SIOM). The design is based on inserting an optical vortex converter consisting of a $q$-plate, two quarter-wave plates, and a polarizer into the CPA laser system.

\subsection{Stimulated Raman amplification of vortex beams}

J. Vieira {\it et al}~\cite{Vieira2016} have shown that stimulated Raman backscattering can generate and amplify vortex beams to petawatt power in plasmas. The analysis was performed using analytical theory and simulations.  
In this scheme, the plasma plays the role of a nonlinear optical medium.  This nonlinear optical medium can be extended to other materials. Stimulated Raman backscattering can be understood from a three-wave mode coupling mechanism where a pump pulse decays into an electrostatic plasma wave and another electromagnetic wave.
Here the pump laser has frequency $\omega_0$ and wavenumber $k_0$, the electrostatic plasma wave has frequency $\omega_p$ and wavenumber $2k_0-\omega_p/c$. The counter-propagating seed laser with frequency $\omega_1 =\omega_0-\omega_p$ and wavenumber $k_1 = \omega_p/c-k_0$ will be amplified. The additional matching conditions ensure that  the AM carried by the pump is conserved when the pump itself decays into a scattered electromagnetic wave and a plasma wave.

A predictive model for the selection rules of stimulated Raman scattering (SRS) OAM can be formulated using the following general equations~\cite{Mendonca2009}: $D_0 {\bf A}_0 = \omega_p^2 \delta n {\bf A}_1$, $D_1 {\bf A}_1 = \omega_p^2 \delta n^* {\bf A}_0$, and $D_p \delta n =  ({\bf A}_0 \cdot {\bf A}_1) e^2 k_p^2 /(2m_e^2)$, where $D_{0} = c^2 (\nabla^2{\perp} + 2ik_{0}\partial_t)$, $D_{1} = c^2 (\nabla^2{\perp} - 2ik_{1}\partial_t)$, and $D_p = 2i\omega_p \partial_t$. 
Here, ${\bf A}_{0}$ and ${\bf A}_{1}$  are the envelopes of the pump and seed lasers, with the complex amplitude given by ${\bf A}_{0, 1}(t, {\bf r}_{\perp})\exp[ik_{0,1}z-i\omega_{0,1}t] + c.c.$, where $t$ is the time and $z$ is the propagation distance. As ${\bf A}_{0,1}$ are arbitrary functions of the transverse coordinate ${\bf r}_{\perp}$, they can be explicitly expressed as ${\bf A}_{0,1} = A_{(0, 1) x} \exp(i l_{(0, 1) x} \phi) {\bf e}_x + A_{(0, 1) y} \exp(i l_{(0, 1) y} \phi) {\bf e}_y$, where $\phi$ is the azimuthal angle in the transverse plane and ${\bf e}_x$ and ${\bf e}_y$ are unit vectors. 
The amplitude of plasma density perturbations is $\delta n(t, {\bf r}_{\perp}) \exp[ik_{p}z-i\omega_{p}t] + c.c.$, with $k_p=\omega_p/c$ being the plasma wavenumber, $\omega_p = \sqrt{e^2 n_0 /\epsilon_0 m_e}$ the plasma frequency, $m_e$ the electron mass, $\epsilon_0$ the vacuum electric permittivity, and $e$ the electron charge. Assuming a long pulse limit and a pump laser with significantly more energy than the seed laser, one can show that the SRS of the seed beam is given by:
\begin{equation}
    {\bf A}_1(t, {\bf r}_{\perp})  = \left( {\bf A}_1(t=0) \cdot \frac{{\bf A}_0^*}{|{\bf A}_0|} \right) \frac{{\bf A}_0}{|{\bf A}_0|} \cosh(\Gamma t) + C,
    \label{eq:sra1}
\end{equation}
where $\Gamma$, defined as
\begin{equation}
    \Gamma^2({\bf r}_{\perp}) = \frac{e^2 k_p^2 \omega_p^2}{8 \omega_p \omega_1 m_e^2} |{\bf A}_0|^2, 
    \label{eq:sra}
\end{equation}
is the growth rate and $C$ is an integration constant. 
By substituting an explicit dependence of  $({\bf A}_1(t=0) \cdot{\bf A}_0^* ){\bf A}_0$ on
$\phi$ into \Cref{eq:sra1}, one can derive specific selection rules for the OAM.

In Ref.~\cite{Vieira2016}, OAM generation and amplification were analyzed for three different setups and the obtained results can be understood with the help of \cref{eq:sra1}. 
In the first setup, a seed in a single state of OAM $l_{1x}$ is amplified in a plasma by a counter-propagating pump with arbitrary OAM $l_{0x}$. The  plasma wave that is excited in the plasma possesses OAM, with $l_{p} = l_{0x} - l_{1x}$. 
In the second setup, a new OAM mode is generated and amplified using a seed of OAM $l_{1x}$ in a plasma pumped by laser with OAM $l_{0x}$ in $x$ and OAM $l_{0y}$ in $y$.
Indeed, it follows from \cref{eq:sra1} that $l_{1y} = l_{1x} + l_{0y} - l_{0x}$. This will be the new azimuthal mode of the electric field after the pump is exhausted.
In the last setup, a high-order TEM seed with no OAM is amplified to obtain a new OAM laser even though the pump laser has no OAM. 
These results for all three setups have been confirmed using 3D PIC simulations. Furthermore, the AM selection rules for the stimulated
Raman backscattering amplification can be explained using vector diagrams in the ($k_{\phi}, k_z$) plane. Here, $k_{\phi}$ is the azimuthal wavenumber corresponding to the OAM and $k_z$ is the wavenumber in the propagation direction. More details can be found in the supplemental section of Ref.~\cite{Vieira2016}.

The discussed simulations results showcase the great potential of stimulated Raman amplification, but this potential is yet to be realized experimentally. According to Ref.~\cite{Vieira2016}, one can expect production of PW-class vortex beams with OAM when using a seed laser with a 1~mm spot size and achieving amplified intensities of $\sim 10^{17} ~\rm{W/cm^{2}}$.
However, no plasma-based Raman amplification beyond 0.1 TW has been achieved experimentally. R.~Trines {\it et al}~\cite{Trines2020} have recently provided specific criteria and recommendations that one must follow in experiments to increase the power of the amplified beam well beyond  the 0.1~TW threshold. 
It is clear that much more targeted experimental work is needed to achieve a high-power vortex beam using the stimulated Raman amplification. 
Finally, it should be pointed out that Y. P. Wu~{\it et al}~\cite{Wu2024} have recently demonstrated amplification of both vortex and vector pulses though  strongly coupled stimulated Brillouin scattering process using analytical theory and 3D PIC simulations.

\subsection{Vortex beams converted from circularly polarized beams}
\label{subsection:cp-vortex}

As discussed in \Cref{sec 1.1}, the total AM of a light beam consists of intrinsic SAM and OAM, so one can also think of generating a vortex beam by converting SAM into OAM. 
Special optical instruments~\cite{Marrucci2006,Slussarenko:11,Gauthier:19} like a q-plate can be used to realize the spin-to-orbital AM conversion. A q-plate is a thin, optical birefringent phase plate with its fast axis perpendicular to the direction of laser propagation. The optical birefringence can result from different dispersion relations of two laser eigenmodes propagating perpendicular to the external magnetic field $B_0$ through a plasma~\cite{Qu2017}. The refractive index is $n_{\parallel}= \sqrt{1 - \omega_p^2/\omega^2}$ when the polarization is parallel to $B_0$ and is  $n_{\perp}=\sqrt{1 - (\omega_p^2/\omega^2)(\omega^2 - \omega_p^2)/(\omega^2 - \omega_c^2 - \omega_p^2)} $ when the polarization is perpendicular to $B_0$.  Here, $\omega$ is the laser frequency, $\omega_p$ is the plasma frequency, and $\omega_c$ is the electron gyrofrequency. K. Qu {\it et al}~\cite{Qu2017} have shown that an axially symmetric magnetic field generated by anti-Helmholtz coils in a plasma can be used to create a vortex beam from a conventional CP beam. The birefringence induced by the field creates a difference in the refractive index, $\Delta n= n_{\parallel} - n_{\perp}$, that, in turn, produces a spatially modulated phase shift. 
The phase shift can be controlled by controlling the plasma length over which the laser is transported. 3D PIC simulations presented in Ref.~\cite{Qu2017} confirm the functionality of this approach. 

A strong nonlinearity of the laser-plasma interaction, such as high-order harmonic generation (HHG), can also cause conversion from SAM to OAM when the laser is circularly polarized. In the HHG process, $n$ photons with fundamental frequency $\omega_0$ can be converted into a single photon with $\omega_n = n \omega_0$. 
The SAM of each CP photon is $\sigma= \pm \hbar$.  The OAM of an HHG photon of order $n$ is expected to be  $l_{n}= (n-1)\sigma$ when the AM conservation rule is followed.
The results presented in Refs.~\cite{10.1063/1.870766,10.1063/1.871619,PhysRevE.74.046404} indicate that a CP laser, when normally incident on a plasma target,
usually does not generate harmonics. This is attributed to the fact that its ponderomotive force lacks the capability to induce an oscillating current on the plasma surface. However, the situation is qualitatively different for a tightly focused CP laser beam, because such a beam has a strong transverse longitudinal electric field~\cite{Li_2020, shi2021electron}. 
For a tightly focused laser beam, the peak force induced by the oscillating longitudinal electric field can be comparable to the ponderomotive force induced by the laser beam in the longitudinal direction.

\begin{figure}[htp]
    \centering
    \includegraphics[width=0.95\columnwidth]{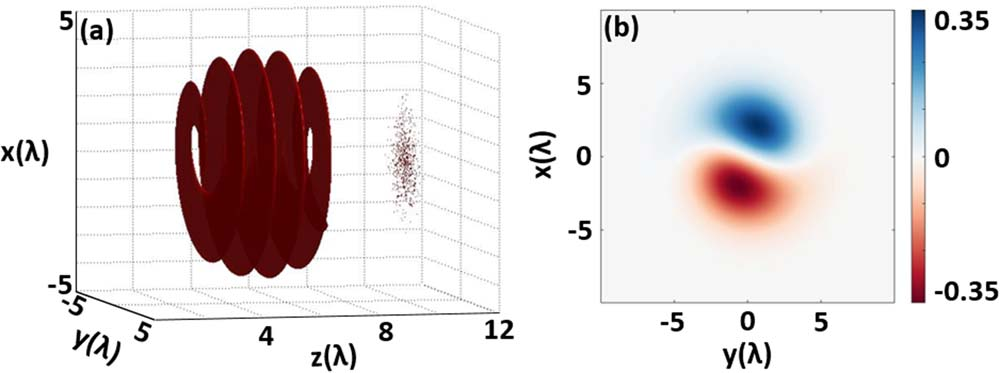}
    \caption{Longitudinal electric field in a tightly focused CP Gaussian laser beam~\cite{Li_2020}: (a) isosurface of the longitudinal field and (b)transverse profile of the longitudinal field in the cross-section of the beam. Copyright 2020 by IOP.}
    \label{fig:cp-lg_Li}
\end{figure}

It should be pointed out that the transverse distribution of the longitudinal field is strongly influenced by the topology of the transverse electric field. 
The longitudinal electric field of a tightly focused CP beam irradiating a sharp surface of a dense plasma drives surface oscillations with a specific phase distribution. The frequency of these oscillations is equal to the laser frequency $\omega_0$, which makes them different from those induced by a linearly polarized (LP) laser. In the case of an LP laser, the oscillations are driven by the ponderomotive force and, therefore, their frequency is 2$\omega_0$.
As a consequence, the tightly focused CP beam drives emission of both odd and even harmonics from the plasma surface, which can be confirmed by PIC simulations. Furthermore, as shown in \cref{fig:cp-lg_Li}, the circular polarization causes the longitudinal electric field to acquire a vortex phase.
This vortex phase influences the phase of the emitted harmonics, leading to the conversion of SAM into OAM.

In an intuitive way, the conversion from SAM to OAM can be explained by examining the longitudinal field of a CP beam. The AM of an electromagnetic field~\cite{Jackson1999} is
\begin{equation} 
    \bm{L} = \varepsilon_0 \int \left( \bm{r} \times \left[ \bm{E} \times \bm{B} \right] \right) d^3 \bm{r},
     \label{eq:Lam}
\end{equation}
where $\varepsilon_0$ is the dielectric permittivity and $\bm{E}$ and $\bm{B}$ are the electric and magnetic fields, respectively. Note that $[\bm{E} \times \bm{B}]$ in~\Cref{eq:Lam} represents the Poynting vector. 
For the CP laser, the AM is due to the vortex distribution of the longitudinal fields. Therefore, the longitudinal fields should play an important role in the AM exchange in the nonlinear laser-plasma interactions.

The discussed HHG from a planar target is based on the well-known relativistic oscillating mirror model. As shown in Ref.~\cite{Wang2019}, the same mechanism can be leveraged to produce a single attosecond pulse with OAM (or a pulse train). There are also alternative approaches to HHG. There is a recently proposed HHG mechanism that relies on light diffraction, termed the ``relativistic oscillating window"~\cite{PhysRevLett.126.134801}. In this process, when a high-power laser beam interacts with a small aperture on a solid foil target, the intense laser field induces a surface plasma oscillation at the periphery of the aperture, effectively functioning as a ``relativistic oscillating window." The diffracted light emerging from this aperture carries high harmonics of the fundamental laser frequency.


\subsection{Summary and discussion}

Opting for a large special mirror which  acts as a phase modulator before focusing could be the most convenient approach for generating a vortex beam. However, the cost of acquiring a large-diameter mirror (approximately $200$ mm in diameter) for a PW-scale laser system or higher might be prohibitively expensive.  
One must also consider that the laser beam may be diffracted by the discontinuity of a helical phase mirror and cause damage to the optics if it propagates a long distance before reaching the target. On the other hand, the helical phase mirror can be damaged by the target debris  if the mirror is too close to the target. Amplification within a CPA system is another option for producing a high quality and high power vortex beam.
Further experimental studies are needed before the generation of high-power vortex beams becomes routine and reliable.  Due to the wavefront distortion, the focusing system is also need to be studied in order to get high intensity spots from high peak power beams. 

When it comes to generating vortex beams at very high power, the use of a `light fan' is probably the best option. These structures can be relatively inexpensive and they are safer to use. 
There can be multiple implementations, with a `light fan' structure in the glass substrate being a popular choice. 
Since the wavefront distortion is always harmful, wavefront optimization of the vortex beam should be considered~\cite{Burger:23}.

Measuring the phase and intensity distributions of an intense vortex beam is a challenge that must be addressed through dedicated studies~\cite{Aboushelbaya2019,Aboushelbaya2020}. The rotational Doppler effect in the scattered laser light~\cite{Zhai2023, Hiroki-MINAGAWA2022} or the electromagnetic field emitted by rotating electrons in an interaction of an intense vortex beam with a plasma can be used as a new diagnostic method.
Finally, discovery and development of novel applications can stimulate further interest in generation of high quality and high power vortex lasers.


\section{Intense vortex interaction with plasma}
\label{section:vortex}

While the generation of intense vortex beams remains a vibrant area of ongoing research, it is noteworthy that experiments with these beams are already feasible. Concurrent theoretical and simulation studies are underway, offering crucial insights into the interactions between intense vortex beams and plasmas. Moreover, these studies play a pivotal role in guiding future experiments. Published findings from these studies underscore that intense vortex beams exhibit a diverse range of phenomena, attributed to distinctive properties like phase front twist (see~\Cref{subsection:hhg-oam} and~\Cref{subsection:oamwave}), orbital angular momentum (OAM) bearing (see~\Cref{subsection:Bfield-oam} and ~\Cref{subsection:gamma-oam}), hollow intensity distribution (see~\Cref{subsection:hollow}), and spatially separated longitudinal fields  (see~\Cref{subsection:longi-field}). These properties have the potential to fundamentally alter the dynamics of laser-plasma interactions, as evidenced by the already available literature. The ongoing synergy between theoretical studies and experimental endeavors promises to deepen our understanding of this intriguing area of research.


\subsection{Generation of high-order vortex harmonics by an intense vortex beam}
\label{subsection:hhg-oam}

In laser-plasma interactions, nonlinear and collective phenomena can often produce higher-frequency radiation. 
It is known that a solid-density plasma driven by an intense infrared laser can generate high harmonics in the extreme ultraviolet (XUV) region~\cite{Teubner2009}. Furthermore, in the scenario where an intense linearly polarized vortex beam with a low-order mode irradiates a thin solid target, X. M. Zhang {\it et al}~\cite{Zhang2015} observed that both the reflected and transmitted light encompass high-order harmonics, and their phases exhibit a correlation with the harmonic order.
In this nonlinear process, the OAM carried by the driving vortex beam is easily transferred to the generated harmonics. The azimuthal mode number of the generated optical vortex beam can easily reach ten and even extend to a few tens. In addition, the plasma as a nonlinear medium shows superiority in the damage threshold compared with conventional optical components. 

\begin{figure}[H]
\centering
\vspace{-5pt}
\includegraphics[scale=0.4]{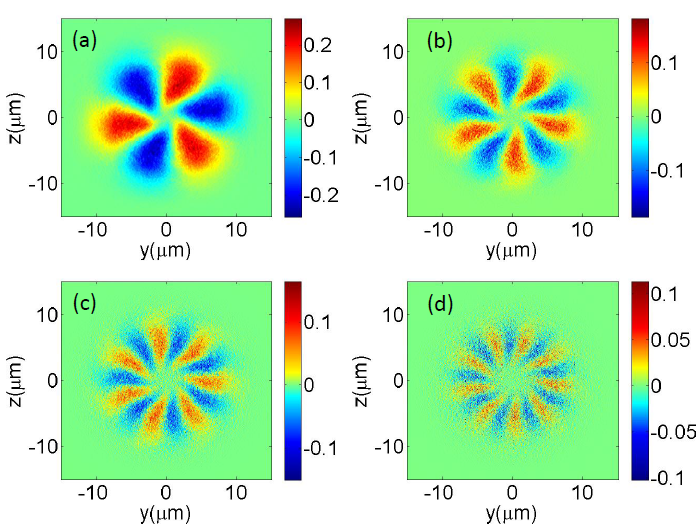}
\vspace{-5pt}
\caption{Generation of high-order vortex harmonics by reflecting an $LG_{10}$ beam off a flat solid target~\cite{Zhang2015}. The panels show $E_y$ of the (a)~third, (b)~fifth, (c)~seventh, and (d)~ninth harmonics in the cross section of the reflected beam. Copyright 2015 by the American Physical Society.}
\label{fig:hhgoam}
\end{figure}

To elucidate the process of harmonic generation by a vortex beam, we employ a simplified model often referred to as the oscillating mirror model. In this model, the incident beam reflects off a surface whose oscillations are created by the laser itself. 
A conventional intense linearly polarized laser beam drives surface oscillations at twice the laser frequency $\omega_0$  because the ponderomotive force induced by the laser and experienced by the surface is proportional to $1 - \cos(2\omega_0t)$. In the case of a vortex beam, the ponderomotive force has additional azimuthal dependence. Consequently, the longitudinal position of the reflecting surface becomes dependent on the azimuthal angle as well. 
The reflected field produced by such a vortex mirror is given by 
	\begin{equation}
	      E \propto \sin[\omega_0 t + l \phi + \kappa \sin(2(\omega_0t + l \phi))], 
	  \label{eq:vortexphase}
	  \end{equation}
where the value of parameter $\kappa$ is set by the amplitude of the oscillations. The $\kappa$-term is the phase shift induced by the oscillating mirror. The field given by \Cref{eq:vortexphase} contains higher harmonics. To show this, we take into account that $E$ is a periodic and antisymmetric function of the phase $\psi = \omega_0t + l \phi$ that can be represent as a Fourier series
\begin{equation} \label{eq:6}
    E \propto  \sum_{m = 1}^{\infty} B_m \sin(m \psi ),
\end{equation}
where $B_m$ are expansion coefficients. A standard calculation of $B_m$ reveals that $B_m = 0$ for all even $m$. We can thus re-write \Cref{eq:6} as
\begin{equation}
	      E \propto \sum_{n=0}^{\infty}\sin[(2n +1)(\omega_0t + l \phi)],
	  \label{eq:hhg}
\end{equation}
making it explicit that the reflected field contains only odd-order harmonics. This is a familiar result from previous studies on HHG from laser-irradiated solid targets. A new aspect in the case of vortex beams is that the twisted phase of the harmonics depends on the harmonic order. As shown in \Cref{fig:hhgoam}, the topological charge of the $m$-th order harmonic is $l_{m} = ml$.
For example, an incident beam with $l = 1$ produces radiation at the third harmonic, with the third harmonic's field resembling  the $LG_{30}$ mode.

Examining the nonlinear process of HHG through the lens of energy and AM conservation provides an alternative perspective. In this context, a photon of the $m$-th order harmonic is generated by the interaction of $m$ photons from the original beam at the fundamental frequency. Each of these $m$ photons carries an OAM equal to $\hbar$ per photon. As a result of this collective interaction, the $m$-th order harmonic inherits an OAM of $m \hbar$ per photon. 
So far we have discussed a scenario where a single beam interacts with a thin target. Incorporating a second vortex beam provides an important advantage. Specifically, using two counter-propagating vortex beams interacting within a solid thin foil, the contribution of each irradiating beam to the generation of harmonics can be distinguished using the conservation of the AM of the photons~\cite{Xiaomei2016}. This is usually impossible for HHG without an optical vortex. 

Approaches for generating high-order vortex harmonics that involve fractional vortex beams~\cite{Li2018} and CP beams~\cite{Li_2020, Wang2019, PhysRevLett.126.134801} have been proposed as well. S.~Li {\it et al}~\cite{Li2018} demonstrated HHG during reflection of fractional vortex beam, observing conservation of average OAM. 
HHG can also be achieved via spin-to-orbital conversion of AM during an interaction of an intense CP beam with a plane foil ~\cite{Li_2020, Wang2019}. In this case, the generation of vortex harmonics is attributed to a vortex-like oscillation of the plasma surface that is driven by the vortex longitudinal electric field of the CP beam~\cite{Li_2020}. Finally, HHG can be achieved using diffraction of a beam that is focused on a foil with a small aperture. L.~Yi~\cite{PhysRevLett.126.134801} observed HHG in this setup that was termed the relativistic oscillating window.
Some of these works have already been discussed in~\Cref{subsection:cp-vortex} within the context of mechanism suitable for the generation of vortex beams. 

The generation of high-order OAM harmonics has also been studied in the context of stimulated Raman backscattering in plasmas.  
Theory and PIC simulations both reveal the capacity to adjust the order of OAM harmonics through a selection rule determined by the initial OAM of the interacting waves. J.~Vieira {\it et al}~\cite{Vieira2016a} examined a specific scenario where, rather than being a single OAM mode, the pump beam is a superposition of two distinct OAM modes. We refer to the corresponding $l$ values of these modes as $l_{00}$ and $l_{01}$, with $\Delta l \equiv l_{00} -l_{01}$. It was shown that in this setup the order of generated high OAM harmonics is given by $l = l_1 + m \Delta l$, where $m$ is an integer and $l_1$ is the $l$-value in the initial vortex seed beam that is assumed to be a single OAM mode.

\begin{figure}[H]
\centering
\vspace{-5pt}
\includegraphics[scale=0.27]{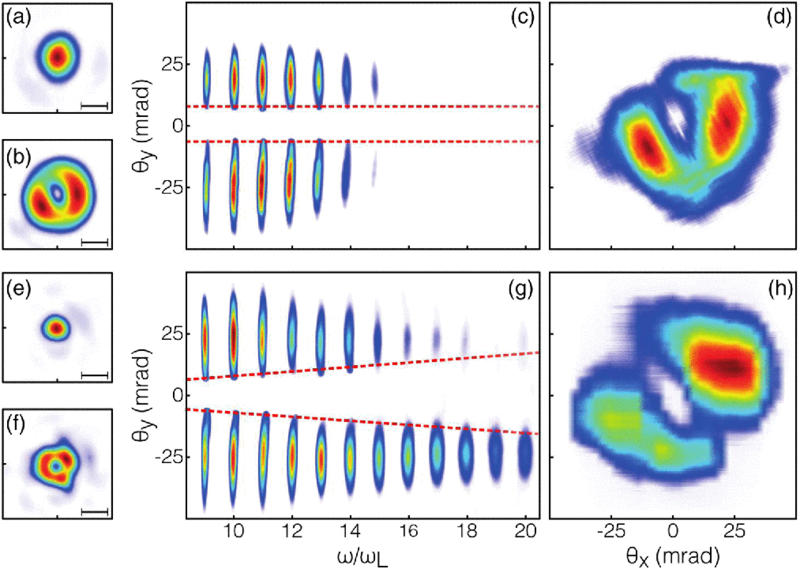}
\vspace{-5pt}
\caption{High-harmonic generation by laser reflecting off a plasma mirror. Left panels: laser focal spots without (a\&e) and with (b\&f) the inclusion of a spiral phase plate in the path of a 25~mm (up) and 45~mm (down) diameter laser beam. The bar scale indicates a length of 10~$\micron$.
Central panels: spectrally resolved 1D angular profiles of the harmonic beams generated by an incident vortex beam. The beam in (c) is generated in the Coherent Wake Emission (CWE) regime, whereas the beam in (g) is generated in the  Relativistic Oscillating Mirror (ROM) regime (g).
Right panels:  reconstructed 2D angular intensity profiles of the 12th harmonic beam in the CWE (d) and ROM (h) regimes~\cite{Denoeud2017}. Reprinted figure with permission from  A. Denoeud {\it et al.}~Phys. Rev. Lett. 118, 033902(2017) . Copyright 2017 by the American Physical Society.}
\label{fig:Denoeud2017prl}
\end{figure}

The harmonic generation process discovered by J.~Vieira {\it et al}~\cite{Vieira2016a} relies on having two distinct OAM modes in the pump beam. In SRS, the seed beam beats with the pump beam, exciting a Langmuir plasma wave that absorbs the OAM difference between the pump and the seed. Initially, the seed has just a single OAM mode, so its beating with the pump produces two plasma waves with $l_{00} - l_1$ and $l_{01} - l_1$. These plasma waves beat with the modes of the pump, producing harmonics. For example, the beating of the plasma wave with $l_{01} - l_1$ and the pump mode with $l_{00}$ produces a new seed beam mode with $l = l_{00} - (l_{01} - l_1) = l_1 + \Delta l$. The pump mode with $l_{01}$ similarly produces a new mode with $l_1 - \Delta l$ by beating with the plasma wave that has $l_{00} - l_1$. After being generated, each of these so-called side-bands beats with the pump, producing high-order OAM modes in the plasma wave. These modes then generate new seed modes and the process continues, leading to the creation of modes with $l_1 \pm m \Delta l$ in the seed beam. This qualitative description has been substantiated by a nalytical theory and PIC simulations that can be found in Ref.~\cite{Vieira2016a}.

{Another work worth mentioning is the generation of mid-infrared vortex lasers, which can be realized by photon deceleration driven by intense vortex lasers. 
The intense laser-plasma interaction can act as a nonlinear transformer, slowing down as well as accelerating photons.  A tailored plasma density structure acting as a photon decelerator of Gaussian beams has been proposed~\cite{Nie2018} and experimentally realized~\cite{Nie2020} by Z.~Nie~{\it et al}.
Using the same mechanism, X.~L.~Zhu~{\it et al}~\cite{Zhu2019, Zhu2020, Zhu2021} proposed to generate single-cycle terawatt vortex pulses driven by intense infrared vortices.}

Some of the first proof-of-principle experiments on HHG in interactions of intense vortex beams with plasma mirrors were carried out by the group of F.~Quéré~\cite{Denoeud2017,Leblanc2017}. A.~Denoeud~{\it et al}~\cite{Denoeud2017} were able to select between two HHG mechanisms - coherent wake emission (CWE) or the relativistic oscillating mirror (ROM) - by controlling the length $L$ of the plasma density gradient and the peak laser intensity $I$. 
To obtain a vortex beam with a central laser wavelength of 800 nm for this experiment, an 8-step silica transmission spiral phase plate with a diameter of 80 mm was inserted into the collimated compressed beam just before focusing, as shown in \cref{fig:Denoeud2017vortex}.

The HHG via CWE was achieved using a relatively moderate laser intensity, $I \approx 5\times 10^{17}~\rm{W/cm^2}$ , and a very steep density gradient, $L < \lambda /20$). In this case, the maximum harmonic order is imposed by the maximum plasma frequency in the target. The HHG via ROM was achieved using a high laser intensity, $I \approx  10^{19}~\rm{W/cm^2}$, and a longer density gradient,  $L \approx \lambda/15$. In this case, the maximum harmonic order depends on the interaction conditions, since $L$ and $I$ determine the energy of the relativistic electrons ejected from the plasma mirror and these electrons are responsible for the ROM harmonic generation. 

\Cref{fig:Denoeud2017prl} shows the results obtained by A.~Denoeud~{\it et al}~\cite{Denoeud2017} with and without the spiral
phase plate for the two HHG mechanisms. These results indicate (1) that the incident beam is indeed a vortex beam and (2) that a part of its OAM has been transferred to the harmonics. It must be pointed out that these results provided the first experimental observation of intense vortex beams at high laser intensities, which effectively demonstrated the feasibility of ultra-high intensity laser-plasma interaction experiments with vortex beams.

From a fundamental point of view, the work by A.~Denoeud~{\it et al}~\cite{Denoeud2017} provided the first experimental confirmation of the OAM conservation rule at ultra-high intensities. Specifically, it showed that the topological charge of the $n$-th harmonic in $n$ times the topological charge of the incident laser.
Related experimental work on conservation of AM was been performed by C.-K.~Huang~{\it et al}in the non-relativistic regime~\cite{Huang2020}. They examined an interaction of a CP vortex beam with an underdense plasma. By measuring the helical phase of the second harmonic generated in the plasma, C.-K.~Huang~{\it et al} observed the conversion of SAM to OAM and verified that the total AM of photons is conserved.
It should be noted that the high-harmonic vortex beams discussed here are essentially XUV vortex beams. They can be used as a powerful tool in the area of optical information to study entanglement states and to enhance atomic transitions in studies of cold atoms. 

In this sub-section, we have reviewed multiple regimes where harmonics are generated upon reflection of a vortex beam off a plasma mirror. A vortex beam carries OAM, which is a specific form of AM. The fact that AM is a pseudo-vector~\cite{Jackson1999}, rather than a regular vector like the momentum, can have profound implications on reflection of vortex beams. For example, L.~Zhang~{\it et al}~\cite{Zhang2016} discovered that an obliquely incident vortex beam no longer follows the optical reflection law. To explain the phenomenon, let us first consider perfect reflection of a normally incident vortex beam from a plasma foil (isotropic in its plane). Because the AM is a pseudo-vector, its direction in the reflected beam is the same as in the incident beam, so the foil gains no AM. In contrast to that, the foil does necessarily gain momentum as a result of the reflection. In the case of oblique incidence, the component of AM normal to the surface again does not change its sign, but this automatically means that the component parallel to the surface flips sign. The implication is that the mirror acquires AM during oblique reflection. This AM leads to an asymmetric deformation of the mirror surface which then causes the beam to be deflected out of the plane of incidence with an experimentally observable deflection angle~\cite{Zhang2016}.
The described mechanism has been illustrated in Ref.~\cite{Zhang2016} using 3D PIC simulations and analytical modeling employing the Maxwell stress tensor. The deformation occurs as a consequence of the foil's rotational symmetry breaking, driven by the asymmetrical shear stress induced by the vortex beam. The shear stress, dependent on the twist index $l$, serves as an intrinsic attribute of the vortex beam and assumes a significant role as a ponderomotive force in the interaction between the relativistic vortex laser and the material.


\subsection{OAM in wave coupling instabilities and twisted plasma waves}
\label{subsection:oamwave}

The question of whether a helical laser beam can transfer some of its OAM to a plasma and whether this transfer can lead to excitation of helical electron plasma waves, i.e., electron plasma waves with azimuthal dependence, has attracted significant attention~\cite{Mendonca2009, Vieira2016, Gao2015, Nuter2022, Ji2023TBD}.

In 2009, T.~Mendonça {\it et al.}~\cite{Mendonca2009} examined the exchange of OAM between electromagnetic and electrostatic waves in the stimulated Raman and Brillouin backscattering processes. This work introduced OAM states for plasmon and phonon fields for the first time~\cite{Mendonca2009}. Subsequently, J.~Vieira {\it et al.}~\cite{Vieira2016} provided a detailed example for the amplification and generation of twisted laser pulses by SRS. At intensities ranging from $10^{14}$ to $10^{16}~\rm{W /cm^{2}}$, which are pertinent to major inertial confinement fusion (ICF) facilities, a vortex laser beam has been shown to selectively transfer its OAM to either electrons or ions within a plasma. This selective transfer occurs through a laser plasma instability (LPI), as outlined by W.~Gao {\it et al.}~\cite{Gao2015} in the context of Brillouin amplification. 
Additionally, the growth of a two-plasmon decay driven by a vortex beam has been explored using both 3D fluid simulations and analytic theory by Y.~Ji {\it et al.}~\cite{Ji2023TBD}.
Under the assumption that the electron plasma waves are locally stimulated by multiple plane-waves, the proposed theory predicts that the maximum growth rate is proportional to the peak amplitude of the pump laser field.

The dispersion relation of electron plasma waves is one of their key characteristics. Its knowledge is critically important, as the dispersion relation influences a number of phenomena ranging from wave-particle interactions to instabilities that involve plasma waves. The derivation is however far from straightforward because of the azimuthal dependence. The first attempt at the derivation using a linearized Vlasov equation was presented by T.~Mendonça {\it et al.}~\cite{Mendonca2009a, Mendonca2012, Mendonca2012a, Mendonca2017, Mendonca2017a}. These influential publications demonstrated that there are indeed qualitative differences between helical and conventional electron plasma waves, which has stimulated further research on the topic. 

The basic idea used by Mendonça~{\it et al.}~\cite{Mendonca2009a, Mendonca2012, Mendonca2012a, Mendonca2017, Mendonca2017a} was to express general solutions of the electrostatic paraxial equation in terms of LG functions. 
The perturbed electron density with the equilibrium value $n_0$ is assumed in the form $\tilde{n}({\bf r}, t) = \tilde{n}_0({\bf r}) \exp(ikz - i \omega t)$, where the amplitude $\tilde{n}_0({\bf r})$ is a slowly varying function of the spatial coordinates and describes the wave profile:
\begin{equation}
    \left( \nabla _{\perp}^{2} + 2ik\frac{\partial}{\partial z} \right) \tilde{n}_0({\bf r})=0,
\end{equation}
where $k$ is the longitudinal wavenumber and  $\nabla_{\perp}^{2} = (1/r) \partial /\partial r (r\partial /\partial r) + (1/r^2)\partial^2 /\partial \phi^2$ is the transverse Laplacian operator in cylindrical coordinates.
It is well known that the LG functions can form a set of orthogonal and possibly normalized functions in the transverse plane ($r, \phi$). Each individual LG mode can then be defined as 
\begin{equation} \label{eq:8}
    \tilde{n}_0(r, \phi, z) = \tilde{n}_0 F_{pl}(r, z)\exp(il\phi),
\end{equation}
where $p$ and $l$ are the radial and angular mode numbers, respectively, $\phi$ is the azimuthal angle. The LG function in \cref{eq:8} is defined as
\begin{equation}\label{eq:lg_func_mendoca}
    F_{pl}(r, z) = \frac{1}{2\sqrt{\pi}}\left[\frac{(l+p)!}{p!}\right]^{1/2}X^{|l|}L_{p}^{|l|}(X)\exp(-X/2),
\end{equation}
where $X=r^2/w^2(z)$, $w(z)$ is the beam waist, and $L_{p}^{|l|}(X)$ is the associated Laguerre polynomial. The electrostatic potential associated with the LG density perturbation $\tilde{n}$ is assumed to be given by $V({\bf r},t) = V_{pl}(r,z)e^{il\phi}\exp(ikz-i\omega t)$. 
The electrostatic field ${\bf E} ({\bf r}, t) $= ($-\partial V/\partial r$, $-r^{-1} \partial V/\partial \phi$, $-\partial V/\partial z$) now has a twist. It is expected that these LG solutions will be able to describe electrostatic waves with finite AM. In contrast to the photon case, plasmons have no spin, so the plasmon AM coincides with its total AM~\cite{Mendonca2009, Mendonca2009a}. The kinetic description of electron plasma waves with OAM is also given for dispersion relation and twisted Landau resonance discussions~\cite{Mendonca2012a}.  All these discussions are valid only in the paraxial approximation. 

Recently, the previous derivation was revisited by D.~Blackman {\it et al.}~\cite{Blackman2019a, Blackman2019b, Blackman2020, Blackman2022} and it was shown that the coupling between helical and radial modes due to transverse electron motion has to be taken into account. The derivation that includes the coupling while assuming that the transverse scale $w_b$ is much greater than the longitudinal wavelength has yielded an explicit closed-form dispersion relation
\begin{equation} \label{eq:disp_max}
    \frac{\omega^2}{\omega_{pe}^2} = 1 + 3k^2\lambda_{D}^2 + \frac{2p+|l|+1}{k^2w_b^2},
\end{equation}
where $\lambda_D$ is the Debye length. The derivation was performed for a Maxwellian electron distribution. It follows from \cref{eq:disp_max} that the helicity ($l$) enhances the qualitatively different dependence on $k$ introduced by the finite transverse scale of the wave, increasing the dispersion. This dispersion relation differs from that derived by T.~Mendonça {\it et al.}~\cite{Mendonca2012a} which, naturally, leads to a different Landau damping rate~\cite{Blackman2019a} for the plasma wave.
{Furthermore, the plasma waves can be structured with more general space-time correlations~\cite{Palastro2024}. }

\begin{figure}[H]
    \centering
    \includegraphics[width=0.99\linewidth]{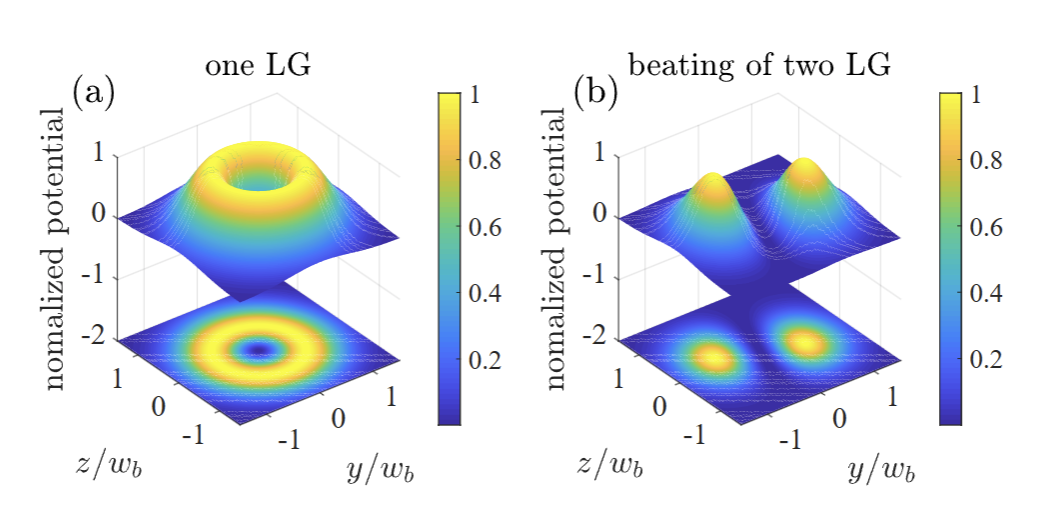}
	\caption{\small Structure of the ponderomotive potential $\varphi_{\text{pond}}$ (in a. u.) of one LG laser pulse (a) and two beating LG laser pulses (b) in the transverse plane ($y,z$). The ponderomotive force ${\bf F}_{\text{pond}} = - \nabla \varphi_{\text{pond}}$ acquires an azimuthal component when two beating waves are present~\cite{shiJUSTC}. }
	\label{fig:twistforce}
\end{figure}

The theory developed by D.~Blackman {\it et al.}~\cite{Blackman2019a, Blackman2019b, Blackman2020, Blackman2022} also provides additional insight into the generation of quasistatic magnetic fields by vortex beams that was first reported by Y. Shi {\it et al.}~\cite{Shi2018, shiJUSTC}.
Specifically, it was discovered that two
co-propagating vortex laser beams with different angular modes, frequencies, and wavelengths generate a quasistatic axial magnetic field while driving a twisted plasma wave carrying OAM.
The effect can be illustrated using a pair of LG waves that have opposite twist indices, $l_1 = - l_2 = l$, and the same radial dependence. The waves are beating with slightly different frequencies: $\omega_1 = \omega_0 + \Delta \omega /2$ and $\omega_2 = \omega_0 - \Delta \omega / 2$, where $\Delta \omega \ll \omega_0$. The amplitude of their superposition now has a multiplier that slowly varies with time and has explicit azimuthal dependence:
\begin{equation}
    E \propto \cos \left( \frac{\Delta \omega}{2} t - \frac{\Delta k}{2} x + l \phi \right),
\end{equation}
where $\Delta k = k_1 - k_2$ is the difference between the wave numbers of the two waves. The ponderomotive potential is then no longer cylindrically symmetric, with
\begin{equation} \label{phi_pond}\varphi_{\text{pond}} \propto \frac{1}{2} \left[ 1 + \cos \left( \Delta \omega t - \Delta k x + 2l \phi \right) \right].
\end{equation}
A snapshot of this potential is shown in ~\cref{fig:twistforce}(b) next to the potential for a single LG beam to highlight the azimuthal dependence. The resulting ponderomotive force, ${\bf F}_{\text{pond}} = - \nabla \varphi_{\text{pond}}$, has an azimuthal component and it is slowly rotating in the cross section of the propagating beams, i.e. in the $(y,z)$-plane. In contrast to this, ${\bf F}_{\text{pond}}$ produced by a superposition of conventional beams without OAM ($l = 0$) has no azimuthal component. Such a twisted ponderomotive force creates a twisted electron density perturbation 
\begin{equation} \label{eq: 14}
    \delta n_e/n_0 = -0.5 \eta N(r)Y(r)\sin \left( \Delta \omega t - \Delta k x + 2 l \phi \right),
\end{equation}
where $N(r)$ and $Y(r)$ represent dimensionless radial shape functions according to the work by Y. Shi~{\it et al.}~\cite{Shi2018}. In \cref{eq: 14}, $\eta$ is a dimensionless parameter that determines the amplitude of the plasma wave and, consequently, the amplitude of particle oscillations. Using the linear fluid theory (LFT), Y. Shi~{\it et al.}~\cite{Shi2018} have shown that $\eta \propto a_0^2 \tau/w^2$, where $a_0$ is the normalized peak amplitude of the driving laser beams, $\tau$ is the laser beam duration, and $w$ is the transverse beam size.

The magnetic field generation is a higher-order effect that requires one to move beyond LFT. In the higher-order analysis, $\eta$ serves as an expansion parameter. In LFT, the electric current density arises from ${\bf j}^e = -en_0 {\bf u}$, while the displacement current density is given by ${\bf j}^{dis}=(\partial {\bf E}/\partial t)/4\pi = en_0 {\bf u}$. The net current, utilized for calculating the magnetic field, is the sum of these currents. Therefore, it vanishes in LFT. However, the second order (in $\eta$) current exists and it can be obtained using the electron density and velocity calculated in LFT. It is given by $j_{\theta} = -en_1 u_{\theta 1}\sin^2 (\Delta \omega t - \Delta k x + 2 l \phi)$, where $n_1$ and $u_{\theta 1}$ represent the amplitudes of first-order electron density and velocity perturbations from LFT.

\begin{figure}[H]
\centering
\vspace{-5pt}
\includegraphics[scale = 0.89]{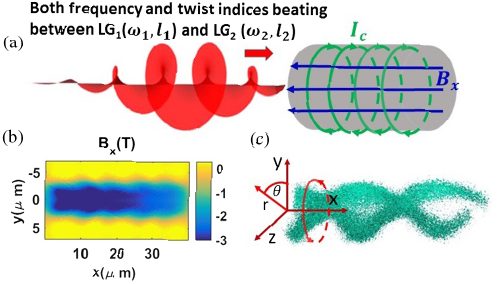}
\vspace{-5pt}
\caption{Generation of a quasi-static magnetic field due to beating of two vortex laser beams in a 3D PIC simulation. (a) Schematic setup for creating plasma waves featuring a helical rotating structure through the interaction of two co-propagating vortex beams, beating in both frequency and twist index. (b) Profile of the resulting axial magnetic field $B_x$ in the ($x,y$) plane at $z$ = 0. (c) A double helical electron density perturbation created by the beating of the two laser beams. The structure is rotating around the $x$-axis. 
The snapshots in panels (b) and (c) are taken 600~fs after the interaction of the vortex laser pulses (160~fs in duration) with the plasma~\cite{Shi2018}.Copyright 2018 by the American Physical Society.}
\label{fig:helicalB}
\end{figure}

The described magnetic field generation mechanism has been confirmed using 3D PIC simulations. The phenomenon is induced by beating of co-propagating LG laser pulses, each possessing OAM, as schematically shown in \cref{fig:helicalB}(a). The simulations showed, for the first time, a helical rotating structure associated with the electron plasma wave. A snapshot of the corresponding density perturbation is given in \cref{fig:helicalB}(c).
Based on the 3D PIC simulations, the mechanism unfolds as follows: Beating LG lasers exert a ponderomotive force with a twisted profile on electrons. After the passage of the laser pulses, electrons are left oscillating on elliptical orbits in the transverse plane, exhibiting an azimuthally dependent phase offset. This collective behavior manifests as a persistent rotating wave structure, accompanied by a nonlinear electrical current resembling a solenoid that gives rise to the magnetic field.

In the simulation illustrated in~\cref{fig:helicalB}, the rotating current density reaches an impressive magnitude of nearly $10^{12} \rm{A/m^2}$. A straightforward calculation indicates that, to achieve such a robust rotating current density, background electrons must rotate at a velocity of $10^6$ m/s—a relatively uncommon occurrence when compared to strong linear current densities. 
The corresponding angular velocity is calculated to be $5 \times 10^{11}$ rad/s, surpassing the corresponding angular velocity observed in the `light fan' phenomenon by three orders of magnitude~\cite{Shi2014}. As shown in \cref{fig:helicalB}(b), the area-averaged axial magnetic field reached nearly 3~T in the described 3D PIC simulations. This magnetic field, generated through the approach introduced by Y. Shi~{\it et al.}~\cite{Shi2018}, outperforms other laser-based methods for magnetic field generation in underdense plasmas, such as the Inverse Faraday effect(IFE). Notably, the scheme of by Y. Shi~{\it et al.}~\cite{Shi2018} yields a much stronger magnetic field that persists for an extended duration even after the laser-plasma interaction. The potential for achieving higher magnetic fields exists and it requires employing  higher intensity laser beams or considering denser plasmas. 
Additional simulations indicate that a quasi-static longitudinal magnetic field can also be driven using an LG and a conventional Gaussian beam via the same mechanism. Such a configuration, rather than using two LG beams,  might be easier to implement in a proof-of-principle experiment. 

\subsection{Magnetic field generation via OAM transfer}
\label{subsection:Bfield-oam}

A distinct mechanism for inducing quasi-static self-generated magnetic fields involves the IFE, where an axial magnetic field spontaneously arises as a laser beam propagating through a plasma  transfers its AM to plasma electrons. Multiple approaches exist for creating axial magnetic fields via the transfer of AM from the laser to the plasma. Initially, IFE observations primarily involved a CP beam passing through an unmagnetized plasma, resulting in the formation of a quasi-static axial magnetic field~\cite{Haines2001}. However, this effect is not limited to CP beams carrying SAM but is also applicable to vortex beams carrying OAM. Theoretical and computational analysis of magnetic field generation using a vortex beam has been undertaken in several prior studies~\cite{Ali2010, Nuter2020, Longman2021}.
There are also other ways to transfer AM from light to plasma, such as using multiple Gaussian laser beams with twisted pointing directions~\cite{Shi2023}.
The generation of a strong magnetic field via the transfer of AM has also been suggested as a collective, macroscopic signature that can be used for detecting the so-called radiation friction~\cite{Liseykina2016}.
A fast rotating plasma environment~\cite{Shi2023, Valenzuela-Villaseca2023} can also be interesting for astrophysics studies and studies of nuclear reactions. Magnetic fields can also be driven in interactions of vortex beams with ionization fronts. Y.~Wu~{\it et al.}~\cite{WuYipeng2023} have recently demonstrated that a sharp relativistic ionization front can generate magnetic fields in the multimegagauss range and create tunable optical vortices when it traverses a relatively long-wavelength vortex.

\subsection{Generation of X-rays and \texorpdfstring{$\gamma$}-rays carrying OAM}
\label{subsection:gamma-oam}

We have already discussed the generation of higher order OAM carrying harmonics by optical lasers. There is also strong interest in hard X-rays and $\gamma$-rays carrying OAM, as such photons can enable novel nuclear physics studies. 
In atomic physics, OAM photons can be used to investigate transitions forbidden by standard selection rules in the electric and magnetic dipole approximation, as discussed by A.~Picon~{\it et al.}~\cite{Picon2010}. Similar research can be performed with nuclei, but requires higher energy photons.
According to recent calculations by Z.-W. Lu~{\it et al.}~\cite{lu2023prl-gammaOAM}, OAM carrying $\gamma$-rays can be used for manipulation of collective excitations of different multipole transitions in even-even nuclei (nuclei with an even number of neutrons and an even number of protons). It has  been suggested by U. Jentschura and V. Serbo~\cite{Jentschura2011} that high-energy photons with large OAM can also be used for fundamental experimental studies of pair production off nuclei. The same authors indicate that one can even potentially study  nuclear fission induced by absorption of one or more OAM-carrying photons at energies below the giant dipole resonances. Potential benefit of such a study could be a fundamental insight into the dynamics of a fast rotating quantum many-body system.

\begin{figure}[H]
\centering
\vspace{-5pt}
\includegraphics[scale=0.35]{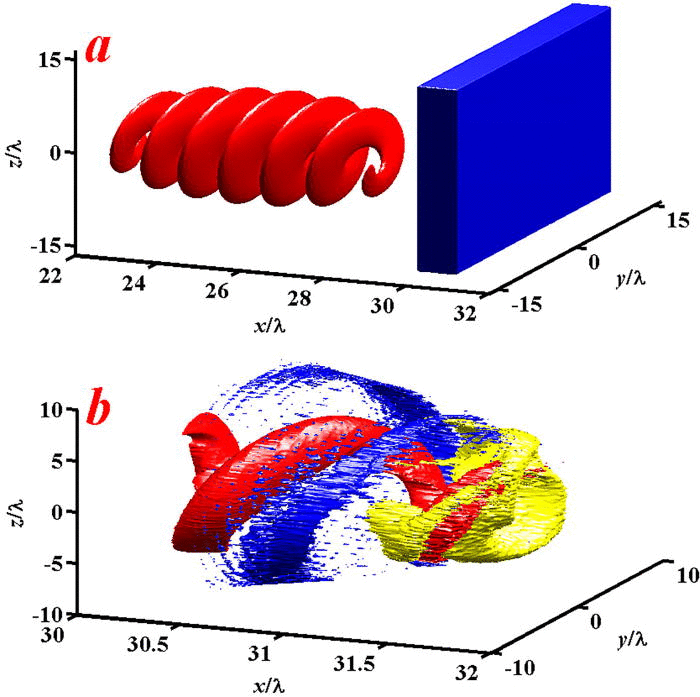}
\vspace{-5pt}
\caption{Illustration depicting the generation of $\gamma$-ray photons with OAM through a laser-plasma interaction. (a) Initial setup with the CP $\rm{LG}_{10}$ laser (in red) and the plasma target (in blue) before the interaction. (b) Snapshot taken after the interaction and showing the laser (red), electrons (blue), and the $\gamma$-ray photons (yellow) emitted during the interaction~\cite{Liu2016}. Copyright 2016 by AIP.}
\label{fig:oamgamma1}
\end{figure}

It has been shown experimentally that OAM can be obtained in the regime of spontaneous emission radiation from variable-polarization undulators and that vortex beams can be generated at short wavelengths with free-electron lasers~\cite{Primoz2014, Sasaki2008, Bahrdt2013, Hemsing2011, Hemsing2012, Hemsing2013,RebernikRibic2017}. At ELI-NP, the plan is to use inverse Compton scattering of short laser pulses on relativistic electron beam pulses to produce brilliant, quasi-monochromatic $\gamma$-ray beams~\cite{ELI-NP_gamma, ELI-NP_nuclear}. It has been shown in multiple theoretical papers that inverse Compton backscattering between a high brightness electron beam and a vortex laser pulse can produce OAM-carrying X-rays and $\gamma$-rays~\cite{Jentschura2011, Jentschura2011a, Stock2015, Petrillo2016}. One can potentially achieve this setup by modifying a conventional setup by inserting a helical mirror or a phase plate into the laser beam before the electron collision to create a high intensity vortex laser beam.
By exploiting the highly nonlinear interactions between electrons and the intense laser beams, the photons carrying OAM can be pushed from the high-intensity extreme to another short-wavelength extreme.

C. Liu~{\it et al.}~\cite{Liu2016} have proposed another scheme for generating OAM-carrying $\gamma$-rays where a high-intensity CP vortex laser beam irradiates a thin plasma target. The setup and the resulting $\gamma$-ray beam (yellow) are shown in \cref{fig:oamgamma1}. First, the SAM and OAM are transferred from the driving laser to the target electrons. The electrons then emit energetic photons in the quantum regime. During the emission process, the absorbed OAM is transferred from the electron to emitted $\gamma$-rays.
Simulations show that the OAM of the $\gamma$-ray photons depends on the overall AM of the driving laser. To enhance the OAM of the $\gamma$-ray photons, two approaches are viable: either increasing the laser intensity or employing a high-order mode vortex laser. In the case of high-order CP vortex lasers, the transverse energy distribution of generated $\gamma$-ray photons has a distinct pattern that consists of several parts whose number is equal to the total normalized angular momentum carried by each photon of the driving laser. 
It is worth pointing out that the OAM-carrying $\gamma$-rays were generated in this work using a quantum radiation model that itself contains simplifications and approximations. {A similar phenomenon can be found in the interaction between an intense vortex and an underdense plasma~\cite{JuLB2019}, a cone-foil target~\cite{Zhu2018}, or near-critical-density plasma~\cite{HuYT2021}.}

As mentioned earlier,  the possibility of generating X-rays or $\gamma$-rays with OAM has been theoretically examined using a setup that involves inverse Compton backscattering between a high-brightness electron beam and a vortex laser pulse. 
V.~Petrillo~{\it et al.}~\cite{Petrillo2016} used the classical electrodynamics retarded fields to evaluate the OAM of the radiation. They chose the linear regime of inverse Compton backscattering, so that quantum corrections can be neglected. Specifically, V.~Petrillo~{\it et al.} used a ps-long electron beam with a charge of 1~nC and energy of 25~MeV. The laser beam is 1~J, 1 ps long with a wavelength of $0.8~\micron$. It is focused in the interaction point to a 20 $\micron$ spot.
The calculations predict $2.5 \times 10^6$ OAM photons at 8 keV~\cite{Petrillo2016}.
The laser and electron beam parameters used by V.~Petrillo~{\it et al.} are relatively modest, suggesting that the same scattering setup can be used to achieve higher energy OAM photons by going to higher electron beam energies and higher laser intensities. It must be pointed out that Y.~Taira~{\it et al.} showed, by examining nonlinear inverse Thomson scattering, that OAM $\gamma$-ray photons can also be produced using a high-intensity CP laser~\cite{Taira2017}, which means that using a vortex laser beam is not a hard requirement.

At higher laser intensities, the electron dynamics experiences a qualitative change due to the so-called radiation reaction or radiation friction. J.-Y.~Wang {\it et al.}~\cite{wang2023radiationreaction} examined the effect of radiation reaction in the nonlinear inverse Thomson
scattering, whereas Y.-Y.~Chen {\it et al.}~\cite{Chen2018,chen2019} examined the effect in the nonlinear Compton scattering. Y.-Y.~Chen {\it et al.} found that the scattering of a strong laser pulse of twisted photons at ultrarelativistic electrons produces a well-collimated $\gamma$-ray beam with very large OAM.
The authors demonstrated numerically the conservation of AM in the quantum radiation-dominated regime by considering laser photons, emitted radiation, and electrons.
Due to the radiation reaction, part of the OAM and SAM of the absorbed laser photons is transferred to the electron beam.
Moreover, it is shown that the accompanying pair production process causes a counterintuitive increase in the OAM of the $\gamma$-beam due to additional absorption of twisted laser photons by secondary particles.

The generation of OAM carrying $\gamma$-rays is a promising, but a relatively new area of research. As such it has a number of open questions. In simulations, the $\gamma$-ray emission is modeled using semiclassical modules that treat the process as incoherent. This approach is extremely valuable, particularly when a full quantum calculation is not feasible. However, it raises a question regarding the type of OAM that is being carried by the emitted photons (e.g., see Ref.~\cite{Bliokh2015} for a discussion of different types of OAM). The concern is that the $\gamma$-rays bearing extrinsic OAM may not be suitable for nuclear physics experiments. New experiments and the development of  advanced modules are likely to provide additional insights.

Finally, we want to mention two publications by M.~Katoh~{\it et al.}~\cite{Katoh2017, Katoh2017b} where it is shown theoretically that a single free electron performing circular motion radiates an electromagnetic wave with a helical phase structure. The emitted radiation can be interpreted as emission of photons carrying well-defined OAM~\cite{Katoh2017b}. The calculation results were confirmed by interference and double-slit diffraction experiments on the radiation of relativistic electrons in spiral motion~\cite{Katoh2017}. These studies highlights possible OAM effects in space plasmas, because they indicate that photons with OAM are likely to be spontaneously generated through cyclotron/synchrotron radiation or Compton scattering in diverse cosmic scenarios. Based on their results, the authors of the studies propose promising laboratory vortex photon sources in various wavelengths ranging from radio waves to $\gamma$-rays~\cite{Katoh2017}.  
The work by M.~Katoh~{\it et al.} may also lead to the consideration of AM in the nonlinear inverse Compton process of CP light~\cite{Taira2017}, which is an example of radiation from an electron performing circular motion.

\subsection{Particle acceleration mechanisms leveraging the hollow-shaped intensity of vortex beams}
\label{subsection:hollow}

The ponderomotive force created by a vortex beam is qualitatively different from that created by a conventional laser beam. The difference arises due to the hollow shape of the laser intensity in a vortex beam.
The direction of the ponderomotive force induced on plasma electrons by a helical beam with electric field amplitude ${\bf E}$ and frequency $\omega$ is
\begin{align} \label{eq:pond_pot}
    {\bf F}_{\text{pond}} = - \nabla \varphi_{\text{pond}}, &\mbox{    }&  \varphi_{\text{pond}} = \frac{e^2}{4 m_e \omega^2}  \left| {\bf E} ( {\bf r}, t) \right|^2,
\end{align}
where $\varphi_{\text{pond}}$ is the ponderomotive potential and $e$ and $m_e$ are the electron charge and mass. If the electric field is an $\rm{LG}_{nm}$ mode given by \Cref{eq:LGnm}, then its amplitude is independent of $\phi$ and it follows from \Cref{eq:pond_pot} that $\varphi_{\text{pond}}$ is also independent of $\phi$, as shown in \cref{fig:twistforce}(a). Therefore, the ponderomotive force ${\bf F}_{\text{pond}}$ of a single LG beam has only a radial component in the cross section of the beam. This feature is common with the conventional laser beams. However, $\varphi_{\text{pond}}$ in \cref{fig:twistforce}(a) has a peak off the laser axis and, as a consequence, the ponderomotive force of a vortex beam is directed inward rather than outward in the region between the beam axis and the peak of $\varphi_{\text{pond}}$.

 The inverted structure of the ponderomotive force has been successfully leveraged to improve laser wakefield acceleration~\cite{Zhang2014, Vieira2014, Mendonca2014, Zhang2016a, Zhang2016b}. 3D PIC simulations and analytical theory show that a vortex laser beam creates a donut-like wakefield~\cite{Zhang2014, Vieira2014, Mendonca2014, Zhang2016a, Zhang2016b}.  
Using 3D PIC simulations,  J. Viera~{\it et al.}~\cite{Vieira2014} demonstrated that the hollow wakefield leads to hollow electron self-injection. This wakefiled also creates a focusing and accelerating structure for a witness positron beam, enabling positron acceleration. 
A unique feature of this regime is that the positron focusing forces in a donut-like wakefield surpass the electron focusing forces in a spherical bubble by more than an order of magnitude. Subsequent research by G.-B.~Zhang~{\it et al.}~has revealed the potential for accelerating ring-shaped electron beams within this specialized wavefield~\cite{Zhang2016a, Zhang2016b}.

\begin{figure}[H]
\centering
\vspace{-1pt}
\includegraphics[scale=0.53]{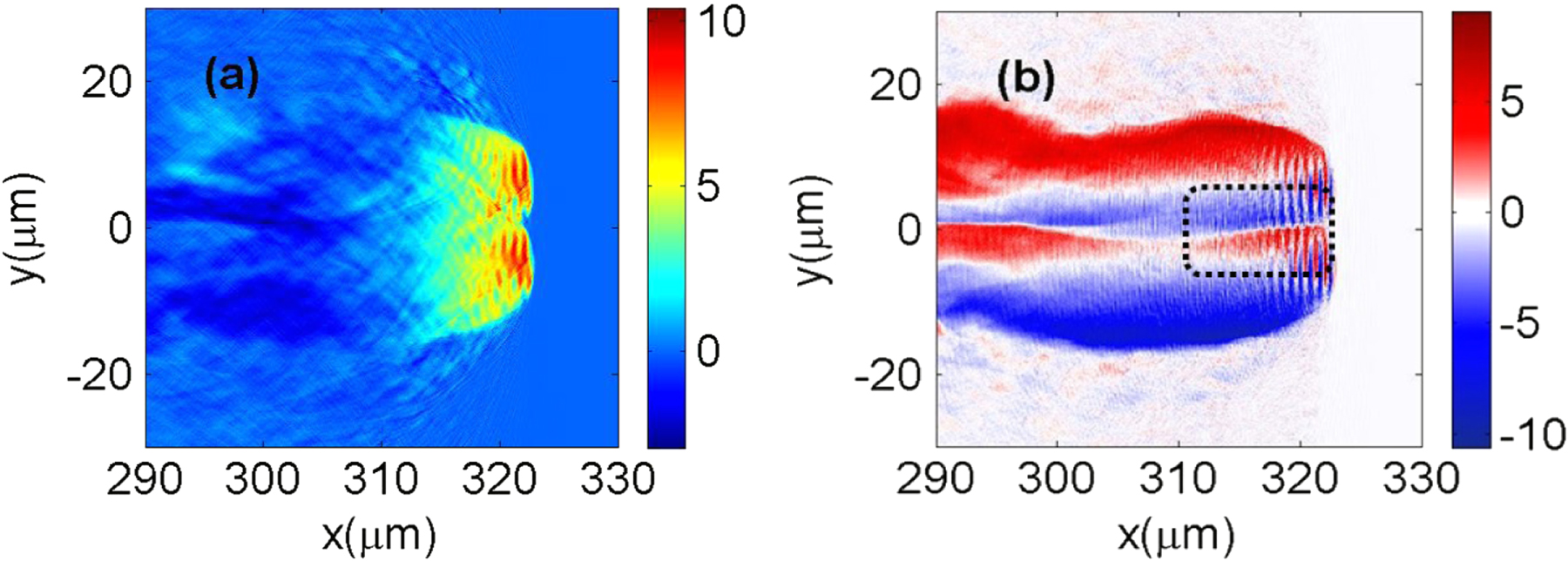}
\vspace{-5pt}
\caption{Field structure created by a vortex beam in an underdense plasma: 
(a) normalized longitudinal electric field (along the laser propagation direction) and (b) normalized transverse electric field. The snapshots are taken 1.12~ps after the start of the 3D~PIC simulation.
The field within the black-dashed box provides transverse focusing to a witness proton beam~\cite{Zhang2014}. All the fields are normalized by $m_e \omega c/ |e|$. Copyright 2014 by IOP.}
\label{fig:Zhang2014NJP-LG}
\end{figure}

Concurrently, X.~M.~Zhang {\it et al.}~\cite{Zhang2014} have explored the use of the donut-like structure produced by a much higher intensity vortex beam, $I \sim 2 \times 10^{22}~\rm{W /cm^{2}}$, for acceleration of a witness proton beam. The 3D~PIC simulations performed by X.~M.~Zhang~{\it et al.}~\cite{Zhang2014} demonstrated the formation of a distinctive bubble featuring an electron column along the axis. \cref{fig:Zhang2014NJP-LG} shows the resulting field configuration. In this configuration, a witness proton beam experiences effective transverse confinement due to the generated transverse focusing field. This enables prolonged acceleration via the longitudinal wakefield. 
Compared to the case with a conventional laser beam, the vortex beam creates a structure that mitigates potential scattering by transverse fields before acceleration, enabling stable and efficient acceleration of protons to multi-GeV energies. 

Several research groups have also examined the use of vortex beams for laser-driven ion acceleration from~\cite{Pae_2020, Camilla2023, Dong2023PoP}. The results for vortex beams were compared to those for conventional laser beams and it was shown using PIC simulations that the use of a vortex beam can significant improve the collimation of the resulting energetic ion beam. K.~Pae~{\it et al.}~\cite{Pae_2020} and H.~Dong~{\it et al.}~\cite{Dong2023PoP} investigated a similar setup where an ultra-high-intensity short pulse vortex laser beam irradiates a thin dense target. H.~Dong~{\it et al.}~\cite{Dong2023PoP} found that the target produces a plasma jet at its rear side, whose formation is aided by the hollow structure of the vortex beam. There, the electrons generate a strong azimuthal magnetic field that increases the transverse confinement time of hot electrons. The collimated electron jet provides collimation for the accelerated ions via the transverse field of its space charge. K.~Pae~{\it et al.}~\cite{Pae_2020} state that the underlying reason for the ion collimation is the inverted ponderomotive force generated by the vortex beam. C.~Willim~{\it et al.}~\cite{Camilla2023} considered a setup with a double-layer target aimed at increasing conversion efficiency of laser energy into the energy of accelerate ions. The target has an extended near-critical layer followed by a dense thin foil. The near-critical layer becomes relativistically transparent during the laser irradiation, which enables the laser to propagate into the target and interact with it volumetrically. The main finding is that a vortex beam experiences reduced relativistic self-focusing compared to a conventional laser beam with the same energy. C.~Willim~{\it et al.}~\cite{Camilla2023} successfully exploited this aspect to reduce the ion beam divergence.
{T.Wilson~{\it et al.}~\cite{Wilson_2023} also studied the self-focusing, compression and collapse of weakly-relativistic vortex laser in near-critical plasma using PIC simulations.}

\begin{figure}[H]
\centering
\includegraphics[scale=0.4]{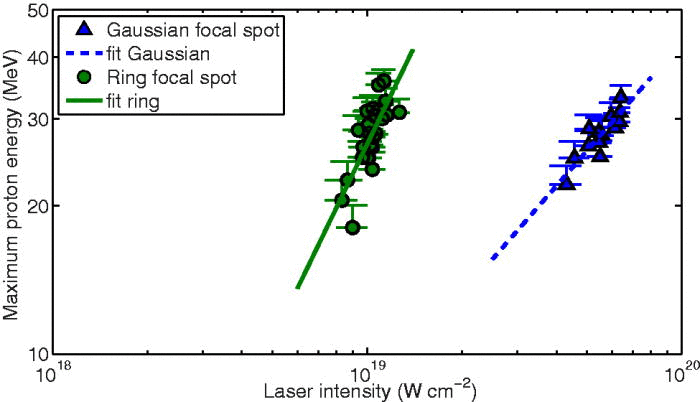}
\caption{Comparison of ion acceleration in the TNSA regime driven by a conventional Gaussian and a vortex laser beam.  C. Brabetz~{\it et al}~\cite{Brabetz2015} observed 
stronger coupling to laser intensity (steeper gradient) in the shots employing a vortex beam (green circles) compared to those with a Gaussian focal spot (blue triangles). Reproduced from C. Brabetz~{\it et al.},~Phys. Plasmas 22 (1): 013105 (2015), with the permission of AIP Publishing. Copyright 2015 by AIP.}
\label{fig:oamtnsa}
\end{figure}

First proof-of-principle experiments on laser-driven ion acceleration using vortex beams were performed by C.~Brabetz {\it et al}~\cite{Brabetz2015} in 2015 at the PHELIX laser facility in Germany. The experiments were carried out in the Target Normal Sheath Acceleration (TNSA) regime.
A spiral phase element was inserted into the laser amplifier to produce a hollow, donut-shaped laser focus profile on the target surface.
The resulting vortex laser had a FWHM pulse duration of 650 fs and a beam energy of 70~J. The focused laser intensity was in the range between $10^{18}~\rm{W / cm^{2}}$ and $10^{20}~\rm{W /cm^{2}}$. 
One of the motivations for using vortex beams was the notion that by the initial emission solid angle of the TNSA ion source can be reduced by flattening the electron sheath at the rear surface of the target~\cite{BUSOLD201494, Brabetz2012}. The experiments confirmed that the use of vortex beams is indeed beneficial in this regard. The proton beam images showed a more uniform spatial distribution when the acceleration was driven by vortex laser beams. More importantly, the experiments revealed that the vortex rather than conventional laser beams produced protons with highest energies. 
As seen in \cref{fig:oamtnsa},  the data points for the shots with a Gaussian focal spot (blue triangles) fit well to a power scaling $E_{p,max}= a I^{b}_{laser}$ with $b=0.73 \pm 0.04$, where $I_{laser}$ is the laser intensity. The data points for the shots using a vortex beam (green circles) fit to a steeper power scaling with $b=1.32 \pm 0.08$, indicating stronger coupling to the laser intensity. 
This dramatic improvement in the intensity scaling of the ion energy remains unexplained. 

Following the work by C.~Brabetz {\it et al}~\cite{Brabetz2015}, two more experimental campaigns on ion acceleration have been performed with vortex beams: one at CoReLS by C.~Jeon~{\it et al}~\cite{Jeon2018} and one at SIOM by W.~P.~Wang~{\it et al}~\cite{wang2020hollow}. C.~Jeon~{\it et al}~\cite{Jeon2018} used a 100~TW LG beam with $l$=1 to drive ion acceleration in the TNSA regime. 
The results showed the same unexplained effect as that observed at the PHELIX facility. As of today, a clear theoretical explanation is still lacking. 
W.~P.~Wang~{\it et al}~\cite{wang2020hollow} used a 30~TW (13~J, 40~fs) vortex laser beam with a peak intensity of $6.3 \times 10^{19}\rm{W / cm^{2}}$.
In these experiments, the beam of accelerated protons had a hollow-shape~\cite{wang2020hollow}. The reason is believed to be the hollow shape of the intensity in the vortex laser. Different from the method used in the work of C.~Brabetz {\it et al}~\cite{Brabetz2015}, W.~P.~Wang~{\it et al} created a high-intensity vortex beam by reflecting a Gaussian laser beam off a high-reflectivity phase plate with 32 steps and then focusing the resulting beam with a parabola mirror.

All three papers we have discussed emphasize the importance of the hollow intensity in a vortex beam~\cite{Brabetz2015, Jeon2018, wang2020hollow}. There are however other ways for achieving a hollow intensity profile. 
A stable mode, such as an LG mode or the radially polarized mode, can produce a hollow intensity that can be transported a long distance.
For example, a radially polarized laser beam also has a hollow intensity profile. Moreover, if the goal is only a hollow intensity profile on the surface of the target, then other methods can be explored as well. For example, a nanowire can be placed at the right location in front of the target, or micro-structured targets can be used to obtain a hollow intensity distribution.  
As we can see, a hollow shape is not an exclusive feature of the vortex beam. 
A hollow Gaussian beam can also maintain zero intensity on the beam axis in the near field region~\cite{Cai2003}.
There seems to be no definitive evidence that an intense vortex beam is the best choice for creating a hollow laser intensity profile in laser-plasma  interactions  aimed at ion acceleration. 
Finally, it should be noted that that the width of the hollow region in a vortex beam becomes wider with the increase of the twist index $l$. In the study of the hollow intensity effect of a vortex, the impact of increasing twist index $l$ on ion acceleration has yet to be explored.
In conclusion, more studies are needed to identify and  illustrate the exclusive advantages of the intense vortex beam.  

\begin{figure}[H]
\centering
\includegraphics[scale=0.25]{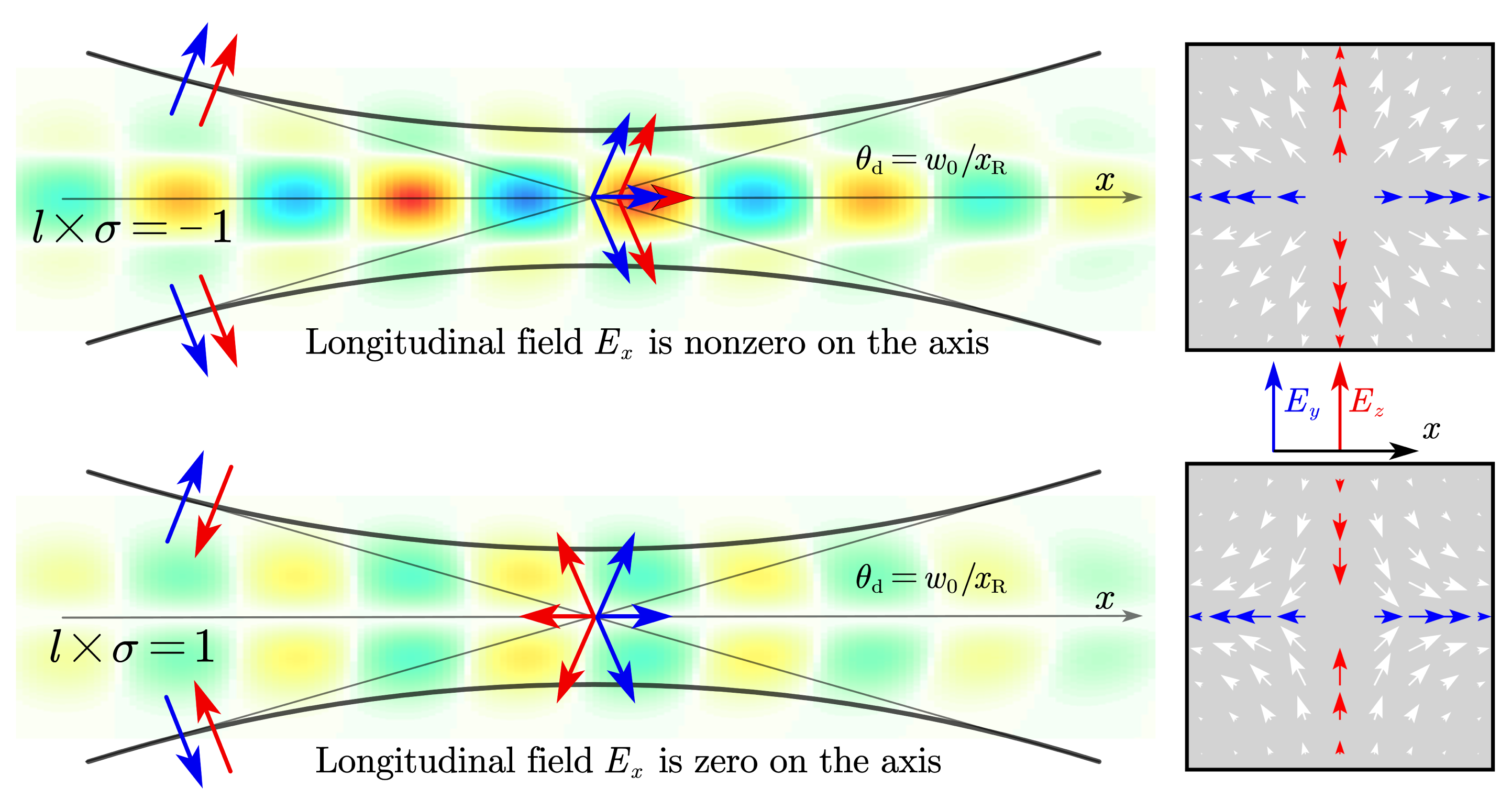}
\caption{Schematic representation of vortex laser beams with and without longitudinal fields on the axis. The transverse electric fields ($E_y, E_z$) are represented by blue and red arrows in the transverse plane (right). The longitudinal section of a propagating beam is shown on the left. The longitudinal fields can be understood as the longitudinal component of $E_y$ and $E_z$, which are not perfectly transverse for a focused beam. The 2D distributions of the longitudinal fields $E_x$ of the focused CP vortex from the PIC simulation are shown in the background. In the colormap, blue represents negative fields and yellow represents positive fields. Top panels: a non-zero longitudinal field on the axis for a vortex beam with $l \times \sigma = -1$. Bottom panels: a zero longitudinal fields on the axis for vortex beam with $l \times \sigma = 1$. Diffraction angle is defined as $\theta_d = w_0/x_R$, where $w_0$ is the beam waist, and $x_R$ is the Rayleigh range.}
\label{fig:Longtitudial field_lgcp}
\end{figure}

\subsection{Electron acceleration by longitudinal fields of an intense vortex beam}
\label{subsection:longi-field}

Tight focusing is essential for achieving high-intensity using a laser beam produced by a high-power laser system. The field structure of a tightly focused laser beam markedly differs from that of a plane wave, giving rise to two significant changes. Firstly, the laser intensity in a tightly focused beam changes significantly in the transverse direction. As a result,  electrons irradiated by such a beam experience a noticeable transverse ponderomotive force. Secondly, longitudinal fields emerge alongside the transverse fields. Despite the ponderomotive force being potentially an order of magnitude smaller than the force associated with the transverse fields of the laser, its non-oscillatory nature can lead to efficient energy and momentum transfer from the laser to matter. The amplitude of the longitudinal fields in a focused beam is typically more than an order of magnitude smaller than the amplitude of the transverse field. However, at relativistic laser intensities, these fields can play a crucial role in electron dynamics and energetics despite their relatively low amplitude. This significance arises because these fields, though smaller in amplitude, are directed along the laser propagation, aligning with the forward motion of laser-accelerated electrons.

By using a vortex beam, one can change both the direction of the ponderomotive force and the topology of the laser fields. While the effect on the ponderomotive force (i.e. its inversion near the axis of the beam) can be viewed as expected, the changes in the field structure are less intuitive and deserve a separate discussion.  
Of particular interest is the topology of the longitudinal electric field, because it is possible to create a peak of the longitudinal field in the region where the transverse field is vanishing small -- the feature that is impossible for a conventional laser beam.

We consider a circularly polarized vortex beam with $|l| = 1$ to illustrate the effect. The amplitude of transverse fields in such a beam peak off the beam axis. The structure of the longitudinal field changes with the change of handedness. The handedness is set by a parameter called $\sigma$ that has two values, $\sigma = \pm 1$, with $E_z = i \sigma E_y$ for a beam propagating in the positive direction along the $x$-axis. \Cref{fig:Longtitudial field_lgcp} shows the structure of $E_{\parallel}$ in vortex beams with $\sigma = -1$ (upper-left panel) and with $\sigma = 1$ (lower-left panel). These two beams have qualitatively different longitudinal field structures: the beam with $l \times \sigma = -1$ has a strong longitudinal field on the axis and the beam with $l \times \sigma = 1$ does not.
Here we provide a qualitative explanation of the effect similar to that given by G.~F.~Quinteiro~{\it et al}~\cite{Quinteiro2017}. The direction of the transverse electric field in the beam cross-section as a function of the azimuthal angle is shown for $\sigma = -1$ and $\sigma = 1$ in the right two panels of \cref{fig:Longtitudial field_lgcp}. In the case of  $\sigma = -1$, the field remains directed towards the axis. In contrast to that, in the case of $\sigma = 1$, the direction of the transverse field changes every time we increase the angle by $\pi/2$. During the beam focusing, wave fronts bend. We can then think of the transverse fields slightly tilting, so that they create a field projection along the axis. This projection is the longitudinal field (this is definitely the case at distances further than the Rayleigh range from the focal plane). In the case of $\sigma = -1$, $E_y$ and $E_z$ create projections that have the same sign. This explains the emergence of strong $E_x$ along the axis of the considered vortex beam. In the case of $\sigma = 1$, $E_y$ and $E_z$ create projections with the opposite sign. These contributions cancel each other out and, as a result, there is no on-axis $E_z$. In \cref{fig:Longtitudial field_lgcp}, $E_y$ and $E_z$ are shown with blue and red arrows to make it easier to distinguish them.
The discussed phenomenon makes it possible to achieve relativistic-scale longitudinal fields using tight focusing of vortex beams with peak power above 50~TW.

\begin{figure}[H]
    \centering
    \includegraphics[scale = 0.9]{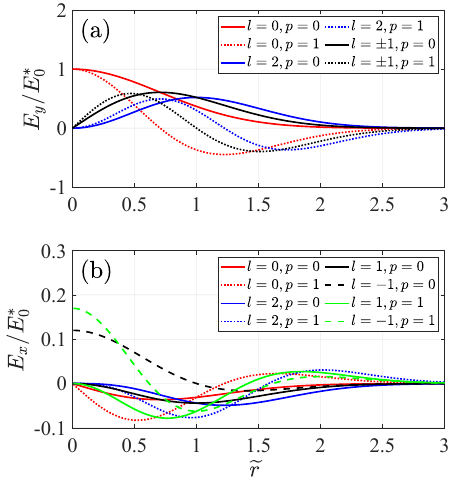}
    \caption{Radial dependence of transverse (a) and longitudinal (b) electric fields in CP-LG laser beams with different azimuthal, $l$, and radial, $p$, mode numbers. All lasers are right-circularly polarized ($\sigma=1 $) and have the same power. The fields are plotted in the focal plane.
    The fields are normalized to $E^*_ 0$, which is the peak electric field amplitude of a Gaussian beam. The laser beams are tightly focused to $k_0 w_0 = 23$ (diffraction angle is $\theta_d = 0.087$), where $k_0$ and $w_0$ are the wavenumber and spot size, respectively.    
    The figure is adapted from Ref.~\cite{shi2021electron}. Copyright 2021 by IOP.}
    \label{fig:lg-topolog}
\end{figure}

\Cref{fig:lg-topolog} shows radial profiles of transverse, $E_y$, and longitudinal, $E_z$, electric fields for different CP-LG modes. These plots indicate that qualitatively different profiles $E_z$ and $E_y$ can be achieved by tuning the twisted index $l$ and the radial index $p$ of the LG beam~\cite{shi2021electron}. The often studied hollow lasers are typically vortex beams with $|l|=1$. As already discussed, such a beam can have a strong on-axis longitudinal electric field. This requires $l \times \sigma=-1$, which is equivalent to a requirement for OAM and SAM to be anti-parallel. The profile of $E_z$ in a conventional CP laser beam is given by curves with $l = 0$, with different curves representing different radial modes. The 3D structure of the same field is shown in~\cref{fig:cp-lg_Li}~\cite{Li_2020}. In experiments, a focused beam is likely to be a mix of several modes rather than a single pure LG mode. Furthermore, nonlinear effects would result in higher-order LG modes even if the driving beam is a pure LG mode.
The vortex beam can then be represented as a superposition of orthogonal LG modes (like the modes shown in \cref{fig:lg-topolog}) and this superposition should then be used to analyze the field topology of the vortex beam.

\begin{figure}[H]
    \centering
    \includegraphics[scale=0.4]{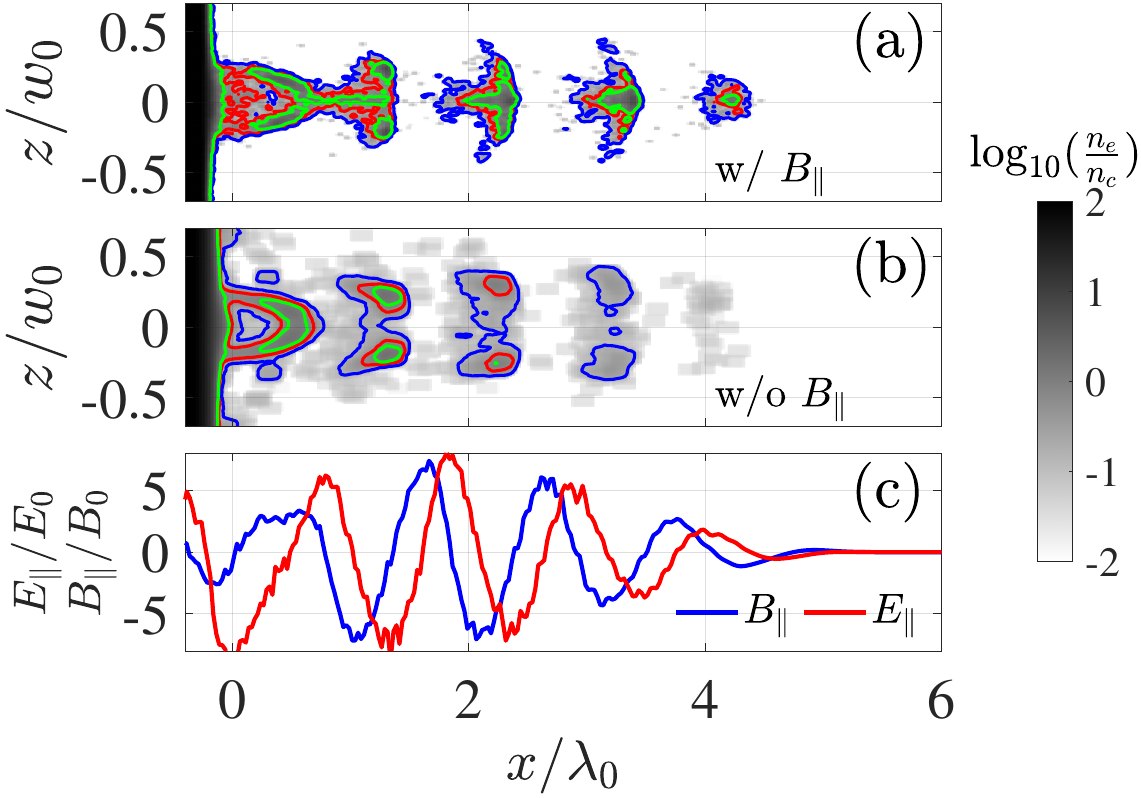}  
\caption{Electron injection into a reflected vortex beam in 3D PIC simulations of Y. Shi~{\it et al.}~\cite{shi2021LGPRL}. The top two panels compare scenarios with (a) and without (b) the inclusion of the laser $B_x$ in the electron equations of motion.  
The density, depicted on a logarithmic scale, reveals dense electron bunches, with the blue, red, and green contours representing $n_e = 0.1n_{c}$, 0.5$n_{c}$, and $n_{c}$, where $n_c$ is the critical density. Additionally, longitudinal electric field profiles, $E_{\parallel}/E_0$, and magnetic field profiles, $B_{\parallel}/B_0$, along the axis are presented in (c), with $E_0 = B_0 \equiv m_e c \omega/|e|$. Copyright 2021 by the American Physical Society.} \label{fig:einject}
\end{figure}

Y. Shi {\it et al.}~\cite{shi2021LGPRL, shi2021electron,shi_blackman_zhu_arefiev_2022} found that the longitudinal electric field of a high-intensity vortex laser beam can be used to directly accelerate electrons in vacuum. Electrons are injected into the laser beam upon its reflection off a sharp plasma-vacuum interface. The injection process was self-consistently simulated using 3D PIC simulations. \Cref{fig:einject}(a) shows electron bunches formed during the injection process. The density is shown on a  logarithmic scale, with the blue, red, and green contours representing $n_e/n_c = 0.1$, $0.5$, and $1.0$, where $n_c$ is the classical critical density. For reference, \cref{fig:einject}(c) shows longitudinal electric, $E_{\parallel}/E_0$, and magnetic, $B_{\parallel}/B_0$, field profiles along the beam axis, with $E_0 = B_0 \equiv m_e c \omega/|e|$.

The injected electrons undergo longitudinal acceleration by $E_{\parallel}$ and, as expected, their phase-slip impacts the energy gain. As the electrons become ultra-relativistic due to their acceleration, they enter a regime where the phase slip is determined primarily by the difference between the laser phase velocity $v_{ph}$ and the speed of light $c$. In what follows, we focus on this regime. It is important that the phase velocity of a tightly focused beam near the focal plane is superluminal due to the Gouy phase shift effect. This causes some of the electrons originally in the accelerating phase to passively phase-slip, and some of the initially accelerated electrons may enter the decelerating phase and give their energy back to the electromagnetic field~\cite{shi2021LGPRL, shi2021electron}. For a beam with $p=0$, the total phase slip for an electron that starts in the focal plane and travels with the laser to $x \gg x_{\rm R}$ is $\Delta \Phi=\pi$, where $x_{\rm R}$ is the Rayleigh range. Therefore, some electrons (depending on their initial phase) can remain in the accelerated phase of the longitudinal electric field until the laser diffracts ($x \gg x_{\rm R}$) and the longitudinal electric field becomes very weak. These electrons retain the energy they gain from the laser field during acceleration.
The described mechanism produces a train-like spatial structure of the electrons in the longitudinal direction since only electrons in the accelerating phase can undergo the prolonged acceleration. 

We have so far discussed a specific case of $l \times \sigma = -1$ and $p = 0$. LG modes with $l \times \sigma = -1$ and $p \geq 1$ also have a strong on-axis $E_{\parallel}$ that can perform electron acceleration. However, the phase velocity has a radial mode dependence~\cite{shi2021electron}. The difference between $v_{ph}$ and $c$ is a manifestation of the transverse beam structure, so it is not surprising that  $v_{ph} -c$ depends on $p$. The implication is that the acceleration and, as a result, the energy gain will differ for different $p$~\cite{shi2021electron}.

Y. Shi {\it et al.}~\cite{shi2021electron} have developed a model for the electron energy gain that accounts for the impact of the radial beam structure on the phase velocity of the accelerating field $E_{\parallel}$. The model considers an ultra-relativistic electron moving along the $x$-axis close to the beam axis and assumes that the phase slip is determined primarily by $v_{ph}$, so that one can set
\begin{equation}
    x = x_0 + c(t-t_0),
\end{equation}
where $x_0$ and $t_0$ are the location and time of the injection. The electron experiences a strong longitudinal electric field of the vortex laser beam ($l \times \sigma = -1$). The field at the electron location is
\begin{equation}
    E_{\parallel} = \frac{E_{\parallel}^{\max} }{1 + x^2/x_R^2} \sin \left( \Delta \Phi + \Phi_0 \right)
\end{equation}
where

\begin{equation}\label{eq:Eparallel}E_{\parallel}^{\max}=2\sqrt{\frac{p+1}{\pi}}\theta_dE_0=\frac{2}{\pi}\sqrt{\frac{p+1}{\pi}}\frac{\lambda_0}{w_0}E_0.
\end{equation}
and
\begin{equation} \label{eq:Delta_Phi}
    \Delta \Phi = 2(p +1) \left[ \tan^{-1} (x_0/x_R) - \tan^{-1} (x/x_R) \right]
\end{equation}
is the phase slip experienced by the electron as it travels from $x_0$ to $x$.
The electrons are injected into the beam close its focal plane, so that $x_0\ll x_R$ and the expression given by \cref{eq:Delta_Phi} can be simplified to $\Delta \Phi = -2(p +1)\tan^{-1} (x/x_R)$. The phase velocity of the laser is superluminal at $x=0$, and then it decays to $c$ at $x \approx x_R$. \Cref{eq:Delta_Phi} explicitly states that the phase slip experienced by the electron is strongly correlated with the magnitude of the radial mode number $p$. In the case of $p=0$, the phase slip of the electron after traveling a distance comparable to the Rayleigh range is about $ \Delta \Phi = -\pi$, which means that a part of the injected electrons can still be in the accelerating phase of the longitudinal field. This is clearly not the case for higher-order radial modes because of the higher phase velocity.
Accelerated electrons will enter the decelerating phase before they reach $x \approx x_R$ due to the large phase slip. As a result, they will experience deceleration that can potentially greatly reduce the terminal energy gain.
In the considered model, the momentum gain of the electrons from the longitudinal laser electric field is given by the following integral
\begin{equation} \label{eq2}
     \Delta p_{\parallel} = \frac{|e| E_{\parallel}^{\max}}{c} \int_{x_0}^{x}  \frac{ \sin (\Delta \Phi + \Phi_0) dx'}{1 + (x')^2 / x_R^2} .
\end{equation}
We again assume injection close to the focal plane ($x_0\ll x_R$), so that the integral yields
\begin{equation} \label{eq:momentum-gain}
\begin{aligned}
    \Delta p_\parallel = -\frac{|e|E_\parallel^{\max} x_R}{2(p+1)c} \Big( &\cos\Phi_0
    -\cos \big[\Phi_0\\
    &-2(p+1) \tan^{-1}(x/x_R) \big] \Big).
    \end{aligned}
\end{equation}
This expression quantifies the impact of the increased phase velocity and faster phase slip at higher $p$ on electron momentum/energy gain. Specifically, it follows from \Cref{eq:momentum-gain} that $\Delta p_\parallel \rightarrow 0$ at $x \gg x_R$ for odd values of $p$. Therefore, an important conclusion from~\Cref{eq:momentum-gain} is that odd radial modes are not able to generate energetic electrons~\cite{shi2021electron}.

The 3D PIC simulations performed by Y. Shi {\it et al.}~\cite{shi2021LGPRL} confirm the feasibility of the discussed acceleration mechanism while also demonstrating the importance of the strong longitudinal laser magnetic field $B_{\parallel}$. By performing simulations with and without $B_{\parallel}$ in the electron equations of motion, it was demonstrated that 
the longitudinal magnetic field of the relativistically strong vortex laser acts as a hoop-like constrictor for electron beams injected into the reflected laser. These two scenarios and the impact of $B_{\parallel}$ on electron injection are illustrated  in~\cref{fig:einject}. The transverse confinement provided by $B_{\parallel}$ plays an important role in the generation of high-quality electron beams~\cite{shi2021LGPRL,shi2021electron}.

In a subsequent publication, Y.~Shi~{\it et al.}~demonstrated using 3D PIC that the circular polarization is not a hard requirement~\cite{shi_blackman_zhu_arefiev_2022}. It was shown that an intense linearly polarized vortex laser beam with $l=1$ reflecting off a dense plasma mirror (i.e. a dense plasma with a sharp plasma-vacuum interface) can also effectively inject electrons into the beam and then accelerate them to ultra-relativistic energies.
The LP laser is a superposition of two CP lasers: one with $\sigma = 1$ and one with $\sigma = -1$. As discussed earlier, only the laser with $l \times \sigma = -1$ has a strong on-axis longitudinal electric field. We note that the power in this CP laser is half of the period-averaged power in the LP laser. Therefore, the on-axis $E_{\parallel}$ in an LP-LG laser that has the same power as the CP-LG laser with $l \times \sigma = -1$ is weaker by a factor of $\sqrt{2}$~\cite{shi_blackman_zhu_arefiev_2022}. The reduction in $E_{\parallel}$ can be viewed as a trade-off for the simplicity of creating an LP-LG vortex beam (compared to a CP-LG beam).

Electron injection during laser reflection off a plasma mirror (i.e. a plasma with a sharp plasma-vacuum interface) is a promising mechanism, but it is sensitive to the difficult to control plasma density gradient. One solution to address this sensitivity is to utilize volume injection that occurs when a vortex laser passes through a low-density target. D.~Blackman~{\it et al.}~\cite{blackman2022electron} demonstrated using 3D PIC simulations that this setup allows a 3 PW vortex laser beam to generate a 50 pc low divergence electron beam with a maximum energy of 1.5~GeV. The unique characteristics of the beam are short acceleration distance ($\sim 100 ~\micron$), compact transverse size, high areal density, and electron bunching~\cite{blackman2022electron}. The nonlinear interaction between the intense laser beam and the plasma representing the target changes the transverse field structure. Even though the incident beam is a single LG mode, the transmitted vortex beam is no longer a single mode. Instead, it becomes a mixture of multiple radial modes each with its own $p$. A detailed mode decomposition performed by D.~Blackman~{\it et al.}~\cite{blackman2022electron} shows that the contribution of high-order radial modes to the transmitted beam increases with target density. Since higher radial modes are less effective in accelerating electrons, their excitation is detrimental. This observation allowed D.~Blackman~{\it et al.}~\cite{blackman2022electron} to formulate an upper limit of the target density.

Near the axis, radially polarized (RP) laser beams also have a strong longitudinal electric field that can be utilized for electron acceleration~\cite{Zaim2017, Zaim2020}, so a comparison between these beams and LG beams is worth discussing. The difference in the field topology can be illustrated by constructing an RP beam using two CP-LG beams of the same amplitude: one with $l = 1$, $\sigma = -1$ and another one, a mirror image of the first, with $l = -1$, $\sigma = 1$. The longitudinal electric fields of the two beams combine near the axis, whereas the longitudinal magnetic fields cancel each other out. Therefore, there is no $B_{\parallel}$. On the other hand, the resulting $E_{\parallel}$ matches that of a CP-LG beam with $l = -1$, $\sigma = 1$, and $p=0$. Note that the transverse field amplitude of this CP-LG beam must be twice the amplitude of the beams used to create the RP beam. 
The absence of $B_\parallel$ has a negative effect on electron acceleration by $E_\parallel$, because there is no mechanism for confining electrons close the axis of the reflected beam. A detailed comparison and additional results can be found in Ref.~\cite{shi2021electron}.

The interaction between a CP-LG laser and a foil target has been claimed to play the role of a vortex cutter in the work of W.~P.~Wang~{\it et al.}~\cite{Wang2018}. The claim is based on a simulation with a low resolution and the text lacks some important information, like the position of the focal plane, needed to facilitate a comparison with other published research. 
To make such a comparison, 
we have performed a 3D PIC simulation with parameters close to those in the paper, but using higher resolution and a bigger longitudinal simulation box size. According to our simulation results and the theory and simulations presented by Y.~Shi~{\it et al.}~\cite{shi2021electron}, the position of the beam waist (focal plane) and the resolution are crucial for the electron acceleration~\cite{shi2021electron}. In direct laser acceleration, electrons may gain energy at the beginning of their interaction with the reflected laser and lose it due to the phase slip after traveling a longer distance~\cite{shi2021electron}. 
We found that the mechanism behind the effect reported in Ref.~\cite{Wang2018} is the acceleration by the longitudinal laser electric field.

There have also been computational studies examining electron acceleration by an intense vortex laser beam in setups where the vortex beam interacts with a highly underdense plasma~\cite{Baumann2018}, a wire~\cite{Hu2018}, or a micro-droplet~\cite{Hu2018a}.
In their work, L.-X.~Hu~{\it et al.}~\cite{Hu2018} provide details regarding an interaction of an intense vortex beam with a thin wire aligned along the axis of the beam. 
As the vortex beam encounters the wire and starts traveling along it, attosecond-duration annular electron bunches are periodically emitted from the wire tip. These annular bunches propagate along the wire while remaining confined near the surface by the radial laser electric field. 
Once the electrons reach the end of the wire, they exit into vacuum. At this stage, the inward-directed transverse ponderomotive force of the vortex beam causes the emission angle of electrons to gradually diminish. As a result, each hollow electron bunch is transformed into an electron disk. Simultaneously, under the influence of the longitudinal electric field, the electrons undergo continuous acceleration, with their energy reaching 100s of MeV. The described acceleration process is accompanied by efficient transfer of laser AM to the AM of the electron bunches.
L.-X.~Hu~{\it et al.}~\cite{Hu2018} have also shown that, using a micro-droplet whose transverse size is smaller than the width of the vortex beam, one can generate dense flying relativistic electron mirrors. 

\begin{figure}[H]
\centering 
\includegraphics[scale=0.8]{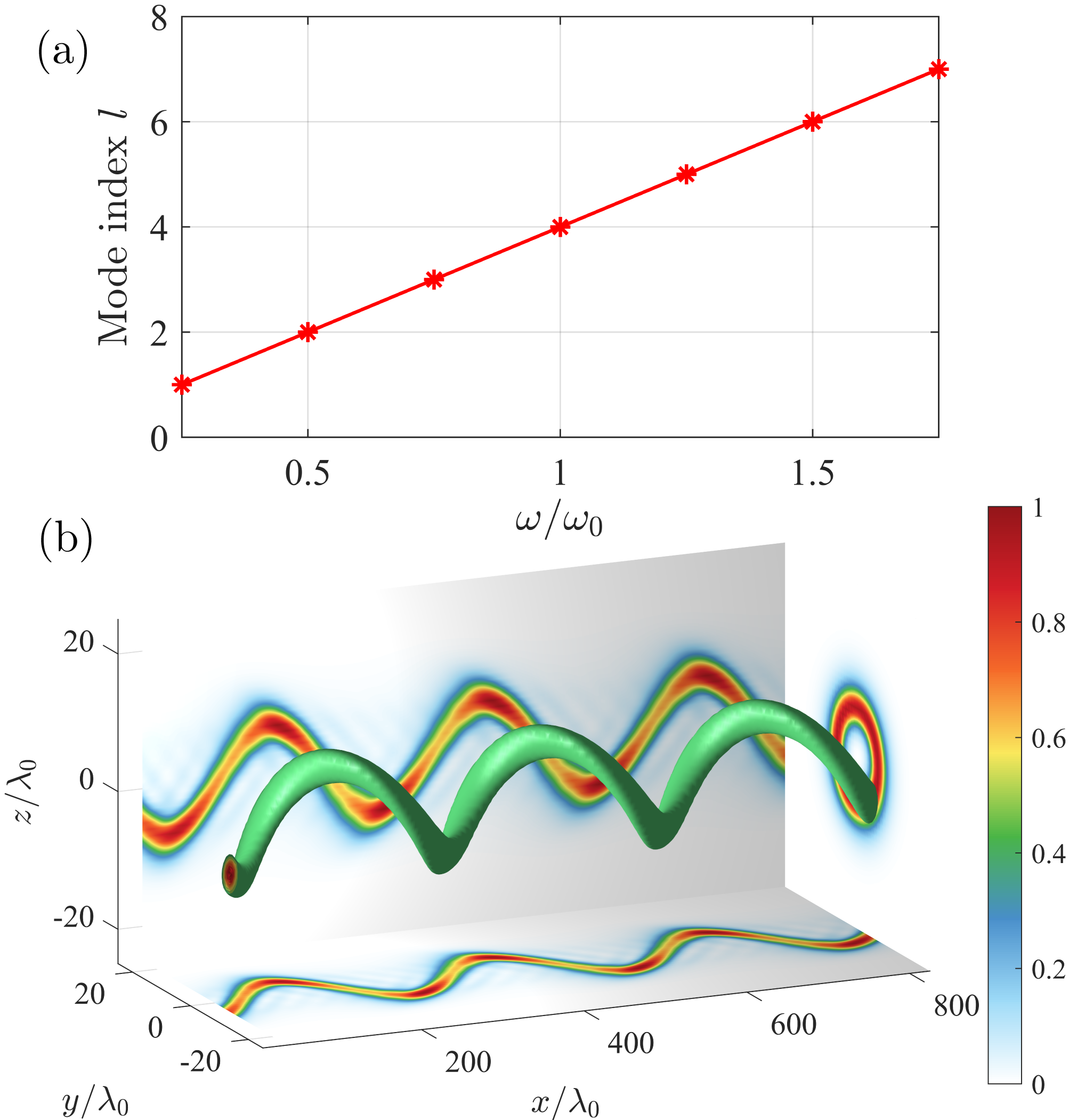} 
\caption{A spatio-temporal light spring created by combining seven LG modes with $1 \leq l \leq 7$ that have a linear relation between the mode frequency and $l$. (a) the linear relationship between the topological charge gap $\Delta l$ and the bandwidth gap $\Delta \omega$: $\Delta l /\Delta \omega = 4/\omega_0 $. (b) the isosurface corresponding to 40\% of the peak laser intensity and showing the typical intensity profile structure of a spatio-temporal light spring.} \label{fig:lightspring}
\end{figure}

\section{Light spring lasers and their interaction with plasmas}
\label{section:lightspring}

In \Cref{section:vortex}, we reviewed new phenomena and physics regimes that can be accessed using LG laser beams. In most of the discussed cases, having a single LG mode is sufficient. However, we have also reviewed regimes where a superposition of two modes can do something that a single LG mode is not able to do. A natural question to ask is whether there is added benefit of combining an even bigger number of LG modes. In what follows, we first show how one can create a so-called spatio-temporally coupled laser beam using multiple specifically selected LG modes and then we explore phenomena that can be accessed using such a beam. Due to the unique structure with a wide OAM bandwidth, interesting phenomena are expected when the intense spatio-temporally coupled laser beams interact with the plasma, such as self-generated magnetic fields, particle acceleration and instability suppression, etc.


\subsection{Spatio-temporal light spring}

A Light Spring (LS) laser is a new type of a spatio-temporally coupled laser beam obtained by superposition of LG beams whose azimuthal mode index $l$ is correlated with their frequency $\omega$. This concept was first introduced theoretically by G. Pariente~{\it et al.}~\cite{Pariente15}. The most striking feature of an LS is the spiral structure of the phase and intensity contours, as shown in~\cref{fig:lightspring}. Unlike the donut-shaped intensity distribution of a single mode LG beam, the high-intensity bright spot of the LS pulse is concentrated within a small arc of the ring, whereas the intensity at other positions on the ring is extremely low (almost zero). A special case of a light spring is the beating of two laser beams in both frequency and twist index. The resulting field structure generates a twisted ponderomotive force and transfers the OAM to the plasma with high efficiency~\cite{Shi2018, shiJUSTC}.

An expression for the field amplitude of an LS pulse can be obtained by combining several LG modes given by \Cref{eq:LGnm}. Our goal here is to provide a simple analysis that provides useful insights. We consider only LG modes with $p = 0$ and restrict our analysis to the area around the beam waist. It follows directly from \Cref{eq:LGnm} that the radial position of the peak intensity in a beam of a given width $w_0$ depends its topological charge $l$: $r = w_0\sqrt{l/2}$.
Therefore, to increase the coherence of the superposed beams and to make the intensity spot more concentrated, we set the width (waist radius) of each LG beam according to its topological charge: $w_{01}\sqrt{l_1}=w_{0n}\sqrt{l_n}$. The maximum intensity of the resulting LS beam is located at $r_* = w_{01}\sqrt{l_1/2}$. We now specifically focus on this radial position and, for simplicity, we consider modes that have the same peak intensity. 
The amplitude of the LS beam is the following superposition of $N$ LG modes with amplitude $a_0$: 
\begin{equation}\label{eq:multi_lg}
    a_{LS}(\theta,t) =a_0 \sum_{n=1}^N \exp \left[ i(l_n\theta-\omega_n (t - x/c)) \right],
\end{equation}
where the topological charge 
\begin{equation}
    l_n=l_1+\Delta l(n-1)
\end{equation}
and the frequency 
\begin{equation}
    \omega_n=\omega_1+\Delta\omega(n-1)
\end{equation}
have a linear relation~\cite{Pariente15, guo2023suppression}. In these expressions, $\omega_1$ is the frequency of the first mode, $l_1$ is the topological charge of the first mode, $\Delta \omega$ is the bandwidth gap, and $\Delta l$ is the topological charge gap. Using the expressions for $l_n$ and $\omega_n$, we obtain
\begin{equation}
    a_{LS}(\theta,t) = a_0 \exp \left[ i l_1 \theta - i \omega_1 (t - x/c) \right]  \sum_{n=1}^N e^{i(n-1) \Delta \psi},
    \label{eq:multi_lg v2}
\end{equation}
where
\begin{equation} \label{Delta psi}
    \Delta \psi = \theta \Delta l  -  (t - x/c) \Delta \omega.
\end{equation}
The sum on the right-hand side has a simple analytical expression
\begin{equation}
    \sum_{n=1}^N e^{i(n-1) \Delta \psi} = \frac{\exp(iN\Delta \psi) -1 }{ \exp(i\Delta \psi) - 1 },
\end{equation}
so that
\begin{equation}
    a_{LS}(\theta,t) = a_0 \frac{\sin (N \Delta \psi /2)}{\sin (\Delta \psi /2)}   e^{i l_1 \theta - i \omega_1 (t - x/c)} e^{ i  [N - 1] \Delta \psi/2}.
    \label{eq:multi_lg v3}
\end{equation}

It is instructive to examine \cref{eq:multi_lg v3} for the LS beam amplitude in the limit of large $N$. The amplitude $a_{LS}$ has a sharp peak at $\Delta \psi = 0$. The condition $\Delta \psi = 0$ describes a rotating helix with
\begin{equation} \label{Delta psi v2}
    \theta = \frac{\Delta \psi}{\Delta l} (t - x/c) .
\end{equation}
The height of the peak is equal to $N a_0$, whereas its width scales as $1/N$. The peak intensity of the constructed LS beam is $I_{LS} = a_{LS} a_{LS}^* \propto N^2 a_0^2$. These results show a clear benefit of coherently combining a large number of LG beams into an LS beam: the intensity has a strong $N^2$ scaling, whereas the angular spread of the intensity peak is reduced as $1/N$. The intensity increase suggests that there might be advantages of using an LS beam for the realization of relativistic effects that require high laser intensity.

LS laser beams have been realized experimentally, albeit at relatively low laser intensity, showcasing intriguing phenomena arising from the correlation between the topological charges and spectral components. 
Utilizing adaptive helical mirrors~\cite{Ghai2011}, one can in principle generate LS beams in experiments. 
Theoretically, an ultrashort LS laser seed can undergo amplification and compression through Raman amplification with a long Gaussian-shaped laser pump~\cite{arteaga2018}.
Current shaping techniques are constrained to narrow topological and/or spectral bands, but a recent work by M.~Piccardo~{\it et al}~\cite{piccardo2023broadband} has provided a much needed  breakthrough. M.~Piccardo~{\it et al} introduced a Fourier space-time shaper capable of handling ultra-broadband pulses covering almost 50\% of the visible spectrum and carrying an extensive range of topological charges, reaching values up to 80~\cite{piccardo2023broadband}. Departing from traditional linear grating geometry, this novel approach employs a diffractive axicon with circular geometry, allowing the imposition of azimuthal phase modulations on beams carrying OAM. The spatio-temporal field was reconstructed using a characterization technique based on hyperspectral off-axis holography. The adjustment of linear topological-spectral correlations provides control over various properties of the wave packets, including chirality, orbital radius, and the number of intertwined helices, while complex correlations enable manipulation of their dynamics. Most recently, Q.~Lin~{\it et al}~\cite{Lin2024} have introduced a new mechanism for direct space-time manipulation of ultrafast light fields. It combines a space-dependent time delay with spatial geometrical transformations and it has been shown to generate high-quality LS in experiments. 


\subsection{Optical control of the topology of laser-plasma accelerators}

Using simulations, J. Vieira {\it et al}~\cite{Vieira2018} demonstrated an important application of an LS beam, showing that it can drive a twisted plasma accelerator. In this work, the LS beam is created by combining two vortex beams with different frequencies and different twist indices~\cite{Vieira2018, Shi2018}. In the simulation, the rotation of the light spring stimulates a twisted plasma wakefield that then accelerates plasma electrons. By tuning the properties of the LS, one can tune the phase velocity of the twisted wakefield. The acceleration by the resulting wakefield not only enhances electron energy gain, but it also improves trapping efficiency beyond what planar wakefields offer. The twisted plasma accelerator has the capacity to produce relativistic electron vortex beams characterized by helical current profiles. The AM of these vortex bunches is quantized. Simulations show that the corresponding azimuthal particle motion dominates the transverse particle dynamics, which leads to spiral particle trajectories around the twisted wakefield~\cite{Vieira2018, LuísMartins2019}.
The helical motion of the electrons in a twisted plasma accelerator driven by the above LS is found to dominate the radiation emission, rather than the betatron oscillation~\cite{LuísMartins2019}.

One of the key results of Ref.~\cite{Vieira2018} is proportionality relation $\Delta L_x/\Delta p_x = l_p/k_p$ for accelerated electrons in the twisted electrostatic wakefield. Here $p_x$ is the longitudinal momentum, $L_x$ is the longitudinal AM, and $l_p$ is the AM acquired by the wakefield of the LS driver. In the considered setup, $l_p = l(\omega + \omega_p) - l(\omega) \approx \omega_p dl/d\omega $.
When the trapped electrons are accelerated, this proportionality relation gives rise to a topological structure that is visible in the simulations. J. Vieira {\it et al}~\cite{Vieira2018} showed that the laser pulse envelope moves relative to LS intensity helix, with the direction of the relative motion determined by the sign of $l^{'} = dl/d\omega$. This relative motion offers a mechanism for adjusting the phase velocity of the twisted wakefield and increasing electron energy gain.
In another work, L.~B.~Ju~{\it et al}~\cite{Ju2018} considered a regime where the discussed acceleration mechanism is supplemented by direct laser acceleration from the transverse laser electric field. The regime makes it possible to manipulate the topological structure of the generated ultrarelativistic electron beams. 

\begin{figure*}
 \centering  
 \includegraphics[scale=1.6]{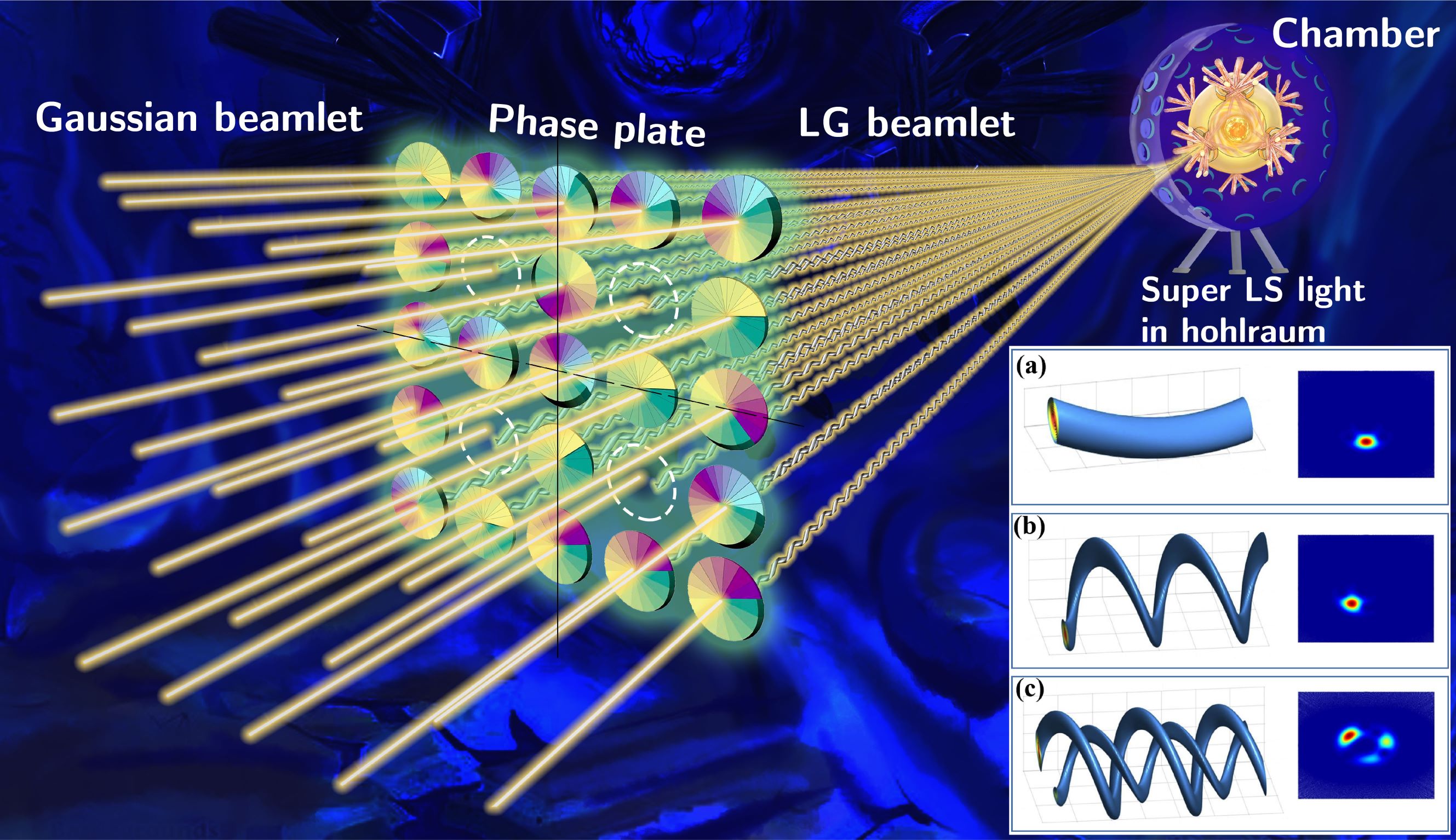} 
\caption{Schematic setup for the generation of a super LS beam within a hohlraum situated at the center of a target chamber. The process involves  combining LG beamlets at varying frequencies, distinct topological charges, and slightly different incidence angles. These beamlets are transformed from Gaussian beamlets using different phase plates. By controlling the phase and frequency of the LG beamlets using the phase plates, Y. Guo~{\it et al}~\cite{guo2023suppression} have achieved three distinct cases at the observation plane: (a) a long pitch LS beam, (b) a short pitch LS beam, and (c) a super LS beam.  Copyright 2023 by AIP.} \label{fig:LS_shen}
\end{figure*}

\subsection{Laser-plasma instability suppression}

Laser-plasma instabilities pose a significant challenge in laser-driven inertial confinement fusion (ICF), impeding the attainment of predictable and reproducible fusion with high gain. Addressing these instabilities is crucial for advancing fusion research. These instabilities can be mitigated through either temporal or spatial incoherence of the driving laser.

Y. Guo~{\it et al}~\cite{guo2023suppression} found that it is also possible to suppress laser-plasma instabilities via angular incoherence when using laser beams carrying AM. Angular coherence is related to AM, whereas temporal and spatial coherence are related to frequency (energy) and momentum. 
In their work, Y.~Guo~{\it et al}~focused on Stimulated Raman Scattering (SRS). SRS is a three-wave coupling process related to the decay of a driving laser into an electron plasma wave and a scattering wave. Therefore, the following frequency and wave number matching conditions should be satisfied: 
$\omega_L=\omega_1+\omega_2$ and $\mathbf{k}_L=\mathbf{k}_1+\mathbf{k}_2$. These conditions are the consequences of the energy and momentum conservation. AM should be conserved as well, but this requirement is typically inconsequential, because conventional laser beams carry no intrinsic AM. The situation however qualitatively changes if the driving beams carry AM. In this case, the conservation of AM requires that $\mathbf{L}_L=\mathbf{L}_1+\mathbf{L}_2$. This condition indicates that angular incoherence can be used to suppress SRS.

Y. Guo~{\it et al}~\cite{guo2023suppression} derived two expressions for the SRS 
growth rate: one considering a frequency spread similar to that present in an LS beam and another one considering a topological charge spread similar to that present in an LS beam. The two corresponding correction terms to the growth rate without the spread reduce the instability growth rate. The correction term that is determined by the total bandwidth $\Delta \omega$ is well known and has been extensively studied in ICF. The other other correction term is  determined by the total topological charge spread $\Delta l$ or the AM spread. Y. Guo~{\it et al} provide a condition on the laser and plasma parameters that guarantees that the $\Delta l$-term is dominant and provide a corresponding condition~\cite{guo2023suppression}. In this regime,  
the AM spread plays a more important role than the bandwidth in suppressing SRS. It is argued that such a regime is easily accessible. 
3D PIC simulations performed by Y. Guo~{\it et al} confirm that the angular incoherence can provide much stronger suppression of the SRS growth rate than the well-known temporal and spatial incoherence that are commonly used in ICF studies.

At a laser facility, LS bundles exhibiting desired temporal, spatial, and angular incoherence can be generated by combining vortex beamlets. A schematic setup is shown in \cref{fig:LS_shen}. By precisely manipulating the phase and frequency of the vortex beamlets before their combination, three distinct types of LS bundles are attainable~\cite{guo2023suppression}: 
1) Long pitch LS beam shown in \cref{fig:LS_shen}(a) and characterized by narrowband features and a single strong point of time-independent intensity;
2) Short pitch LS beam shown in \cref{fig:LS_shen}(b) and characterized by broadband characteristics and a single strong point of time-dependent intensity;
3) Super LS beam shown in \cref{fig:LS_shen}(c) and characterized by multiple strong points of time-dependent intensity, attributed to random phase distribution.
Simulation results for these cases indicate that angular incoherence significantly reduces the growth rate of SRS. Notably, the suppression effect when using the super LS beam is particularly pronounced, surpassing even the suppression achieved by temporal incoherence. Therefore, LS beams can be central to creating a low-LPI laser system.  The demonstrated ability of an LS pulse with angular incoherence to suppress SRS can potentially be extrapolated to the suppression of Stimulated Brillouin Scattering (SBS) and other parametric instabilities, holding substantial significance for laser-driven ICF.

\section{Spatiotemporal vortex beam interaction with plasma}
\label{section:STOV}

\begin{figure}[htp]
 \centering  
 \includegraphics[scale=0.8]{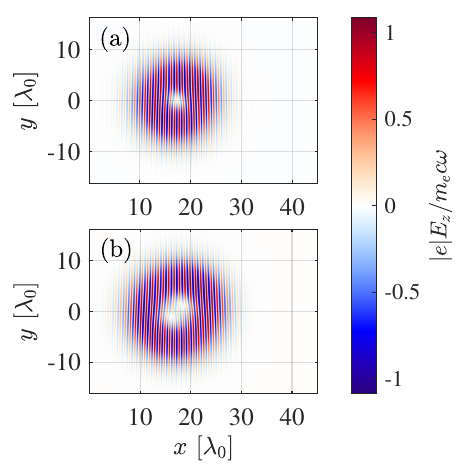} 
\caption{Transverse field distribution of  ST vortex beams polarized along the $z$-axis is shown in the $x-y$ plane for (a) $l = 1$ and (b) $l = 2$. The ST vortex beams are propagating along the $x$ axis. In contrast to conventional vortex beams whose OAM is along the propagation direction, the ST vortex beams in (a) and (b) carry transverse OAM.} \label{fig:STOV}
\end{figure}

A spatial vortex beam has a spiral phase in the $x-y$ plane of the form $\exp(i l \phi)$ aligned with the beam axis with a phase singularity of zero intensity at the center. Such a phase possesses purely spatial singularity and provides the beam with a transverse donut-intensity distribution and longitudinal OAM that is parallel to its propagation direction~\cite{Yao2011,Shen2019_Light}. 
Apart from the longitudinal OAM, theoretical predictions ~\cite{10.1117/12.623906,PhysRevA.86.033824,BLIOKH20151, Hancock2021} and experiments~\cite{Jhajj2016,Hancock:19,Hancock2024,chong2020generation} in optics indicate that light can possess a transverse OAM which is orthogonal to the propagation direction. For easy understanding, we can imagine that the spatial vortex beam is rotated by $90^{\circ}$ with respect to $x$ axis, resulting in a 3D wave packet with a transverse OAM as shown in~\cref{fig:STOV}. Such wave packet has a spiral phase in the spatiotemporal plane ($x$-$t$ plane in ~\cref{fig:STOV}). In contrast to the conventional monochromatic spatial vortex beam which has to be described in 3D geometry, such light is essentially polychromatic with a phase singularity of zero intensity at the center in the spatiotemporal domain, and therefore called as the spatiotemporal optical vortex (STOV)~\cite{Jhajj2016}. The spatial and temporal profiles of the STOV pulse are strongly coupled, but its description can be simplified in 2D geometry. Especially, the accurate theoretical analysis of the STOV pulse by constructing spatiotemporal Bessel-type solutions using full vector description has also been provided~\cite{PhysRevLett.126.243601}.
The special structure of STOV pulses, as well as the transverse OAM, is expected to greatly expand the applications of wave vortices. For example, a spatio-temporal wave packet with orientation-controllable OAM can be generated by assembling transverse OAM and longitudinal OAM, which can be applied in optical spanners with arbitrary three-dimensional orientation~\cite{10.1093/nsr/nwab149,WAN20201334,Wang:21}. 

The quasi-monochromatic spatiotemporal vortex beam can be simply expressed as follows,
\begin{equation}\label{eq:E-STOV}
\begin{aligned}
E(x,y,t)=~&E_0 \left(\hat y+i\sigma\hat z \right) \left(\frac{\sqrt{2} \widetilde{r}} {\omega_0} \right)^{\vert l \vert}  \\
    &\times \exp\left(\frac{-\widetilde{r}^2} {\omega_0^2} \right)\exp\left[i\left(k_0\xi +l \widetilde{\varphi} \right)\right],
\end{aligned}
\end{equation}
where $E_0$ is the amplitude of the laser electric field, $\omega_0 = 2 \pi c/\lambda_0$ is the frequency, and $k_0 = \omega_0 /c$ is the wave number. 
The phase of the spatiotemporal vortex is described by $l\widetilde{\varphi}$, where $l$ is its topological charge, $\widetilde{\varphi}$ =arctan$(y / \xi)$ is the azimuth angle in the spatiotemporal domain, $\xi =x-ct=\widetilde{r} \cos{\varphi}$ is its spatiotemporal coupling coordinate, and $\widetilde{r}=\sqrt{\xi^2+y^2} )$, $(r, \varphi)$ is the polar coordinate of the $(\xi, y)$ plane.
In recent years, the generation, propagation, reflection and refraction of STOV beams have been studied in detail. For example, when a STOV beam is reﬂected and refracted at a plane surface, novel types of transverse pulse shifts and time delays can be induced~\cite{MazanovSugicAlonsoNoriBliokh+2022+737+744}. This phenomenon is quite different from the spatial vortex case. In addition, the nonlinear effect such as the generation of high harmonics carrying TOAM from crystal or gaseous target has also been investigated~\cite{Hancock:21,Gui2021,PhysRevLett.127.273901}. It should be noted that all of the above studies are performed with intense STOV beam of intensity $<10^{18}\rm{W/cm}^2$, which is relatively easy to realize in experiments, considering the limitations of the media damage thresholds. To improve the STOV intensity, it has been demonstrated that STOV beams can be compressed to femtosecond pulse durations and be focused to subwavelength spatial sizes, showing the ability to yield high-intensity STOV pulses~\cite{Chen:20,Rui:22}. In addition, the coherent beam combing technique by superposing intense plane waves with different wave vectors is an alternative approach to produce the high-intensity STOV pulses. Plasma-based methods have also been proposed to produce high-intensity vortex beams with tilted or transverse OAM~\cite{Qiu_2019,Chen2022,Zhang_2022}. In this regard, it is reasonable to expect that the STOV beam with higher intensity can be applied in broader fields, so that the intense (relativistic) STOV beam generation and interaction with plasma has been attempted in recent years.

\begin{figure}[H]
 \centering  
 \includegraphics[scale=0.95]{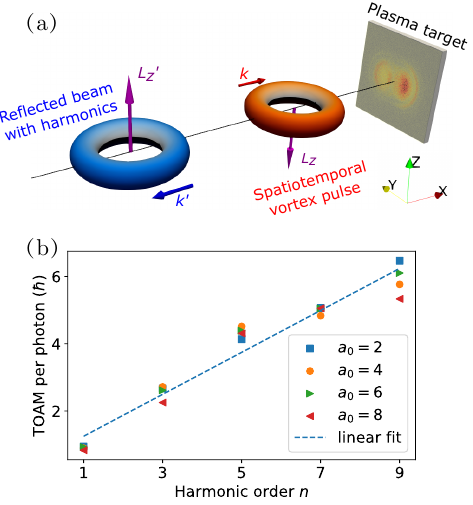} 
\caption{ (a) A spatiotemporal optical vortex (STOV) pulse, linearly polarized along the $z$-axis ($z$-polarized), exhibiting purely transverse orbital angular momentum (TOAM) denoted by the red torus, is directed onto a solid plasma target. Harmonics are anticipated to be generated in the reflected beam, represented by the blue torus.
(b) The graph illustrates the TOAM values ($L_z$) per photon of harmonics corresponding to STOV drivers with different normalized vector potentials ($a_0$). Four scenarios are considered with $a_0$ values of 2 (square), 4 (circle), 6 (right triangle), and 8 (left triangle). The blue dashed line signifies a linear fit depicting the average TOAM for each harmonic order~\cite{LGzhang2022}.} \label{fig:STOV_LGzhang}
\end{figure}

Spatiotemporal high harmonics carrying TOAM generation driven by a linearly polarized relativistic STOV pulse irradiating a solid plasma target has been demonstrated by L. G. Zhang~{\it et al.}~\cite{LGzhang2022}, as shown in~\cref{fig:STOV_LGzhang}. By combining an ultraintense Gaussian `pump' pulse and a weak STOV `seed' pulse with orthogonal polarizations, a scheme has been proposed to increase the energy conversion efficiency~\cite{Wu:23}.
It has been found that an interesting deflection effect deviating from the optical reflection law occurs when an intense LG pulse obliquely impinges onto a solid target~\cite{Zhang2016}.
For the STOV pulse, a deflection effect deviating from the optical reflection law also occurs~\cite{Guo:23}. The noticeable deflection angle increases with the intensity and topological charge of the STOV beam. Unlike the shear press of the vortex beam, which occurs only at oblique incidence, this phenomenon induced by the STOV pulse also exists at normal incidence. Recently, M. S. Le~{\it et al.}~\cite{le2024selffocused} give a new understanding of self-focused propgation of intense laser in nonlinear medium from the formation and evolution of STOVs.

\section{Conclusions and future prospects}

Technological innovations in laser beam control have enabled transformative developments and breakthroughs in the field of laser-plasma interactions. Examples include the invention of chirped pulse amplification that enabled reaching ultra-high laser intensities~\cite{cpa1985}, the development of ultra-broadband lasers that enabled mitigation of laser-plasma instabilities, the development of controlable tilt or curvature of pulse-front ~\cite{Lichaoyang2024} that enabled isolated attosecond pulses generation~\cite{Vincenti2012} and dephasingless laser wakefield acceleration~\cite{Palastro2020}, and the development of the flying focus technique that enabled precise spatiotemporal manipulation of laser intensity~\cite{Froula2018}. The generation of vortex beams that carry OAM~\cite{Allen1992} is another important example. These beams have attracted significant interest and stimulated extensive research. In this paper, we provide an up-to-date overview of the field. In \Cref{section:vortex-generation}, we review various methodologies for vortex beam generation. Despite remarkable progress, further efforts are required to reliably achieve complex and intense vortex beams carrying OAM, such as quality LG beams with a high twist index, light spring beams, and spatiotemporal optical vortex beams. Additionally, novel diagnostic methods are imperative to  probe the structure and behavior of high-power vortex beams and to elucidate their interactions with plasmas.

Intense laser beams exhibit notable properties such as high energy density, high momentum density, and high AM density. The energy, momentum, and AM must be conserved during laser-plasma interactions. Therefore, laser-plasma interactions can be viewed within the context of these conservation laws. Efficient energy transfer during laser-plasma interactions facilitates the creation of high-energy-density plasmas, unlocking numerous research opportunities in the field of high energy density science. Efficient momentum transfer from the laser beam to particles during laser-plasma interactions enables the development of compact laser-driven accelerators, unlocking numerous research opportunities associated with energetic particle beams. Similarly, one can expect new phenomena and new research opportunities associated with efficient AM transfer from vortex laser beams to the irradiated plasma. For example, it is intuitively reasonable to expect an environment with high rotation speed or strong magnetic fields. The publications discussed in this review clearly illustrate that vortex beams provide access to previously inaccessible regimes involving high density AM. Based on the material discussed in
\Cref{section:vortex}, \Cref{section:lightspring}, and \Cref{section:STOV}, we can identify four promising general research areas involving vortex beams. In what follows, we briefly summarize some of the key ideas in each area.

\begin{itemize}
    \item {\bf Particle acceleration with low divergence.} Vortex laser beams offer a customizable field structure that can be used to improve electron acceleration. Presented simulation studies show that the divergence of electrons driven directly by the laser may be reduced due to the spatial separation of longitudinal and transverse fields of a tightly focused vortex beam, the self-generated axial magnetic fields caused by OAM absorption, and the reduced relativistic self-focusing caused by the hollow intensity profile. In most cases, plasma ions are accelerated by hot electrons, so it is reasonable to expect that reduced electron divergence can benefit laser-driven ion acceleration. Future studies coupling simulations and dedicated experiments can be particularly insightful. The reduced electron divergence is also viewed as a benefit in the context of laser-driven radiation source and fast ignition. Therefore, exploring electron acceleration by vortex beams in the context of these specific applications might be particularly promising.
    
    \item {\bf High-energy photons carrying OAM.} X-ray and $\gamma$-ray beams with OAM can be impactful for atomic and nuclear physics studies. One way for generating these beams that takes advantage of high-power optical vortex beams is the inverse Compton scattering in a collision of a vortex laser with an electron beam. Although the AM of a vortex beam with a high-twist index can significantly exceed that of a regular CP beam with the same power, the production of high power vortices with large twist indeces poses serious challenges. When comparing an $|l| = 1$ vortex beam  with a CP laser beam, the ease of producing X-rays and $\gamma$-rays with large twist indeces using these beams must be an important consideration. 
    Additional studies that directly assess advantages of using intense optical vortex beams can be particularly impactful in steering future experimental research in this area. 
      
    \item {\bf Strong magnetic field generation.} Macroscopic magnetic fields whose strength greatly exceeds that of a conventional static magnetic field can find impactful applications in high energy density science and laboratory astrophysics. High-intensity vortex lasers can transfer their axial OAM to the plasma, thus offering a tool for generating strong plasma magnetic fields via the Inverse Faraday effect. Even with energy conversion efficieny around 1\%, the amplitude of the quasi-static plasma magnetic field can reach 10\% of the laser amplitude. In the case of intense vortex and CP beams, the plasma magnetic field may impact the laser-plasma interactions. Other methods of producing magnetic fields, such as laser-driven coils and multiple laser beams, exist and they need to be evaluated and compared to the approaches offered by vortex beams.
    
    \item {\bf Instability suppression.}  It is well known that temporal or spatial incoherence of a driving laser can suppress  instabilities in laser-driven ICF. A light spring can be used to itroduce angular incoherence, which can be seen as a new control knob in spatial incoherence and a potential avenue for instability suppression. The rotating effect linked to axial OAM of the plasma can also contribute to instability suppression. Studies of possible instability suppression in a laser interaction with near-critical density plasma and during radiation pressure acceleration are likely to provide valuable insights regarding vortex laser beams.
 
\end{itemize}

The research involving vortex beams is inderdisciplinary, conencting several diverse fields of research. Vortex-related research can be found in atomic and molecular physics~\cite{Hernandez-Garcia2013}, quantum electrons~\cite{Bliokh2007}, astrophysics~\cite{Martin2003, Thidé2011, Tamburini2011}, and plasma physics~\cite{Stenzel2015}. Vortex-related physics also plays a role in studies of elastic photon-photon scattering in vacuum~\cite{Aboushelbaya2019}, in studies of gratival waves~\cite{Iwo2016}, in the development of plasma diagnostics~\cite{Tsujimura2021, Zhangsiming2022}, and in studies of wave propagation past partial and total obstructions~\cite{Rop_2012}. The terminology varies from field to field. Different terms like vortex beams, twisted beams, Laguerre-Gaussian beams, helical beams, structured beams, and helicons are used in different research areas. However, an important unifying feature of these waves is the OAM. 
In this paper, we emphasize the relativistic vortex laser interaction with plasma, where many nonlinear effects occur when the speed of electrons is close to the speed of light.
Despite our best efforts, we might have unintentionally missed some relevant contributions or connections when preparing this review. If this is the case, we hope that this review will prompt future publications estabilishing new connections not covered here, linking the laser-plasma community with other research communities. Of particular importance is the development of new approaches for utilizing the unique features offered by intense vortex beams, as this will likely offer new research opportunities.

\section*{Acknowledgments}
Yin Shi acknowledges the support by the National Natural Science Foundation of China (Grant No. 12322513), USTC Research Funds of the Double First-Class Initiative, CAS Project for Young Scientists in Basic Research (Grant No. YSBR060). 
Yin Shi also acknowledge Rui Yan and Yi Cai for enthusiastic discussions. Baifei Shen acknowledges the support by the National Natural Science Foundation of China (Grant No. 11935008). Alexey Arefiev was supported by the US DOE Office of Fusion Energy Sciences under Award No. DE-SC0023423.




\bibliographystyle{ieeetr}

\begin{thebibliography}{185}

\bibitem{cpa1985}
D.~Strickland and G.~Mourou, ``Compression of amplified chirped optical pulses,'' {\em Optics Communications}, vol.~56, no.~3, pp.~219--221, 1985.

\bibitem{Jackson1999}
J.~D. Jackson, {\em Classical electrodynamics}.
\newblock New York, {NY}: Wiley, 3rd ed.~ed., 1999.

\bibitem{Allen1992}
L.~Allen, M.~W. Beijersbergen, R.~J.~C. Spreeuw, and J.~P. Woerdman, ``Orbital angular momentum of light and the transformation of laguerre-gaussian laser modes,'' {\em Physical Review A}, vol.~45, no.~11, pp.~8185--8189, 1992.
\newblock PRA.

\bibitem{Yao2011}
A.~M. Yao and M.~J. Padgett, ``Orbital angular momentum: origins, behavior and applications,'' {\em Advances in Optics and Photonics}, vol.~3, no.~2, pp.~161--204, 2011.

\bibitem{Shen2019_Light}
Y.~Shen, X.~Wang, Z.~Xie, C.~Min, X.~Fu, Q.~Liu, M.~Gong, and X.~Yuan, ``Optical vortices 30 years on: Oam manipulation from topological charge to multiple singularities,'' {\em Light: Science {\&} Applications}, vol.~8, p.~90, Oct 2019.

\bibitem{Mair2001}
A.~Mair, A.~Vaziri, G.~Weihs, and A.~Zeilinger, ``Entanglement of the orbital angular momentum states of photons,'' {\em Nature}, vol.~412, p.~313, 2001.

\bibitem{Mann2018}
A.~Mann, ``Core concept: {\textquotedblleft}twisted{\textquotedblright} light beams promise an optical revolution,'' {\em Proceedings of the National Academy of Sciences}, vol.~115, no.~22, pp.~5621--5623, 2018.

\bibitem{Padgett2011}
M.~Padgett and R.~Bowman, ``Tweezers with a twist,'' {\em Nature Photonics}, vol.~5, p.~343, 2011.

\bibitem{Zurch2012}
M.~Zurch, C.~Kern, P.~Hansinger, A.~Dreischuh, and C.~Spielmann, ``Strong-field physics with singular light beams,'' {\em Nature Physics}, vol.~8, no.~10, pp.~743--746, 2012.
\newblock 10.1038/nphys2397.

\bibitem{Primoz2014}
P.~R. Ribi\ifmmode~\check{c}\else \v{c}\fi{}, D.~Gauthier, and G.~De~Ninno, ``Generation of coherent extreme-ultraviolet radiation carrying orbital angular momentum,'' {\em Physical Review Letters}, vol.~112, p.~203602, May 2014.

\bibitem{Pariente15}
G.~Pariente and F.~Qu\'{e}r\'{e}, ``Spatio-temporal light springs: extended encoding of orbital angular momentum in ultrashort pulses,'' {\em Opt. Lett.}, vol.~40, pp.~2037--2040, May 2015.

\bibitem{Aiello2015}
A.~Aiello, P.~Banzer, M.~Neugebauer, and G.~Leuchs, ``From transverse angular momentum to photonic wheels,'' {\em Nature Photonics}, vol.~9, pp.~789--795, Dec 2015.

\bibitem{Mourou2006}
G.~A. Mourou, T.~Tajima, and S.~V. Bulanov, ``Optics in the relativistic regime,'' {\em Reviews of Modern Physics}, vol.~78, no.~2, pp.~309--371, 2006.
\newblock RMP.

\bibitem{Esarey2009}
E.~Esarey, C.~B. Schroeder, and W.~P. Leemans, ``Physics of laser-driven plasma-based electron accelerators,'' {\em Reviews of Modern Physics}, vol.~81, pp.~1229--1285, 2009.

\bibitem{Macchi2013}
A.~Macchi, M.~Borghesi, and M.~Passoni, ``Ion acceleration by superintense laser-plasma interaction,'' {\em Reviews of Modern Physics}, vol.~85, no.~2, pp.~751--793, 2013.
\newblock RMP.

\bibitem{Teubner2009}
U.~Teubner and P.~Gibbon, ``High-order harmonics from laser-irradiated plasma surfaces,'' {\em Reviews of Modern Physics}, vol.~81, no.~2, pp.~445--479, 2009.
\newblock RMP.

\bibitem{Corde2013}
S.~Corde, K.~Ta~Phuoc, G.~Lambert, R.~Fitour, V.~Malka, A.~Rousse, A.~Beck, and E.~Lefebvre, ``Femtosecond x rays from laser-plasma accelerators,'' {\em Reviews of Modern Physics}, vol.~85, pp.~1--48, 2013.

\bibitem{danson2015}
C.~Danson, D.~Hillier, N.~Hopps, and D.~Neely, ``Petawatt class lasers worldwide,'' {\em High Power Laser Science and Engineering}, vol.~3, p.~e3, 2015.

\bibitem{elinp10pw2022}
C.~Radier, O.~Chalus, M.~Charbonneau, S.~Thambirajah, G.~Deschamps, S.~David, J.~Barbe, E.~Etter, G.~Matras, S.~Ricaud, and et~al., ``10 pw peak power femtosecond laser pulses at eli-np,'' {\em High Power Laser Science and Engineering}, p.~1–5, 2022.

\bibitem{Li2022}
Z.~Li, Y.~Leng, and R.~Li, ``Further development of the short-pulse petawatt laser: Trends, technologies, and bottlenecks,'' {\em Laser \& Photonics Reviews}, vol.~17, no.~1, p.~2100705, 2022.

\bibitem{Wang2022sel}
X.~Wang, X.~Liu, X.~Lu, J.~Chen, Y.~Long, W.~Li, H.~Chen, X.~Chen, P.~Bai, Y.~Li, Y.~Peng, Y.~Liu, F.~Wu, C.~Wang, Z.~Li, Y.~Xu, X.~Liang, Y.~Leng, and R.~Li, ``13.4~fs, 0.1~hz opcpa front end for the 100 pw-class laser facility,'' {\em Ultrafast Science}, vol.~2022, 2022.

\bibitem{Liang2020sulf}
X.~Liang, Y.~Leng, R.~Li, and Z.~Xu, ``Recent progress on the shanghai superintense ultrafast laser facility (sulf) at siom,'' in {\em OSA High-brightness Sources and Light-driven Interactions Congress 2020 (EUVXRAY, HILAS, MICS)}, p.~HTh2B.2, Optica Publishing Group, 2020.

\bibitem{Forbes2021}
A.~Forbes, M.~de~Oliveira, and M.~R. Dennis, ``Structured light,'' {\em Nature Photonics}, vol.~15, pp.~253--262, Apr 2021.

\bibitem{Shi2014}
Y.~Shi, B.~Shen, L.~Zhang, X.~Zhang, W.~Wang, and Z.~Xu, ``Light fan driven by a relativistic laser pulse,'' {\em Physical Review Letters}, vol.~112, no.~23, p.~235001, 2014.
\newblock PRL.

\bibitem{Denoeud2017}
A.~Denoeud, L.~Chopineau, A.~Leblanc, and F.~Quéré, ``Interaction of ultraintense laser vortices with plasma mirrors,'' {\em Physical Review Letters}, vol.~118, no.~3, p.~033902, 2017.
\newblock PRL.

\bibitem{porat2022spiral}
E.~Porat, S.~Lightman, I.~Cohen, and I.~Pomerantz, ``Spiral phase plasma mirror,'' {\em Journal of Optics}, vol.~24, no.~8, p.~085501, 2022.

\bibitem{Longman2017}
A.~Longman and R.~Fedosejevs, ``Mode conversion efficiency to laguerre-gaussian oam modes using spiral phase optics,'' {\em Opt. Express}, vol.~25, pp.~17382--17392, Jul 2017.

\bibitem{BAE2020103499}
J.~Y. Bae, C.~Jeon, K.~H. Pae, C.~M. Kim, H.~S. Kim, I.~Han, W.-J. Yeo, B.~Jeong, M.~Jeon, D.-H. Lee, D.~U. Kim, S.~Hyun, H.~Hur, K.-S. Lee, G.~H. Kim, K.~S. Chang, I.~W. Choi, C.~H. Nam, and I.~J. Kim, ``Generation of low-order laguerre-gaussian beams using hybrid-machined reflective spiral phase plates for intense laser-plasma interactions,'' {\em Results in Physics}, vol.~19, p.~103499, 2020.

\bibitem{wang2020hollow}
W.~P. Wang, C.~Jiang, H.~Dong, X.~M. Lu, J.~F. Li, R.~J. Xu, Y.~J. Sun, L.~H. Yu, Z.~Guo, X.~Y. Liang, Y.~X. Leng, R.~X. Li, and Z.~Z. Xu, ``Hollow plasma acceleration driven by a relativistic reflected hollow laser,'' {\em Phys. Rev. Lett.}, vol.~125, p.~034801, Jul 2020.

\bibitem{Longman2022}
A.~Longman and R.~Fedosejevs, ``{Modeling of high intensity orbital angular momentum beams for laser–plasma interactions},'' {\em Physics of Plasmas}, vol.~29, p.~063109, 06 2022.

\bibitem{Burger:23}
M.~Burger, J.~Murphy, L.~Finney, N.~Peskosky, J.~Nees, K.~Krushelnick, and I.~Jovanovic, ``Wavefront uniformity optimization of laguerre-gaussian ultrafast beams,'' in {\em Optica Nonlinear Optics Topical Meeting 2023}, p.~M2B.2, Optica Publishing Group, 2023.

\bibitem{Leblanc2017}
A.~Leblanc, A.~Denoeud, L.~Chopineau, G.~Mennerat, P.~Martin, and F.~Quere, ``Plasma holograms for ultrahigh-intensity optics,'' {\em Nature Physics}, vol.~13, pp.~440--443, 2017.

\bibitem{Sueda2004}
K.~Sueda, G.~Miyaji, N.~Miyanaga, and M.~Nakatsuka, ``Laguerre-gaussian beam generated with a multilevel spiral phase plate for high intensity laser pulses,'' {\em Opt. Express}, vol.~12, pp.~3548--3553, Jul 2004.

\bibitem{Brabetz2012}
C.~Brabetz, U.~Eisenbarth, O.~Kester, T.~Stoehlker, T.~E. Cowan, B.~Zielbauer, and V.~Bagnoud, ``Hollow beam creation with continuous diffractive phase mask at phelix,'' in {\em Conference on Lasers and Electro-Optics 2012}, p.~JTu1K.5, Optica Publishing Group, 2012.

\bibitem{Brabetz2015}
C.~Brabetz, S.~Busold, T.~Cowan, O.~Deppert, D.~Jahn, O.~Kester, M.~Roth, D.~Schumacher, and V.~Bagnoud, ``Laser-driven ion acceleration with hollow laser beams,'' {\em Physics of Plasmas}, vol.~22, no.~1, p.~013105, 2015.

\bibitem{chen2022forty}
Z.~Chen, S.~Zheng, X.~Lu, X.~Wang, Y.~Cai, C.~Wang, M.~Zheng, Y.~Ai, Y.~Leng, S.~Xu, {\em et~al.}, ``Forty-five terawatt vortex ultrashort laser pulses from a chirped-pulse amplification system,'' {\em High Power Laser Science and Engineering}, vol.~10, p.~e32, 2022.

\bibitem{pan2020generation}
W.~Pan, X.~Liang, L.~Yu, A.~Wang, J.~Li, and R.~Li, ``Generation of terawatt-scale vortex pulses based on optical parametric chirped-pulse amplification,'' {\em IEEE Photonics Journal}, vol.~12, no.~3, pp.~1--8, 2020.

\bibitem{Feng2023}
R.~Feng, J.~Qian, Y.~Peng, Y.~Li, W.~Li, Y.~Leng, and R.~Li, ``Terawatt-class few-cycle short-wave infrared vortex laser,'' {\em Ultrafast Science}, vol.~3, p.~0039, 2023.

\bibitem{Vieira2016}
J.~Vieira, R.~M. G.~M. Trines, E.~P. Alves, R.~A. Fonseca, J.~T. Mendonça, R.~Bingham, P.~Norreys, and L.~O. Silva, ``Amplification and generation of ultra-intense twisted laser pulses via stimulated raman scattering,'' {\em Nature Communications}, vol.~7, p.~10371, 2016.

\bibitem{Mendonca2009}
J.~T. Mendonça, B.~Thidé, and H.~Then, ``Stimulated raman and brillouin backscattering of collimated beams carrying orbital angular momentum,'' {\em Physical Review Letters}, vol.~102, no.~18, p.~185005, 2009.
\newblock PRL.

\bibitem{Trines2020}
R.~M. G.~M. Trines, E.~P. Alves, E.~Webb, J.~Vieira, F.~Fi{\'u}za, R.~A. Fonseca, L.~O. Silva, R.~A. Cairns, and R.~Bingham, ``New criteria for efficient raman and brillouin amplification of laser beams in plasma,'' {\em Scientific Reports}, vol.~10, p.~19875, Nov 2020.

\bibitem{Wu2024}
Y.~Wu, C.~Zhang, Z.~Nie, M.~Sinclair, A.~Farrell, K.~A. Marsh, E.~P. Alves, F.~Tsung, W.~B. Mori, and C.~Joshi, ``Efficient generation and amplification of intense vortex and vector laser pulses via strongly-coupled stimulated brillouin scattering in plasmas,'' {\em Communications Physics}, vol.~7, p.~18, Jan 2024.

\bibitem{Marrucci2006}
L.~Marrucci, C.~Manzo, and D.~Paparo, ``Optical spin-to-orbital angular momentum conversion in inhomogeneous anisotropic media,'' {\em Physical Review Letters}, vol.~96, no.~16, p.~163905, 2006.
\newblock PRL.

\bibitem{Slussarenko:11}
S.~Slussarenko, A.~Murauski, T.~Du, V.~Chigrinov, L.~Marrucci, and E.~Santamato, ``Tunable liquid crystal q-plates with arbitrary topological charge,'' {\em Opt. Express}, vol.~19, pp.~4085--4090, Feb 2011.

\bibitem{Gauthier:19}
D.~Gauthier, S.~Kaassamani, D.~Franz, R.~Nicolas, J.-T. Gomes, L.~Lavoute, D.~Gaponov, S.~F\'{e}vrier, G.~Jargot, M.~Hanna, W.~Boutu, and H.~Merdji, ``Orbital angular momentum from semiconductor high-order harmonics,'' {\em Opt. Lett.}, vol.~44, pp.~546--549, Feb 2019.

\bibitem{Qu2017}
K.~Qu, Q.~Jia, and N.~J. Fisch, ``Plasma $q$-plate for generation and manipulation of intense optical vortices,'' {\em Physical Review E}, vol.~96, no.~5, p.~053207, 2017.
\newblock PRE.

\bibitem{10.1063/1.870766}
S.~V. Bulanov, N.~M. Naumova, and F.~Pegoraro, ``{Interaction of an ultrashort, relativistically strong laser pulse with an overdense plasma},'' {\em Physics of Plasmas}, vol.~1, pp.~745--757, 03 1994.

\bibitem{10.1063/1.871619}
R.~Lichters, J.~Meyer‐ter‐Vehn, and A.~Pukhov, ``{Short‐pulse laser harmonics from oscillating plasma surfaces driven at relativistic intensity},'' {\em Physics of Plasmas}, vol.~3, pp.~3425--3437, 09 1996.

\bibitem{PhysRevE.74.046404}
T.~Baeva, S.~Gordienko, and A.~Pukhov, ``Theory of high-order harmonic generation in relativistic laser interaction with overdense plasma,'' {\em Phys. Rev. E}, vol.~74, p.~046404, Oct 2006.

\bibitem{Li_2020}
S.~Li, X.~Zhang, W.~Gong, Z.~Bu, and B.~Shen, ``Spin-to-orbital angular momentum conversion in harmonic generation driven by intense circularly polarized laser,'' {\em New Journal of Physics}, vol.~22, p.~013054, jan 2020.

\bibitem{shi2021electron}
Y.~Shi, D.~R. Blackman, and A.~Arefiev, ``Electron acceleration using twisted laser wavefronts,'' {\em Plasma Physics and Controlled Fusion}, vol.~63, no.~12, p.~125032, 2021.

\bibitem{Wang2019}
J.~W. Wang, M.~Zepf, and S.~G. Rykovanov, ``Intense attosecond pulses carrying orbital angular momentum using laser plasma interactions,'' {\em Nature Communications}, vol.~10, p.~5554, Dec 2019.

\bibitem{PhysRevLett.126.134801}
L.~Yi, ``High-harmonic generation and spin-orbit interaction of light in a relativistic oscillating window,'' {\em Phys. Rev. Lett.}, vol.~126, p.~134801, Mar 2021.

\bibitem{Aboushelbaya2019}
R.~Aboushelbaya, K.~Glize, A.~F. Savin, M.~Mayr, B.~Spiers, R.~Wang, J.~Collier, M.~Marklund, R.~M. G.~M. Trines, R.~Bingham, and P.~A. Norreys, ``Orbital angular momentum coupling in elastic photon-photon scattering,'' {\em Phys. Rev. Lett.}, vol.~123, p.~113604, Sep 2019.

\bibitem{Aboushelbaya2020}
R.~Aboushelbaya, K.~Glize, A.~F. Savin, M.~Mayr, B.~Spiers, R.~Wang, N.~Bourgeois, C.~Spindloe, R.~Bingham, and P.~A. Norreys, ``{Measuring the orbital angular momentum of high-power laser pulses},'' {\em Physics of Plasmas}, vol.~27, p.~053107, 05 2020.

\bibitem{Zhai2023}
Y.~Zhai, J.~Fan, H.~Qiao, T.~Zhou, J.~Wu, and Q.~Dai, ``The rotational doppler effect of twisted photons in scattered fields,'' {\em Laser \& Photonics Reviews}, vol.~17, no.~10, p.~2201022, 2023.

\bibitem{Hiroki-MINAGAWA2022}
H.~MINAGAWA, S.~YOSHIMURA, K.~TERASAKA, and M.~ARAMAKI, ``Analysis of azimuthal doppler shift of anisotropically absorbed laguerre-gaussian beam propagating in transverse flow,'' {\em Plasma and Fusion Research}, vol.~17, pp.~1401099--1401099, 2022.

\bibitem{Zhang2015}
X.~Zhang, B.~Shen, Y.~Shi, X.~Wang, L.~Zhang, W.~Wang, J.~Xu, L.~Yi, and Z.~Xu, ``Generation of intense high-order vortex harmonics,'' {\em Physical Review Letters}, vol.~114, no.~17, p.~173901, 2015.
\newblock PRL.

\bibitem{Xiaomei2016}
X.~Zhang, B.~Shen, Y.~Shi, L.~Zhang, L.~Ji, X.~Wang, Z.~Xu, and T.~Tajima, ``Intense harmonics generation with customized photon frequency and optical vortex,'' {\em New Journal of Physics}, vol.~18, no.~8, p.~083046, 2016.

\bibitem{Li2018}
S.~Li, B.~Shen, X.~Zhang, Z.~Bu, and W.~Gong, ``Conservation of orbital angular momentum for high harmonic generation of fractional vortex beams,'' {\em Optics Express}, vol.~26, pp.~23460--23470, Sep 2018.

\bibitem{Vieira2016a}
J.~Vieira, R.~Trines, E.~P. Alves, R.~A. Fonseca, J.~T. Mendonça, R.~Bingham, P.~Norreys, and L.~O. Silva, ``High orbital angular momentum harmonic generation,'' {\em Physical Review Letters}, vol.~117, no.~26, p.~265001, 2016.
\newblock PRL.

\bibitem{Nie2018}
Z.~Nie, C.-H. Pai, J.~Hua, C.~Zhang, Y.~Wu, Y.~Wan, F.~Li, J.~Zhang, Z.~Cheng, Q.~Su, S.~Liu, Y.~Ma, X.~Ning, Y.~He, W.~Lu, H.-H. Chu, J.~Wang, W.~B. Mori, and C.~Joshi, ``Relativistic single-cycle tunable infrared pulses generated from a tailored plasma density structure,'' {\em Nature Photonics}, vol.~12, pp.~489--494, Aug 2018.

\bibitem{Nie2020}
Z.~Nie, C.-H. Pai, J.~Zhang, X.~Ning, J.~Hua, Y.~He, Y.~Wu, Q.~Su, S.~Liu, Y.~Ma, Z.~Cheng, W.~Lu, H.-H. Chu, J.~Wang, C.~Zhang, W.~B. Mori, and C.~Joshi, ``Photon deceleration in plasma wakes generates single-cycle relativistic tunable infrared pulses,'' {\em Nature Communications}, vol.~11, p.~2787, Jun 2020.

\bibitem{Zhu2019}
X.-L. Zhu, M.~Chen, S.-M. Weng, P.~McKenna, Z.-M. Sheng, and J.~Zhang, ``Single-cycle terawatt twisted-light pulses at midinfrared wavelengths above 10 \textmu{}m,'' {\em Phys. Rev. Appl.}, vol.~12, p.~054024, Nov 2019.

\bibitem{Zhu2020}
X.-L. Zhu, S.-M. Weng, M.~Chen, Z.-M. Sheng, and J.~Zhang, ``Efficient generation of relativistic near-single-cycle mid-infrared pulses in plasmas,'' {\em Light: Science {\&} Applications}, vol.~9, p.~46, Mar 2020.

\bibitem{Zhu2021}
X.-L. Zhu, W.-Y. Liu, S.-M. Weng, M.~Chen, Z.-M. Sheng, and J.~Zhang, ``{Generation of single-cycle relativistic infrared pulses at wavelengths above 20 µm from density-tailored plasmas},'' {\em Matter and Radiation at Extremes}, vol.~7, p.~014403, 12 2021.

\bibitem{Huang2020}
C.-K. Huang, C.~Zhang, Z.~Nie, K.~A. Marsh, C.~E. Clayton, and C.~Joshi, ``Conservation of angular momentum in second harmonic generation from under-dense plasmas,'' {\em Communications Physics}, vol.~3, p.~213, Nov 2020.

\bibitem{Zhang2016}
L.~Zhang, B.~Shen, X.~Zhang, S.~Huang, Y.~Shi, C.~Liu, W.~Wang, J.~Xu, Z.~Pei, and Z.~Xu, ``Deflection of a reflected intense vortex laser beam,'' {\em Physical Review Letters}, vol.~117, no.~11, p.~113904, 2016.
\newblock PRL.

\bibitem{Gao2015}
W.~Gao, C.~Mu, H.~Li, Y.~Yang, and Z.~Zhu, ``{Parametric amplification of orbital angular momentum beams based on light-acoustic interaction},'' {\em Applied Physics Letters}, vol.~107, p.~041119, 07 2015.

\bibitem{Nuter2022}
R.~Nuter, P.~Korneev, and V.~T. Tikhonchuk, ``{Raman scattering of a laser beam carrying an orbital angular momentum},'' {\em Physics of Plasmas}, vol.~29, p.~062101, 06 2022.

\bibitem{Ji2023TBD}
Y.~Ji, C.-W. Lian, Y.~Shi, R.~Yan, S.~Cao, C.~Ren, and J.~Zheng, ``Generating axial magnetic fields via two plasmon decay driven by a twisted laser,'' {\em Phys. Rev. Res.}, vol.~5, p.~L022025, May 2023.

\bibitem{Mendonca2009a}
J.~T. Mendonca, S.~Ali, and B.~Thidé, ``Plasmons with orbital angular momentum,'' {\em Physics of Plasmas (1994-present)}, vol.~16, no.~11, p.~112103, 2009.

\bibitem{Mendonca2012}
J.~T. Mendonça, ``Twisted waves in a plasma,'' {\em Plasma Physics and Controlled Fusion}, vol.~54, no.~12, p.~124031, 2012.

\bibitem{Mendonca2012a}
J.~T. Mendonça, ``Kinetic description of electron plasma waves with orbital angular momentum,'' {\em Physics of Plasmas (1994-present)}, vol.~19, no.~11, p.~112113, 2012.

\bibitem{Mendonca2017}
J.~T. Mendonca and P.~S.~B. Joao, ``Twisted waves in a magnetized plasma,'' {\em Plasma Physics and Controlled Fusion}, vol.~59, no.~5, p.~054003, 2017.

\bibitem{Mendonca2017a}
J.T.Mendonça, ``Emission of twisted photons from quantum vacuum,'' {\em EPL (Europhysics Letters)}, vol.~120, no.~6, p.~61001, 2017.

\bibitem{Blackman2019a}
D.~R. Blackman, R.~Nuter, P.~Korneev, and V.~T. Tikhonchuk, ``Kinetic plasma waves carrying orbital angular momentum,'' {\em Physical Review E}, vol.~100, p.~013204, Jul 2019.

\bibitem{Blackman2019b}
D.~R. Blackman, R.~Nuter, P.~Korneev, and V.~T. Tikhonchuk, ``Twisted kinetic plasma waves,'' {\em Journal of Russian Laser Research}, vol.~40, pp.~419--428, Sep 2019.

\bibitem{Blackman2020}
D.~R. Blackman, R.~Nuter, P.~Korneev, and V.~T. Tikhonchuk, ``Nonlinear landau damping of plasma waves with orbital angular momentum,'' {\em Phys. Rev. E}, vol.~102, p.~033208, Sep 2020.

\bibitem{Blackman2022}
D.~R. Blackman, R.~Nuter, P.~Korneev, A.~Arefiev, and V.~T. Tikhonchuk, ``{Kinetic phenomena of helical plasma waves with orbital angular momentum},'' {\em Physics of Plasmas}, vol.~29, p.~072105, 07 2022.

\bibitem{Palastro2024}
J.~P. Palastro, K.~G. Miller, R.~K. Follett, D.~Ramsey, K.~Weichman, A.~V. Arefiev, and D.~H. Froula, ``Space-time structured plasma waves,'' {\em Phys. Rev. Lett.}, vol.~132, p.~095101, Feb 2024.

\bibitem{Shi2018}
Y.~Shi, J.~Vieira, R.~M. G.~M. Trines, R.~Bingham, B.~F. Shen, and R.~J. Kingham, ``Magnetic field generation in plasma waves driven by copropagating intense twisted lasers,'' {\em Physical Review Letters}, vol.~121, p.~145002, Oct 2018.

\bibitem{shiJUSTC}
Y.~Shi, D.~R. Blackman, R.~J. Kingham, and A.~Arefiev, ``Twisted plasma waves driven by twisted ponderomotive force,'' {\em JUSTC}, vol.~53, no.~1, p.~3, 2023.

\bibitem{Haines2001}
M.~G. Haines, ``Generation of an axial magnetic field from photon spin,'' {\em Physical Review Letters}, vol.~87, no.~13, p.~135005, 2001.
\newblock PRL.

\bibitem{Ali2010}
S.~Ali, J.~R. Davies, and J.~T. Mendonca, ``Inverse faraday effect with linearly polarized laser pulses,'' {\em Phys. Rev. Lett.}, vol.~105, p.~035001, Jul 2010.

\bibitem{Nuter2020}
R.~Nuter, P.~Korneev, E.~Dmitriev, I.~Thiele, and V.~T. Tikhonchuk, ``Gain of electron orbital angular momentum in a direct laser acceleration process,'' {\em Phys. Rev. E}, vol.~101, p.~053202, May 2020.

\bibitem{Longman2021}
A.~Longman and R.~Fedosejevs, ``Kilo-tesla axial magnetic field generation with high intensity spin and orbital angular momentum beams,'' {\em Phys. Rev. Res.}, vol.~3, p.~043180, Dec 2021.

\bibitem{Shi2023}
Y.~Shi, A.~Arefiev, J.~X. Hao, and J.~Zheng, ``Efficient generation of axial magnetic field by multiple laser beams with twisted pointing directions,'' {\em Phys. Rev. Lett.}, vol.~130, p.~155101, Apr 2023.

\bibitem{Liseykina2016}
T.~V. Liseykina, S.~V. Popruzhenko, and A.~Macchi, ``Inverse faraday effect driven by radiation friction,'' {\em New Journal of Physics}, vol.~18, p.~072001, jul 2016.

\bibitem{Valenzuela-Villaseca2023}
V.~Valenzuela-Villaseca, L.~G. Suttle, F.~Suzuki-Vidal, J.~W.~D. Halliday, S.~Merlini, D.~R. Russell, E.~R. Tubman, J.~D. Hare, J.~P. Chittenden, M.~E. Koepke, E.~G. Blackman, and S.~V. Lebedev, ``Characterization of quasi-keplerian, differentially rotating, free-boundary laboratory plasmas,'' {\em Phys. Rev. Lett.}, vol.~130, p.~195101, May 2023.

\bibitem{WuYipeng2023}
Y.~Wu, X.~Xu, C.~Zhang, Z.~Nie, M.~Sinclair, A.~Farrell, K.~A. Marsh, J.~Hua, W.~Lu, W.~B. Mori, and C.~Joshi, ``Efficient generation of tunable magnetic and optical vortices using plasmas,'' {\em Phys. Rev. Res.}, vol.~5, p.~L012011, Jan 2023.

\bibitem{Picon2010}
A.~Picón, A.~Benseny, J.~Mompart, J.~R.~V. de~Aldana, L.~Plaja, G.~F. Calvo, and L.~Roso, ``Transferring orbital and spin angular momenta of light to atoms,'' {\em New Journal of Physics}, vol.~12, no.~8, p.~083053, 2010.

\bibitem{lu2023prl-gammaOAM}
Z.-W. Lu, L.~Guo, Z.-Z. Li, M.~Ababekri, F.-Q. Chen, C.~Fu, C.~Lv, R.~Xu, X.~Kong, Y.-F. Niu, and J.-X. Li, ``Manipulation of giant multipole resonances via vortex $\ensuremath{\gamma}$ photons,'' {\em Phys. Rev. Lett.}, vol.~131, p.~202502, Nov 2023.

\bibitem{Jentschura2011}
U.~D. Jentschura and V.~G. Serbo, ``Generation of high-energy photons with large orbital angular momentum by compton backscattering,'' {\em Physical Review Letters}, vol.~106, p.~013001, Jan 2011.

\bibitem{Liu2016}
C.~Liu, B.~Shen, X.~Zhang, Y.~Shi, L.~Ji, W.~Wang, L.~Yi, L.~Zhang, T.~Xu, Z.~Pei, and Z.~Xu, ``Generation of gamma-ray beam with orbital angular momentum in the qed regime,'' {\em Physics of Plasmas}, vol.~23, no.~9, p.~093120, 2016.

\bibitem{Sasaki2008}
S.~Sasaki and I.~McNulty, ``Proposal for generating brilliant x-ray beams carrying orbital angular momentum,'' {\em Physical Review Letters}, vol.~100, p.~124801, Mar 2008.

\bibitem{Bahrdt2013}
J.~Bahrdt, K.~Holldack, P.~Kuske, R.~M\"uller, M.~Scheer, and P.~Schmid, ``First observation of photons carrying orbital angular momentum in undulator radiation,'' {\em Physical Review Letters}, vol.~111, p.~034801, Jul 2013.

\bibitem{Hemsing2011}
E.~Hemsing, A.~Marinelli, and J.~B. Rosenzweig, ``Generating optical orbital angular momentum in a high-gain free-electron laser at the first harmonic,'' {\em Physical Review Letters}, vol.~106, p.~164803, Apr 2011.

\bibitem{Hemsing2012}
E.~Hemsing and A.~Marinelli, ``Echo-enabled x-ray vortex generation,'' {\em Physical Review Letters}, vol.~109, p.~224801, 2012.

\bibitem{Hemsing2013}
E.~Hemsing, A.~Knyazik, M.~Dunning, D.~Xiang, A.~Marinelli, C.~Hast, and J.~B. Rosenzweig, ``Coherent optical vortices from relativistic electron beams,'' {\em Nature Physics}, vol.~9, no.~9, pp.~549--553, 2013.

\bibitem{RebernikRibic2017}
P.~Rebernik~Ribič, B.~Rösner, D.~Gauthier, E.~Allaria, F.~Döring, L.~Foglia, L.~Giannessi, N.~Mahne, M.~Manfredda, C.~Masciovecchio, R.~Mincigrucci, N.~Mirian, E.~Principi, E.~Roussel, A.~Simoncig, S.~Spampinati, C.~David, and G.~De~Ninno, ``Extreme-ultraviolet vortices from a free-electron laser,'' {\em Physical Review X}, vol.~7, no.~3, p.~031036, 2017.
\newblock PRX.

\bibitem{ELI-NP_gamma}
C.~A. Ur, ``{Gamma beam system at ELI-NP},'' {\em AIP Conference Proceedings}, vol.~1645, pp.~237--245, 02 2015.

\bibitem{ELI-NP_nuclear}
F.~Negoita, M.~Roth, P.~Thirolf, S.~Tudisco, F.~Hannachi, S.~Moustaizis, I.~Pomerantz, P.~Mckenna, J.~Fuchs, K.~Sphor, G.~Acbas, A.~Anzalone, P.~Audebert, S.~Balascuta, F.~Cappuzzello, M.~Cernaianu, S.~Chen, I.~Dancus, R.~Freeman, H.~Geissel, P.~Ghenuche, L.~Gizzi, F.~Gobet, G.~Gosselin, M.~Gugiu, D.~Higginson, E.~d'Humi{\^e}res, C.~Ivan, D.~Jaroszynski, S.~Kar, L.~Lamia, V.~Leca, L.~Neagu, G.~Lanzalone, V.~Meot, S.~Mirfayzi, I.~Mitu, P.~Morel, C.~Murphy, C.~Petcu, H.~Petrascu, C.~Petrone, P.~Raczka, M.~Risca, F.~Rotaru, J.~Santos, D.~Schumacher, D.~Stutman, M.~Tarisien, M.~Tataru, B.~Tatulea, I.~Turcu, M.~Versteegen, D.~Ursescu, S.~Gales, and N.~Zamfir, ``{Laser driven nuclear physics at ELI-NP},'' {\em Romanian Reports in Physics}, vol.~68, pp.~S37--S144, 2016.

\bibitem{Jentschura2011a}
U.~D. Jentschura and V.~G. Serbo, ``Compton upconversion of twisted photons: backscattering of particles with non-planar wave functions,'' {\em The European Physical Journal C}, vol.~71, p.~1571, Mar 2011.

\bibitem{Stock2015}
S.~Stock, A.~Surzhykov, S.~Fritzsche, and D.~Seipt, ``Compton scattering of twisted light: Angular distribution and polarization of scattered photons,'' {\em Physical Review A}, vol.~92, p.~013401, Jul 2015.

\bibitem{Petrillo2016}
V.~Petrillo, G.~Dattoli, I.~Drebot, and F.~Nguyen, ``Compton scattered x-gamma rays with orbital momentum,'' {\em Physical Review Letters}, vol.~117, no.~12, p.~123903, 2016.

\bibitem{JuLB2019}
L.~Ju, C.~Zhou, T.~Huang, K.~Jiang, C.~Wu, T.~Long, L.~Li, H.~Zhang, M.~Yu, and S.~Ruan, ``Generation of collimated bright gamma rays with controllable angular momentum using intense laguerre-gaussian laser pulses,'' {\em Phys. Rev. Appl.}, vol.~12, p.~014054, Jul 2019.

\bibitem{Zhu2018}
X.-L. Zhu, M.~Chen, T.-P. Yu, S.-M. Weng, L.-X. Hu, P.~McKenna, and Z.-M. Sheng, ``{Bright attosecond $\gamma$-ray pulses from nonlinear Compton scattering with laser-illuminated compound targets},'' {\em Applied Physics Letters}, vol.~112, p.~174102, 04 2018.

\bibitem{HuYT2021}
Y.-T. Hu, J.~Zhao, H.~Zhang, Y.~Lu, W.-Q. Wang, L.-X. Hu, F.-Q. Shao, and T.-P. Yu, ``{Attosecond $\gamma$-ray vortex generation in near-critical-density plasma driven by twisted laser pulses},'' {\em Applied Physics Letters}, vol.~118, p.~054101, 02 2021.

\bibitem{Taira2017}
Y.~Taira, T.~Hayakawa, and M.~Katoh, ``Gamma-ray vortices from nonlinear inverse thomson scattering of circularly polarized light,'' {\em Scientific Reports}, vol.~7, p.~5018, Jul 2017.

\bibitem{wang2023radiationreaction}
J.-Y. Wang, Q.~Zhao, M.~Ababekri, and J.-X. Li, ``Radiation-reaction effects on the production of twisted photon in the nonlinear inverse thomson scattering,'' {\em arXiv preprint arXiv:2312.05580}, 2023.

\bibitem{Chen2018}
Y.~Chen, J.~Li, K.~Z. Hatsagortsyan, and C.~H. Keitel, ``$\ensuremath{\gamma}$-ray beams with large orbital angular momentum via nonlinear compton scattering with radiation reaction,'' {\em Physical Review Letters}, vol.~121, no.~7, p.~074801, 2018.
\newblock PRL.

\bibitem{chen2019}
Y.-Y. Chen, K.~Z. Hatsagortsyan, and C.~H. Keitel, ``Generation of twisted $\gamma$-ray radiation by nonlinear thomson scattering of twisted light,'' {\em Matter and Radiation at Extremes}, vol.~4, no.~2, p.~024401, 2019.

\bibitem{Bliokh2015}
K.~Y. Bliokh, F.~J. Rodr{\'i}guez-Fortu{\~{n}}o, F.~Nori, and A.~V. Zayats, ``Spin--orbit interactions of light,'' {\em Nature Photonics}, vol.~9, pp.~796--808, Dec 2015.

\bibitem{Katoh2017}
M.~Katoh, M.~Fujimoto, N.~S. Mirian, T.~Konomi, Y.~Taira, T.~Kaneyasu, M.~Hosaka, N.~Yamamoto, A.~Mochihashi, Y.~Takashima, K.~Kuroda, A.~Miyamoto, K.~Miyamoto, and S.~Sasaki, ``Helical phase structure of radiation from an electron in circular motion,'' {\em Scientific Reports}, vol.~7, no.~1, p.~6130, 2017.

\bibitem{Katoh2017b}
M.~Katoh, M.~Fujimoto, H.~Kawaguchi, K.~Tsuchiya, K.~Ohmi, T.~Kaneyasu, Y.~Taira, M.~Hosaka, A.~Mochihashi, and Y.~Takashima, ``Angular momentum of twisted radiation from an electron in spiral motion,'' {\em Physical Review Letters}, vol.~118, no.~9, p.~094801, 2017.
\newblock PRL.

\bibitem{Zhang2014}
X.~Zhang, B.~Shen, L.~Zhang, J.~Xu, X.~Wang, W.~Wang, L.~Yi, and Y.~Shi, ``Proton acceleration in underdense plasma by ultraintense laguerre–gaussian laser pulse,'' {\em New Journal of Physics}, vol.~16, no.~12, p.~123051, 2014.

\bibitem{Vieira2014}
J.~Vieira and J.~Mendonca, ``Nonlinear laser driven donut wakefields for positron and electron acceleration,'' {\em Physical Review Letters}, vol.~112, no.~21, p.~215001, 2014.
\newblock PRL.

\bibitem{Mendonca2014}
J.~T. Mendonça and J.~Vieira, ``Donut wakefields generated by intense laser pulses with orbital angular momentum,'' {\em Physics of Plasmas (1994-present)}, vol.~21, no.~3, p.~033107, 2014.

\bibitem{Zhang2016a}
G.~Zhang, M.~Chen, C.~B. Schroeder, J.~Luo, M.~Zeng, F.~Li, L.~Yu, S.~Weng, Y.~Ma, T.~Yu, Z.~Sheng, and E.~Esarey, ``Acceleration and evolution of a hollow electron beam in wakefields driven by a laguerre-gaussian laser pulse,'' {\em Physics of Plasmas}, vol.~23, no.~3, p.~033114, 2016.

\bibitem{Zhang2016b}
G.~Zhang, M.~Chen, J.~Luo, M.~Zeng, T.~Yuan, J.~Yu, Y.~Ma, T.~Yu, L.~Yu, S.~Weng, and Z.~Sheng, ``Acceleration of on-axis and ring-shaped electron beams in wakefields driven by laguerre-gaussian pulses,'' {\em Journal of Applied Physics}, vol.~119, no.~10, p.~103101, 2016.

\bibitem{Pae_2020}
K.~H. Pae, H.~Song, C.-M. Ryu, C.~H. Nam, and C.~M. Kim, ``Low-divergence relativistic proton jet from a thin solid target driven by an ultra-intense circularly polarized laguerre–gaussian laser pulse,'' {\em Plasma Physics and Controlled Fusion}, vol.~62, p.~055009, mar 2020.

\bibitem{Camilla2023}
C.~Willim, J.~Vieira, V.~Malka, and L.~O. Silva, ``Proton acceleration with intense twisted laser light,'' {\em Phys. Rev. Res.}, vol.~5, p.~023083, May 2023.

\bibitem{Dong2023PoP}
H.~Dong, W.~P. Wang, J.~Z. He, Z.~Y. Shi, Y.~X. Leng, R.~X. Li, and Z.~Z. Xu, ``{Self-generated magnetic collimation mechanism driven by ultra-intense LG laser},'' {\em Physics of Plasmas}, vol.~30, p.~083101, 08 2023.

\bibitem{Wilson_2023}
T.~C. Wilson, Z.-M. Sheng, P.~McKenna, and B.~Hidding, ``Self-focusing, compression and collapse of ultrashort weakly-relativistic laguerre–gaussian lasers in near-critical plasma,'' {\em Journal of Physics Communications}, vol.~7, p.~035002, mar 2023.

\bibitem{BUSOLD201494}
S.~Busold, A.~Almomani, V.~Bagnoud, W.~Barth, S.~Bedacht, A.~Blažević, O.~Boine-Frankenheim, C.~Brabetz, T.~Burris-Mog, T.~Cowan, O.~Deppert, M.~Droba, H.~Eickhoff, U.~Eisenbarth, K.~Harres, G.~Hoffmeister, I.~Hofmann, O.~Jaeckel, R.~Jaeger, M.~Joost, S.~Kraft, F.~Kroll, M.~Kaluza, O.~Kester, Z.~Lecz, T.~Merz, F.~Nürnberg, H.~Al-Omari, A.~Orzhekhovskaya, G.~Paulus, J.~Polz, U.~Ratzinger, M.~Roth, G.~Schaumann, P.~Schmidt, U.~Schramm, G.~Schreiber, D.~Schumacher, T.~Stoehlker, A.~Tauschwitz, W.~Vinzenz, F.~Wagner, S.~Yaramyshev, and B.~Zielbauer, ``Shaping laser accelerated ions for future applications – the light collaboration,'' {\em Nuclear Instruments and Methods in Physics Research Section A: Accelerators, Spectrometers, Detectors and Associated Equipment}, vol.~740, pp.~94--98, 2014.
\newblock Proceedings of the first European Advanced Accelerator Concepts Workshop 2013.

\bibitem{Jeon2018}
C.~Jeon, S.~G. Lee, H.~W. Lee, J.~H. Sung, S.~K. Lee, I.~W. Choi, and C.~H. Nam, ``Towards the manipulation of relativistic laguerre-gaussian laser pulse,'' {\em Conference on Lasers and Electro-Optics}, p.~JTu2A.140, 2018.

\bibitem{Cai2003}
Y.~Cai, X.~Lu, and Q.~Lin, ``Hollow gaussian beams and their propagation properties,'' {\em Opt. Lett.}, vol.~28, pp.~1084--1086, Jul 2003.

\bibitem{Quinteiro2017}
G.~F. Quinteiro, F.~Schmidt-Kaler, and C.~T. Schmiegelow, ``Twisted-light--ion interaction: The role of longitudinal fields,'' {\em Phys. Rev. Lett.}, vol.~119, p.~253203, Dec 2017.

\bibitem{shi2021LGPRL}
Y.~Shi, D.~Blackman, D.~Stutman, and A.~Arefiev, ``Generation of ultrarelativistic monoenergetic electron bunches via a synergistic interaction of longitudinal electric and magnetic fields of a twisted laser,'' {\em Phys. Rev. Lett.}, vol.~126, p.~234801, Jun 2021.

\bibitem{shi_blackman_zhu_arefiev_2022}
Y.~Shi, D.~R. Blackman, P.~Zhu, and A.~Arefiev, ``Electron pulse train accelerated by a linearly polarized laguerre–gaussian laser beam,'' {\em High Power Laser Science and Engineering}, vol.~10, p.~e45, 2022.

\bibitem{blackman2022electron}
D.~R. Blackman, Y.~Shi, S.~R. Klein, M.~Cernaianu, D.~Doria, P.~Ghenuche, and A.~Arefiev, ``Electron acceleration from transparent targets irradiated by ultra-intense helical laser beams,'' {\em Communications Physics}, vol.~5, no.~1, p.~116, 2022.

\bibitem{Zaim2017}
N.~Za\"{\i}m, M.~Th\'evenet, A.~Lifschitz, and J.~Faure, ``Relativistic acceleration of electrons injected by a plasma mirror into a radially polarized laser beam,'' {\em Phys. Rev. Lett.}, vol.~119, p.~094801, Aug 2017.

\bibitem{Zaim2020}
N.~Za\"{\i}m, D.~Gu\'enot, L.~Chopineau, A.~Denoeud, O.~Lundh, H.~Vincenti, F.~Qu\'er\'e, and J.~Faure, ``Interaction of ultraintense radially-polarized laser pulses with plasma mirrors,'' {\em Phys. Rev. X}, vol.~10, p.~041064, Dec 2020.

\bibitem{Wang2018}
W.~P. Wang, C.~Jiang, B.~F. Shen, F.~Yuan, Z.~M. Gan, H.~Zhang, S.~H. Zhai, and Z.~Z. Xu, ``New optical manipulation of relativistic vortex cutter,'' {\em Physical Review Letters}, vol.~122, p.~024801, Jan 2019.

\bibitem{Baumann2018}
C.~Baumann and A.~Pukhov, ``Electron dynamics in twisted light modes of relativistic intensity,'' {\em Physics of Plasmas}, vol.~25, no.~8, p.~083114, 2018.

\bibitem{Hu2018}
L.~Hu, T.~Yu, Y.~Lu, G.~Zhang, D.~Zou, H.~Zhang, Z.~Ge, Y.~Yin, and F.~Shao, ``Dynamics of the interaction of relativistic laguerre-gaussian laser pulses with a wire target,'' {\em Plasma Physics and Controlled Fusion}, vol.~61, p.~025009, 2018.

\bibitem{Hu2018a}
L.~Hu, T.~Yu, H.~Li, Y.~Yin, P.~McKenna, and F.~Shao, ``Dense relativistic electron mirrors from a laguerre gaussian laser-irradiated micro-droplet,'' {\em Optics Letters}, vol.~43, pp.~2615--2618, Jun 2018.

\bibitem{guo2023suppression}
Y.~Guo, X.~Zhang, D.~Xu, X.~Guo, B.~Shen, and K.~Lan, ``Suppression of stimulated raman scattering by angularly incoherent light, towards a laser system of incoherence in all dimensions of time, space, and angle,'' {\em Matter and Radiation at Extremes}, vol.~8, no.~3, p.~035902, 2023.

\bibitem{Ghai2011}
D.~P. Ghai, ``Generation of optical vortices with an adaptive helical mirror,'' {\em Applied Optics}, vol.~50, pp.~1374--1381, Apr 2011.

\bibitem{arteaga2018}
J.~A. Arteaga, A.~Serbeto, K.~H. Tsui, and J.~T. Mendonça, ``Light spring amplification in a multi-frequency raman amplifier,'' {\em Physics of Plasmas}, vol.~25, no.~12, p.~123111, 2018.

\bibitem{piccardo2023broadband}
M.~Piccardo, M.~de~Oliveira, V.~R. Policht, M.~Russo, B.~Ardini, M.~Corti, G.~Valentini, J.~Vieira, C.~Manzoni, G.~Cerullo, {\em et~al.}, ``Broadband control of topological--spectral correlations in space--time beams,'' {\em Nature Photonics}, vol.~17, pp.~822--828, 2023.

\bibitem{Lin2024}
Q.~Lin, F.~Feng, Y.~Cai, X.~Lu, X.~Zeng, C.~Wang, S.~Xu, J.~Li, and X.~Yuan, ``Direct space--time manipulation mechanism for spatio-temporal coupling of ultrafast light field,'' {\em Nature Communications}, vol.~15, p.~2416, Mar 2024.

\bibitem{Vieira2018}
J.~Vieira, J.~T. Mendonça, and F.~Quere, ``Optical control of the topology of laser-plasma accelerators,'' {\em Physical Review Letters}, vol.~121, no.~5, p.~054801, 2018.
\newblock PRL.

\bibitem{LuísMartins2019}
J.~Lu{\'i}s~Martins, J.~Vieira, J.~Ferri, and T.~F{\"u}l{\"o}p, ``Radiation emission in laser-wakefields driven by structured laser pulses with orbital angular momentum,'' {\em Scientific Reports}, vol.~9, p.~9840, Jul 2019.

\bibitem{Ju2018}
L.~Ju, C.~Zhou, K.~Jiang, T.~Huang, H.~Zhang, T.~Cai, J.~Cao, B.~Qiao, and S.~Ruan, ``Manipulating the topological structure of ultrarelativistic electron beams using laguerre–gaussian laser pulse,'' {\em New Journal of Physics}, vol.~20, no.~6, p.~063004, 2018.

\bibitem{10.1117/12.623906}
A.~P. Sukhorukov and V.~V. Yangirova, ``{Spatio-temporal vortices: properties, generation and recording},'' in {\em Nonlinear Optics Applications} (M.~A. Karpierz, A.~D. Boardman, and G.~I. Stegeman, eds.), vol.~5949, p.~594906, International Society for Optics and Photonics, SPIE, 2005.

\bibitem{PhysRevA.86.033824}
K.~Y. Bliokh and F.~Nori, ``Spatiotemporal vortex beams and angular momentum,'' {\em Phys. Rev. A}, vol.~86, p.~033824, Sep 2012.

\bibitem{BLIOKH20151}
K.~Y. Bliokh and F.~Nori, ``Transverse and longitudinal angular momenta of light,'' {\em Physics Reports}, vol.~592, pp.~1--38, 2015.
\newblock Transverse and longitudinal angular momenta of light.

\bibitem{Hancock2021}
S.~W. Hancock, S.~Zahedpour, and H.~M. Milchberg, ``Mode structure and orbital angular momentum of spatiotemporal optical vortex pulses,'' {\em Phys. Rev. Lett.}, vol.~127, p.~193901, Nov 2021.

\bibitem{Jhajj2016}
N.~Jhajj, I.~Larkin, E.~W. Rosenthal, S.~Zahedpour, J.~K. Wahlstrand, and H.~M. Milchberg, ``Spatiotemporal optical vortices,'' {\em Phys. Rev. X}, vol.~6, p.~031037, Sep 2016.

\bibitem{Hancock:19}
S.~W. Hancock, S.~Zahedpour, A.~Goffin, and H.~M. Milchberg, ``Free-space propagation of spatiotemporal optical vortices,'' {\em Optica}, vol.~6, pp.~1547--1553, Dec 2019.

\bibitem{Hancock2024}
S.~W. Hancock, S.~Zahedpour, A.~Goffin, and H.~M. Milchberg, ``Spatiotemporal torquing of light,'' {\em Phys. Rev. X}, vol.~14, p.~011031, Feb 2024.

\bibitem{chong2020generation}
A.~Chong, C.~Wan, J.~Chen, and Q.~Zhan, ``Generation of spatiotemporal optical vortices with controllable transverse orbital angular momentum,'' {\em Nature Photonics}, vol.~14, no.~6, pp.~350--354, 2020.

\bibitem{PhysRevLett.126.243601}
K.~Y. Bliokh, ``Spatiotemporal vortex pulses: Angular momenta and spin-orbit interaction,'' {\em Phys. Rev. Lett.}, vol.~126, p.~243601, Jun 2021.

\bibitem{10.1093/nsr/nwab149}
C.~Wan, J.~Chen, A.~Chong, and Q.~Zhan, ``{Photonic orbital angular momentum with controllable orientation},'' {\em National Science Review}, vol.~9, p.~nwab149, 08 2021.

\bibitem{WAN20201334}
C.~Wan, J.~Chen, A.~Chong, and Q.~Zhan, ``Generation of ultrafast spatiotemporal wave packet embedded with time-varying orbital angular momentum,'' {\em Science Bulletin}, vol.~65, no.~16, pp.~1334--1336, 2020.

\bibitem{Wang:21}
H.~Wang, C.~Guo, W.~Jin, A.~Y. Song, and S.~Fan, ``Engineering arbitrarily oriented spatiotemporal optical vortices using transmission nodal lines,'' {\em Optica}, vol.~8, pp.~966--971, Jul 2021.

\bibitem{MazanovSugicAlonsoNoriBliokh+2022+737+744}
M.~Mazanov, D.~Sugic, M.~A. Alonso, F.~Nori, and K.~Y. Bliokh, ``Transverse shifts and time delays of spatiotemporal vortex pulses reflected and refracted at a planar interface,'' {\em Nanophotonics}, vol.~11, no.~4, pp.~737--744, 2022.

\bibitem{Hancock:21}
S.~W. Hancock, S.~Zahedpour, and H.~M. Milchberg, ``Second-harmonic generation of spatiotemporal optical vortices and conservation of orbital angular momentum,'' {\em Optica}, vol.~8, pp.~594--597, May 2021.

\bibitem{Gui2021}
G.~Gui, N.~J. Brooks, H.~C. Kapteyn, M.~M. Murnane, and C.-T. Liao, ``Second-harmonic generation and the conservation of spatiotemporal orbital angular momentum of light,'' {\em Nature Photonics}, vol.~15, pp.~608--613, Aug 2021.

\bibitem{PhysRevLett.127.273901}
Y.~Fang, S.~Lu, and Y.~Liu, ``Controlling photon transverse orbital angular momentum in high harmonic generation,'' {\em Phys. Rev. Lett.}, vol.~127, p.~273901, Dec 2021.

\bibitem{Chen:20}
J.~Chen, C.~Wan, A.~Chong, and Q.~Zhan, ``Subwavelength focusing of a spatio-temporal wave packet with transverse orbital angular momentum,'' {\em Opt. Express}, vol.~28, pp.~18472--18478, Jun 2020.

\bibitem{Rui:22}
G.~Rui, B.~Yang, X.~Ying, B.~Gu, Y.~Cui, and Q.~Zhan, ``Numerical modeling for the characteristics study of a focusing ultrashort spatiotemporal optical vortex,'' {\em Opt. Express}, vol.~30, pp.~37314--37322, Oct 2022.

\bibitem{Qiu_2019}
J.~Qiu, B.~Shen, X.~Zhang, Z.~Bu, L.~Yi, L.~Zhang, and Z.~Xu, ``Vortex beam of tilted orbital angular momentum generated from grating,'' {\em Plasma Physics and Controlled Fusion}, vol.~61, p.~105001, aug 2019.

\bibitem{Chen2022}
W.~Chen, X.~Zhang, D.~Xu, X.~Guo, and B.~Shen, ``Reflection of vortex beam from relativistic flying mirror,'' {\em Scientific Reports}, vol.~12, p.~12524, Jul 2022.

\bibitem{Zhang_2022}
X.~Zhang, L.~Zhang, and B.~Shen, ``Generation of isolated intense vortex laser with transverse angular momentum,'' {\em New Journal of Physics}, vol.~24, p.~113041, nov 2022.

\bibitem{LGzhang2022}
L.~Zhang, L.~Ji, and B.~Shen, ``Intense harmonic generation driven by a relativistic spatiotemporal vortex beam,'' {\em High Power Laser Science and Engineering}, vol.~10, p.~06000e46, 6 2022.

\bibitem{Wu:23}
Y.~Wu, Z.~Nie, F.~Li, C.~Zhang, K.~A. Marsh, W.~B. Mori, and C.~Joshi, ``Spatial and spatiotemporal vortex harmonics carrying controllable orbital angular momentum generated by plasma mirrors,'' in {\em Optica Nonlinear Optics Topical Meeting 2023}, p.~M2B.6, Optica Publishing Group, 2023.

\bibitem{Guo:23}
X.~Guo, L.~Zhang, X.~Zhang, and B.~Shen, ``Deflection of a reflected intense spatiotemporal optical vortex beam,'' {\em Opt. Lett.}, vol.~48, pp.~1610--1613, Apr 2023.

\bibitem{le2024selffocused}
M.~S. Le, G.~A. Hine, A.~Goffin, J.~P. Palastro, and H.~M. Milchberg, ``Self-focused pulse propagation is mediated by spatiotemporal optical vortices,'' {\em arXiv preprint arXiv:2403.04669}, 2024.

\bibitem{Lichaoyang2024}
{Zhaoyang, Li and Yuxin, Leng and Ruxin Li}, ``From pulse-front distortions of ultra-intense ultrashort lasers to group-velocity controls of x-shape optical wave-packets,'' {\em Laser \& Optoelectronics Progress}, vol.~61, no.~5, p.~0500001, 2024.

\bibitem{Vincenti2012}
H.~Vincenti and F.~Qu\'er\'e, ``Attosecond lighthouses: How to use spatiotemporally coupled light fields to generate isolated attosecond pulses,'' {\em Phys. Rev. Lett.}, vol.~108, p.~113904, Mar 2012.

\bibitem{Palastro2020}
J.~P. Palastro, J.~L. Shaw, P.~Franke, D.~Ramsey, T.~T. Simpson, and D.~H. Froula, ``Dephasingless laser wakefield acceleration,'' {\em Phys. Rev. Lett.}, vol.~124, p.~134802, Mar 2020.

\bibitem{Froula2018}
D.~H. Froula, D.~Turnbull, A.~S. Davies, T.~J. Kessler, D.~Haberberger, J.~P. Palastro, S.-W. Bahk, I.~A. Begishev, R.~Boni, S.~Bucht, J.~Katz, and J.~L. Shaw, ``Spatiotemporal control of laser intensity,'' {\em Nature Photonics}, vol.~12, pp.~262--265, May 2018.

\bibitem{Hernandez-Garcia2013}
C.~Hernández-García, A.~Picón, J.~San~Román, and L.~Plaja, ``Attosecond extreme ultraviolet vortices from high-order harmonic generation,'' {\em Physical Review Letters}, vol.~111, no.~8, p.~083602, 2013.
\newblock PRL.

\bibitem{Bliokh2007}
K.~Y. Bliokh, Y.~P. Bliokh, S.~Savel’ev, and F.~Nori, ``Semiclassical dynamics of electron wave packet states with phase vortices,'' {\em Physical Review Letters}, vol.~99, no.~19, p.~190404, 2007.
\newblock PRL.

\bibitem{Martin2003}
H.~Martin, ``Photon orbital angular momentum in astrophysics,'' {\em The Astrophysical Journal}, vol.~597, no.~2, p.~1266, 2003.

\bibitem{Thidé2011}
B.~Thidé, N.~M. Elias~II, F.~Tamburini, S.~M. Mohammadi, and J.~T. Mendonça, {\em Applications of Electromagnetic OAM in Astrophysics and Space Physics Studies}, ch.~9, pp.~155--178.
\newblock John Wiley \& Sons, Ltd, 2011.

\bibitem{Tamburini2011}
F.~Tamburini, B.~Thid{\'e}, G.~Molina-Terriza, and G.~Anzolin, ``Twisting of light around rotating black holes,'' {\em Nature Physics}, vol.~7, pp.~195--197, Mar 2011.

\bibitem{Stenzel2015}
R.~L. Stenzel and J.~M. Urrutia, ``Helicons in unbounded plasmas,'' {\em Phys. Rev. Lett.}, vol.~114, p.~205005, May 2015.

\bibitem{Iwo2016}
B.~Iwo and B.~Zofia, ``Gravitational waves carrying orbital angular momentum,'' {\em New Journal of Physics}, vol.~18, no.~2, p.~023022, 2016.

\bibitem{Tsujimura2021}
T.~I. Tsujimura and S.~Kubo, ``{Propagation properties of electron cyclotron waves with helical wavefronts in magnetized plasma},'' {\em Physics of Plasmas}, vol.~28, p.~012502, 01 2021.

\bibitem{Zhangsiming2022}
S.~Zhang, H.~Li, J.~He, S.~Wu, B.~Xu, and L.~Bai, ``{Scattering by a low-velocity charge with applied magnetic fields in a Laguerre–Gaussian beam},'' {\em Physics of Plasmas}, vol.~29, p.~123504, 12 2022.

\bibitem{Rop_2012}
R.~Rop, I.~A. Litvin, and A.~Forbes, ``Generation and propagation dynamics of obstructed and unobstructed rotating orbital angular momentum-carrying helicon beams,'' {\em Journal of Optics}, vol.~14, p.~035702, feb 2012.

\end{thebibliography}

\end{document}